\documentclass{jfm}

\usepackage{amsmath}
\usepackage{mathtools}
\usepackage[dvipsnames]{xcolor}
\usepackage{lineno}
\usepackage{booktabs}

\usepackage[plain]{algorithm}
\usepackage{algpascal}

\usepackage{siunitx}

\usepackage[colorlinks=true,linkcolor=blue,citecolor=blue]{hyperref}%

\usepackage{natbib}
\usepackage{subcaption}

    \makeatother
    
\usepackage{titlesec}
\titlespacing*{\section}{0pt}{0.7\baselineskip}{0.5\baselineskip}
\titlespacing*{\subsection}{0pt}{0.8\baselineskip}{0.5\baselineskip}

\begin{document}

\shorttitle{Dynamics of the Jet Wiping Process via Integral Models} 
\shortauthor{M. A. Mendez, A. Gosset, B. Scheid, M. Balabane, J.-M. Buchlin} 

\title{Dynamics of the Jet Wiping Process\\ via Integral Models}

\author
 {
 M.A. Mendez\aff{1}
  \corresp{\email{mendez@vki.ac.be}},
 A. Gosset\aff{2},
 B. Scheid\aff{3},\\
 M. Balabane\aff{4} \and 
  J.-M. Buchlin\aff{1}
  }

\affiliation
{
\aff{1}
von Karman Institute for Fluid Dynamics, \\Environmental and Applied Fluid Dynamics Department
\aff{2}
Naval and Industrial Engineering Department, University of A Coru\~{n}a
\aff{3}
Transfers, Interfaces and Processes (TIPs), Universit\'{e} libre de Bruxelles
\aff{4}
 Laboratoire Analyse, G\'{e}om\'{e}trie et Applications, Universit\'{e} Paris 13
}

\maketitle

\begin{abstract}
The jet wiping process is a cost-effective coating technique that uses impinging gas jets to control the thickness of a liquid layer dragged along a moving strip. This process is fundamental in various coating industries (mainly in hot-dip galvanizing) and is characterized by an unstable interaction between the gas jet and the liquid film that results in wavy final coating films. To understand the dynamics of the wave formation, we extend classic laminar boundary layer models for falling films to the jet wiping problem, including the self-similar integral boundary layer (IBL) and the weighted integral boundary layer (WIBL) models. Moreover, we propose a transition and turbulence model (TTBL) to explore modelling extensions to larger Reynolds numbers and to analyze the impact of the modelling strategy on the liquid film dynamics.
{The validity of the long-wave formulation was first analyzed on a simpler problem, consisting of a liquid film falling over an upward-moving wall, using Volume Of Fluid (VOF) simulations. This validation proved the robustness of the integral formulation in conditions that are well outside their theoretical limits of validity. Finally, the three models were used to study the response of the liquid coat to harmonic and non-harmonic oscillations and pulsations in the impinging jet}. The impact of these disturbances on the average coating thickness and wave amplitude is analyzed, and the range of dimensionless frequencies yielding maximum disturbance amplification is presented.
\end{abstract}

\section{\label{sec:level1}Introduction}

Integral boundary layer models for falling liquid films have been extensively used to study flow configurations that are encountered in many coating, chemical, heat and mass transfer processes. 

These models, also referred to as low dimensional models, reduce the number of variables governing the problem by eliminating the velocity and pressure fields from the full set of Navier-Stokes equations, thus describing the dynamics of the liquid film as a function of thickness and flow rate. 
{From the pioneering 2D formulations proposed by \citet{Ka1,Ka2,Shkadov1971} and extended by \cite{Ruyer-Quil2000}, to the three-dimensional models firstly proposed by \citet{Demekhin1985} and extended by \cite{SCHEID2006}, the literature on the topic is vast and discussed in various monographs
\citep{Kalliadasis2012,Alekseen1994,Hen-hongChang2002} and reviews \citep{Chang1996,Craster2009,Ruyer-Quil2014}.}

\textcolor{black}{The capability} of low dimensional models to describe the dynamics of the liquid interface has been largely demonstrated for the fundamental case of a gravity-driven isothermal film \citep{Alekseenko1985,Ruyer-Quil2000,Ruyer-Quil2002,SCHEID2006}, for which many experimental \citep{Liu1994,Liu1995,Alekseenko1996,Nosoko1996,Tihon2006,DIETZE2009,Mendez2017} and numerical \citep{Salamon1994,Gao2003,Nosoko2004,Malamataris2008,Meza2008,Doro2013,DIETZE2008,Dietze2014} investigations have been carried out. Because of their minor computational cost, if compared to full simulations, and because of the analytic insights they enable, these models have been largely used in more complex configurations. {These include, among others, liquid films in the presence of interface shear stress \citep{Samanta2014,Frank2008,Frank2006,Gatapova2008,Vellingiri2013, Lavalle2017}, for which the linear stability analysis based on the full Orr-Sommerfeld equations is presented by \cite{Lavalle2018}.

Moreover, despite the restrictive hypotheses in their derivation, low order formulations have been successfully validated with experimental and numerical data in operating conditions that are well outside their theoretical range of validity \citep{Denner2018}. This has made integral model reliable tools to explore complex phenomena such as the origin of capillary ripples \citep{Dietze2016}, {the onset of circulating waves
and flow reversal in liquid films \citep{Rohlfs2014}, the effect of co-flowing turbulent gas on the interface dynamics} \citep{Vellingiri2013}, or the formulation of active feedback flow control methods to suppress interface instabilities \citep{Thompson2015,Thompson2016,Tomlin2019}.

This work extends the classical integral boundary layer models for liquid films to a flow configuration {for which these have never been used: the jet wiping process}. This process consists in using an impinging gas jet to control the thickness of a coating film on a {vertical} moving substrate, and it is characterized by an unstable dynamics \citep{ECS2017}, recently investigated experimentally by the authors \citep{Gosset2019,Mendez2019}. In particular, it has been shown that instabilities on the gas jet can propagate to the impinged liquid and produce a nonuniform coating distribution referred to as \emph{undulation}. Although several working hypotheses have been proposed, the mechanisms through which unsteadiness in the jet propagates to the liquid film are still not fully understood and are explored in this work.

{The modelling of this configuration presents two distinctive features that have not been considered in the liquid film modelling literature: 1) the upward motion of the vertical substrate and 2) the simultaneous presence of time-dependent sources of shear stress and pressure gradient.} Simplified theoretical modeling of the jet wiping have been proposed by \citet{Thornton1976,Tuck1983,Tuck1984,Tu1986,Ellen1984}, and \citet{Buchlin}; experimental and numerical validation have been provided by \citet{Lacanette2006} and \citet{Gosset2007}. These first formulations aimed at describing the mean thickness distribution of the liquid film under the action of the pressure gradient and the shear stress produced by the jet impingement, and thus at predicting the final coating thickness as a function of all the operating parameters. 

The first works discussing the stability of the problem have been presented by \citet{TUSTAB} and  \citet{Tuck1983}. The first presented a linear stability analysis; the second discussed the possible evolution of kinematic waves on the liquid coat. Since then, most of the investigations on the process have been based on high fidelity numerical simulations, combining Large Eddy Simulation (LES) of the gas jet with Volume of Fluid (VOF) treatment of the liquid film \citep{Myrillas2009,Myrillas2013,Pfeiler2017,Esl2017,Aniszewski2019}. While these simulations can potentially provide a complete picture of the unstable interaction between the jet and the gas flow, their computational cost remains prohibitively large for analyzing configurations of industrial interest, as  discussed by \citet{Aniszewski2019}. 


A theoretical analysis of the stability of the process has been proposed by \citet{Hocking2010}, who used a quasi-steady formulation to study the evolution of liquid film disturbances. Hocking and coworkers concluded that the coating film is neutrally stable and incapable of producing the undulation patterns observed in the wiping lines without the presence of disturbances produced by the gas jet. Using the same quasi-steady formulation, \citet{Johnstone2019} investigated the response of the liquid coat to a set of possible unsteady behavior of the impinging jet. The formulation presented in these works neglects the role of inertia in the liquid film, disregarding the nonlinear contribution of advection. 

In this work, the extension of more advanced integral models to the jet wiping process is used to study the dynamic response of the liquid film to various disturbances on the gas jet. These include localized perturbation, simulated by pulsation of the wiping actuators, and various kinds of harmonic and nonharmonic oscillations.
The general form of these models is presented in section \ref{Model}, while section \ref{Laminar}
and section \ref{TBL} provide the details of the laminar and turbulent models. Appendix \ref{A1} provides complementary material for the full derivation of the models. 

Among the laminar models in section \ref{Laminar}, this work covers the zero-order (ZO) formulation of the jet wiping, the extension of Integral Boundary Layer model from  Kapitza-Skhadov (IBL, \citealp{Shkadov1971,Shkadov2017}) and the extension of the Weighted IBL from C. Ruyer-Quil and P. Manneville (WIBL, \citealp{Ruyer-Quil2000,Ruyer-Quil2002}). The proposed transition and turbulence model in section \ref{TBL} combines ideas from mixing length formulation \citep{DRIEST1956,King1966,Geshev2014} and shallow-water formulations \citep{James2019,Vita2020}.
 
Section \ref{JETU} reviews the implementation of the wiping actuators in the models. Section \ref{NUM} presents the numerical methods, including the Finite Volume (FV) solver implemented to validate the integral models (in \ref{SOLV1}) and the DNS simulations using OpenFoam (in \ref{OPEN}) that were implemented to set up simple validation test cases. The results are presented in section \ref{Res}, including the numerical validation (in \ref{RES_A}), the relative weight of all the forces governing the process (in \ref{RES_B}), and the frequency response of the coating film (in \ref{RES_C}). Finally, the impact of the modelling strategy on the identified wave formation mechanisms is discussed in section \ref{RES_D}. Conclusions and perspectives are presented in section \ref{Conclu}.

\section{The Integral Formulation for the Jet Wiping}\label{Model}

The jet wiping process is represented schematically in figure \ref{Fig1}. A liquid film is dragged along a vertical plate moving upward at a constant velocity $U_p$ and is impinged upon by a gas jet. The configuration is assumed two-dimensional, with incompressible liquid flow bounded by the plate at $y=0$, and the dynamic liquid interface at $y=h(x,t)$.

\begin{figure}
\centering
\includegraphics[width=8.2cm]{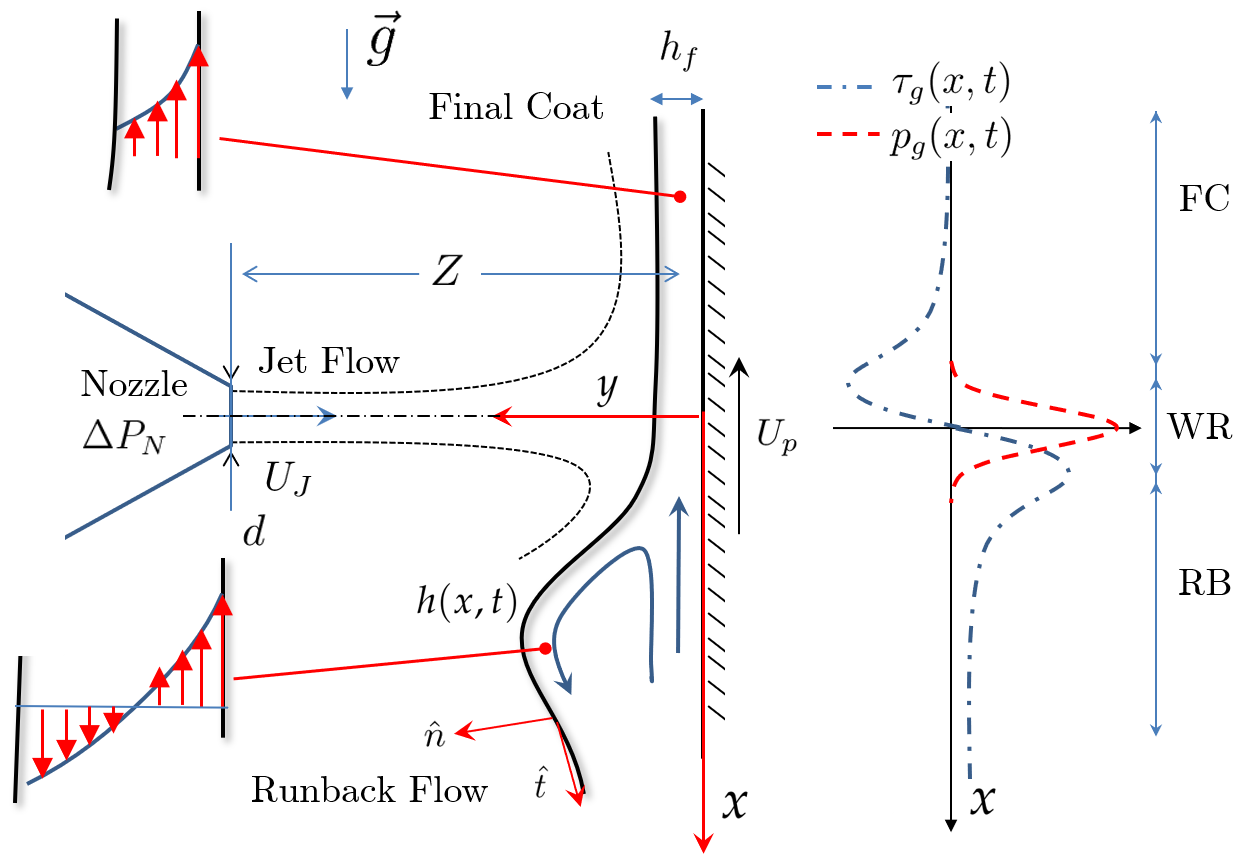}
\caption{Schematic of the jet wiping process: a nozzle with an opening $d$ and a stagnation pressure $\Delta P_N$ releases a jet flow at a distance $Z$ from a dip-coated substrate moving at a speed $U_p$. {The impingement produces a wiping meniscus (region WR). This forces a run-back film to flow backward (region RB) and leaves a thinner liquid film downstream (region FC) before solidification takes place. } }
\label{Fig1}
\end{figure}

The origin $x=0$ is located at the nozzle axis, and the streamwise coordinate $x$ is oriented in the direction of gravity and counter to the substrate velocity. The impinging jet flow produces a pressure $p_g(x,t)$ and a shear stress distribution $\tau_g(x,t)$ that identify three areas, qualitatively pictured in Figure \ref{Fig1} on the right. In the wiping region (indicated as WR), the pressure gradient imposed by the gas jet forces part of the liquid to reverse direction resulting in a wiping meniscus. The falling liquid forms the runback flow in the region $x\rightarrow \infty$ {(indicated as RB)}; the remaining liquid evolves upward in the final coating region $x\rightarrow -\infty$ {(indicated as FC)}. 
 
In a one-way coupling formulation, it is assumed that the presence of the liquid film does not influence the gas jet. {This assumption has been extensively validated for the prediction of the averaged final coating thickness \citep{Lacanette2006,Gosset2007}, but it is certainly not able to simulate the complex interaction between the two flows analyzed by \cite{Gosset2019,Mendez2019}. In this work, this formulation is used to de-couple the dynamics of the liquid from the one of the gas jet and to analyze the liquid film frequency response and possible mechanisms of undulation formation.} We thus assume that both the pressure and the shear stress produced by the jet solely depends on the nozzle gauge stagnation pressure $\Delta P_N$, the nozzle opening $d$, its discharge coefficient $C_d$ and the standoff distance $Z$. This dependency is described in section \ref{JETU}.  

All the integral models investigated in this work rely on the Navier-Stokes equation and the related boundary conditions in the `long-wave' formulation. This formulation is derived by scaling the cross streamwise direction with a reference length $[h]$, which is much smaller than the streamwise reference length $[x]$. Appendix \ref{A1} summarizes this derivation and presents the rationale behind the choice of the reference quantities used to scale the problem. For conciseness, these reference quantities are listed in table \ref{Scaling_Table}, and the focus is here kept on the derivation of the various integral models. {In what follows, dimensionless variables scaled with respect to the quantities in table \ref{Scaling_Table} are indicated with a hat (e.g. $\hat{h}=h/[h]$).}

In the long-wave formulation of the problem, the dimensionless continuity and the momentum equation in the $x$ and the $y$ directions reduce to the boundary layer equations:

\begin{subequations}
\begin{align}
\partial_{\hat{x}} \hat{u}+ \partial_{\hat y} \hat{v}&=0\,,
\label{C}\\
\varepsilon\,Re\Bigl(\partial_{\hat t}\hat{u}+\hat {u}\, \partial_{\hat x} \hat{u} + \hat {v}\, \partial_{\hat y} \hat{u}\Bigr) &=- \partial_{\hat x}\, \hat{p}_l+\partial_{\hat y \hat y} \hat{u} +1\label{Mx}\,,\\
0&= \partial_{\hat y } \hat{p}_l\label{My}\,,
\end{align}
\end{subequations}
where $\hat{p}_l$ is the pressure in the liquid, $\hat{u}$ and $\hat{v}$ are the streamwise and cross-stream velocity components, $\varepsilon=[h]/[x]=Ca^{1/3}$ is the film parameter, with $Ca=\mu_l\,U_p/\sigma$ the capillary number and $Re=[u][h]/\nu_l=(U_p^3/g\nu_l)^{1/2}$ is the \emph{global} Reynolds number of the process, to be distinguished form other Reynolds numbers that will be introduced later.

\begin{table}
\small\addtolength{\tabcolsep}{1.5pt}
\renewcommand{\arraystretch}{1.15}
\centering
\begin{tabular}{c|c|c}
Reference Quantity & Definition & Expression \\
$[h]$ & $(\nu_l\,[u]/g)^{1/2}$ & $(\nu_l\,U_p/g)^{1/2}$ \\
$[x]$ & $[h]/\varepsilon$ & $(\nu_l\,U_p/g)^{1/2}\,Ca^{-1/3}$ \\
$[u]$ & $U_p$ & $U_p $ \\
$[v]$ & $\varepsilon U_p$ & $U_p\,Ca^{1/3} $ \\
$[p]$ & $\rho_l\,g\,[x]$ & $(\mu_l\,\rho_l\,g\,U_p)^{1/2}\, Ca^{-1/3}$\\
$[\tau]$ & $\mu_l\,[u]/[h]$ & $(\mu_l\,\rho_l\,g\,U_p)^{1/2}$\\
$[t]$ & $[x]/[u]$ &  $(\nu_l/U_p\,g)^{1/2}\,Ca^{-1/3}$
\end{tabular}
\caption{Reference quantities for the Shkadov-like scaling, for which $\varepsilon=Ca^{1/3}$, with $Ca=\mu_l\,U_p/\sigma$ the capillary number.}
\label{Scaling_Table}
\end{table}

Since the proposed scaling laws hold for $\varepsilon\ll1$, the long wavelength formulation presented in this work is valid for $Ca^{1/3}\ll1$. This generally occurs in galvanizing conditions, where typically $\mu_l\approx 0.003$\,\SI{}{Pa \cdot s}, $\sigma\approx 0.8$\,\SI{}{N/m}, and $U_p=1-2$ \SI{}{m/s} and hence $Ca\approx 0.004-0.008$. The formulation proposed in this work is also valid for the experimental conditions encountered in the Essor laboratory (see \cite{Buchlin}) developed at the von Karman Institute (VKI), operating with water ($\mu_l\approx 0.001$\,\SI{}{Pa \cdot s}, $\sigma\approx 0.07$\,\SI{}{N/m} at $U_p=0.2-2$ \SI{}{m/s} (i.e. $Ca=0.003-0.03$). On the other hand, the experiment recently conducted in the VKI Ondule laboratory \citep{Gosset2019,Mendez2019} using Dipropilene glycole ($\mu_l \approx 0.1$\,\SI{}{Pa\cdot s}, $\sigma \approx 0.03$\,\SI{}{N/m} at $U_p=0.2-0.4$ \SI{}{m/s}, i.e. $Ca\approx 0.6-1.3$), and hence requires a different scaling strategy. While the previous experimental work of the authors has focused on the dynamics of very viscous flows (see also \citealt{Mendez2017}), this work focuses on the low $Ca$ limit, in which surface tension plays a more important role.

The kinematic boundary conditions at the wall and at the gas-liquid interface set:

\begin{equation}
\label{BCKin}
\begin{cases}
\hat{\textbf{v}}=(\hat u,\hat v)=(-1,0)  &\mbox{in } \hat{y}=0\,, \\
\hat v= \partial_{\hat t} {h}+\hat{u}\, \partial_{\hat x} \hat h & \mbox{in } \hat{y}=\hat{h}. \end{cases} 
\end{equation} 

The dynamic boundary conditions (see \eqref{NN} and \eqref{TT}) at the interface simplify to:

\begin{subequations}
\begin{align}
\hat p_l\Bigl|_{\hat h}=\hat p_g(\hat x,\hat t)-\partial_{\hat x \hat x}\hat h \quad \mbox{along } \hat{\mathbf{n}}\quad \mbox{in} \quad \hat{y}=\hat{h}\,,\label{P_BC} \\
\partial_{y} \hat u\Bigl|_{\hat h}=\hat{\tau}_g (\hat x,\hat t) \quad \mbox{along } \hat{\mathbf{t}} \quad \mbox{in} \quad \hat{y}=\hat{h} \, \label{Tau_BC}
\end{align}
\end{subequations} 

All integral models reduce the modeling complexity by rendering the problem 1D.
Integrating \eqref{C},\eqref{Mx},\eqref{My} across the film thickness using Leibniz integral rule together with the boundary conditions \eqref{P_BC}-\eqref{Tau_BC} gives:

\begin{subequations}
\label{INT_GEN}
\begin{align}
\partial_{\hat t} \hat h+\partial_{\hat x }\hat q&=0\,, \label{I1}\\
\varepsilon Re \biggl(\partial_{\hat t} \hat q+ \partial_{\hat x} {\mathcal{F}}\biggr)&=\hat h\biggl (1-\partial_{\hat x} \hat{p}_g+\partial_{\hat x\hat x\hat x}\hat h\biggr) +{\Delta \hat \tau}\,, \label{I2}
\end{align}
\end{subequations} where $\hat q$ is the volumetric flow rate per unit width, and

\begin{equation}
\label{DEFs}
\mathcal{F}=\int_0^{{\hat h}}\hat{u}^2\,d{\hat y} \quad \quad \mbox{and} \quad\quad\Delta \hat \tau  \equiv \hat{\tau}_g-\hat{\tau}_w=\hat{\tau}_g-\partial_{\hat y} \hat u\Bigl|_{\hat y=0}\,,
\end{equation} are the advection and the shear stress terms respectively and $\hat{\tau}_w$ is the wall shear stress. To determine the functional forms of these (and thus to close integral models), some assumptions on the velocity profile are required. 

In this work, we assume that the velocity profile is the superposition of three terms:

\begin{equation}
    \label{U_Three}
    \hat{u}(\hat{x},\hat{y},\hat{t})=\hat{u}_{F}(\hat{x},\hat{y},\hat{t})+\hat{u}_{C}(\hat{x},\hat{y},\hat{t})+\hat{u}_{P}=\hat{u}_{F}(\hat{x},\hat{y},\hat{t})+\hat{\tau}_g(\hat{x},\hat{t})\hat y -1\,.
\end{equation}

The term $\hat{u}_{F}$ accounts for the contributions of gravity, viscous stresses, surface tension and pressure gradient. The term $\hat{u}_{C}=\hat{\tau}_g\hat y$ accounts for the shear stress produced at the gas liquid interface while $\hat{u}_{P}=-1$ accounts for the motion of the substrate. This kinematic decomposition satisfies the boundary conditions if $\hat{u}_{F}=0$ at $\hat{y}=0$ and $\partial_{\hat{y}}\hat{u}_{F}=0$ at $\hat{y}=\hat{h}$. 
For later convenience, it is interesting to identify the reference velocity and the associated Reynolds number for each of the three contributions. The flow rate per unit width can be split accodringly as

\begin{equation}
\label{q_Three}
    \hat q\equiv\int_0^{\hat h} \hat u \, d \hat y=\hat{q}_F+\hat{q}_C+\hat{q}_P=\hat{q}_F+\frac{1}{2}{\hat{\tau}_g \hat {h}^2}-\hat{h}\,,
\end{equation} from which the associated \emph{local} {(i.e. function of $\hat{x}$)} Reynolds numbers are:

\begin{equation}
\label{Reynoldss}
Re_F=\frac{q_F}{\nu}=|\hat{q}_F| Re\quad;\quad Re_{\tau}=\frac{1}{2}{\hat h^2 |\hat{\tau}_g}| Re \quad;\quad  Re_h=\hat{h} Re .
\end{equation}

It is worth noticing that the term $\hat{q}_F$ corresponds to a falling film flow in the absence of the other terms, but it can eventually lead to a negative contribution $\hat{q}_F<0$ in a strongly dominated shear stress flow (if  $\hat{h}^2 |\hat{\tau}_g|\gg |\hat{q}_F|$) as it is the case in the run-back flow region (more about this in section \ref{Tau_CLO}).
The models developed in the following sections only differ in the treatment of this term.

\section{Laminar Film Models}\label{Laminar}

The term $\hat{u}_{F}$ in \eqref{U_Three} is decomposed in a series of basis functions as 

\begin{equation}
\label{Vel_PROF}
\hat{u}_{F}=\sum^{N}_{j=0} a_j(\hat{x},\hat{t})\,f_j\Biggl(\frac{\hat y}{\hat h(\hat x,\hat t)}\Biggr)\,,
\end{equation} {where $a_0$ is of order $\mathcal{O}(1)$ and $a_j$ with $j>0$ are corrections of order $\mathcal{O}(\varepsilon)$}. Following \citet{Ruyer-Quil2000,Ruyer-Quil2002,Ruyer-Quil2014}, the basis functions $f_j$ are taken as:

\begin{equation}
f_j=\overline{y}^{j+1}-\frac{j+1}{j+2}\,\overline{y}^{j+2}\,\,,
\end{equation} where $\overline{y}=\hat{y}/\hat{h}(x,t)$ is the reduced coordinate. This choice was introduced for falling liquid films, imposing that each of the basis function satisfies the boundary conditions. This enables reduced-order models based on Galerkin projections \citep{Kalliadasis2012}. The flow rate per unit width, highlighting the contributions in \eqref{q_Three}, becomes:

\begin{equation}
\label{Q}
\hat q_F\equiv\int_0^{\hat h} \hat u_F \, d \hat y=\sum^{N}_{j=0}\,\frac{2}{(j+2)(j+3)}a_j.
\end{equation}

In all the laminar models presented in this work, valid at $\mathcal{O}(\varepsilon)$, only the first term ($j=0$) contributes to the advection term $\mathcal{F}$, since this is multiplied by $\varepsilon$ in \eqref{INT_GEN}. The advection term, using \eqref{U_Three} and \eqref{Vel_PROF}, reads: 

\begin{equation}
\label{adve}
\hat{\mathcal{F}}\equiv\int^{\hat{h}}_{0} \hat{u}^{2} d\hat y =\frac{1}{3} \hat{\tau}^{2}_g \hat h^3+\frac{5}{12} \hat{\tau}_g a_0  \hat {h}^2 -{\hat{\tau}_g} h^2 +\frac{2}{15} h\,{a^{2}_0}-\frac{2 }{3}a_0\,\hat h+\hat h. 
\end{equation}

The shear stress contribution is solely linked to the first coefficient $a_0$ regardless of the number of terms included in the expansion of the velocity profile. The shear stress term, using \eqref{U_Three} and \eqref{Vel_PROF}, reads: 

\begin{equation}
\label{D_Tau}
\Delta \hat \tau\equiv\hat{\tau}_g-\partial_{\hat y} \hat u\Bigl|_{\hat y=0}=-\hat{\tau}_{wF}=-\frac{a_0 }{\hat h }\,,
\end{equation} where $\hat{\tau}_{wF}$ is the wall shear stress produced by the $\hat{u}_F$ portion of the velocity profile.

Before presenting the derivation of the complete Weighted Integral Boundary Layer (WIBL) model at $\mathcal{O}(\varepsilon)$, it is worth introducing the Zero-Order (ZO) formulation and the Integral Boundary Layer (IBL) formulation.

\subsection{Zero-Order (Inertialess) Formulation (ZO)}\label{QS}

This model is based on two assumptions. First, only the first $j=0$ term of the velocity profile is relevant, that is $N=0$ in \eqref{Vel_PROF}. Second, the inertial effects can be neglected, that is $\varepsilon Re\sim 0$: the LHS in \eqref{I2} vanishes. The velocity profile is parabolic:

\begin{equation}
\label{uQS}
\hat u=a_0(\hat x,\hat t)\,\Bigl[\Bigl(\frac{\hat y}{\hat h}\Bigr)-\frac{1}{2}\Bigl( \frac{\hat y}{\hat h}\Bigr)^2\Bigr]+\hat \tau_g\,\hat y-1\,,
\end{equation} and the coefficient $a_0$ can be derived from the flow rate definition: 
\begin{equation}
\label{qQS}
 \hat q\equiv\int_0^{ \hat h}\, \hat u\,d \hat y=\frac{a_0\, \hat h}{3}+\frac{ \hat h^2\, \hat \tau_g}{2}- \hat h\,\rightarrow a_0=\frac{3 \hat{q}_F}{\hat h}= \frac{3\,\hat q}{\hat h}-\frac{3}{2}\,\hat h\,\hat \tau_g+3\,.
\end{equation} 

Using \eqref{qQS} in \eqref{D_Tau}, the shear stress term becomes

\begin{equation}
\label{DeltaT}
\Delta\,\hat \tau\equiv  -\hat{\tau}_{wF}=\frac{3}{2} \hat\tau_g-\frac{3\, \hat q}{ \hat h^2}-\frac{3}{ \hat h}\,.
\end{equation}

Introducing this into \eqref{I2}, and recalling that the LHS is set to zero, the flow rate is:

\begin{equation}
\label{Q_SIMPLE}
 \hat q=\frac{ \hat h^3}{3}\,\Bigl(1-\partial_{ \hat x}\,{\hat p}_g+\partial_{ \hat x \hat x  \hat x} \hat  h \Bigr)+\frac{1}{2} \hat \tau_g\, \hat h^2\,-\hat h.
\end{equation}

Introducing \eqref{Q_SIMPLE} in \eqref{I1} yields a single equation governing the film dynamics:

\begin{equation}
\label{Benney}
\partial_{ \hat t} \, \hat h+\partial_{\hat  x}\Bigl[\frac{\hat h^3}{3}\,\Bigl(1-\partial_{\hat x}\,\hat p_g+\partial_{\hat x\hat x\hat x} \hat h \Bigr)+\frac{1}{2}\hat \tau_g\,\hat h^2\,-\hat h\Bigr]=0\,.
\end{equation}

 This model has been widely used in the literature of the jet wiping for linear stability analysis \citep{Tu1986,Tuck1983,AnneThesis} or sensitivity studies similar to those performed in this work: \citet{Hocking2010} used this formulation to study the evolution of liquid disturbances for an ideally stationary jet; \citet{Johnstone2019} used it to study the response of the liquid film to an oscillating jet.

Since this work focuses on more advanced formulations, the time dependent simulation of the ZO model is not investigated further. It is nevertheless interesting to use this model to illustrate the basic features of a steady-state solution and the propagation of small flow disturbances, for which one could expect inertia to play a negligible role. In steady-state conditions ($\partial_{ \hat t}\hat h=-\partial_{\hat x}\hat q=0$), neglecting the contribution of the surface tension term $\partial_{ \hat x \hat x \hat x} \hat h=0$ (which is known to have little influence on the final thickness in case of strong wiping, as shown by \citealt{Yoneda, Tuck1984,Buchlin}), \eqref{Q_SIMPLE} reduces to a cubic polynomial in $\hat{h}$.

At each location $\hat x$ (hence given $\partial_{\hat {x}} \hat{p}_g (\hat{x}),\hat{\tau}_g(\hat{x})$), and for a given flow rate $\hat {q}<0$, this polynomial admits a negative solution (of no interest) and two positive solutions $\hat h^{+}( \hat x)$ and $\hat h^{-}( \hat x)$.
 These branches of the positive solution give the liquid thickness in the final coating region $\hat h^{-}( \hat x)\rightarrow h_f$ for $\hat x \rightarrow -\infty$ and the run back region $\hat h^{+}( \hat x)\rightarrow h_R$ for $\hat  x \rightarrow \infty$. The admissible flow rate $\hat {q}$ can thus be computed by imposing that the two branches of solutions meet (see \citealt{Hocking2010,Tuck1984}) in a critical point $\hat{x}=\hat{x}_c$ to form a continuous thickness profile:

\begin{equation}
\hat h(\hat{x})=
\begin{cases}
\hat{h}^{+}(\hat{x}) \quad \mbox{for } \hat{x}\leq\hat{x}_c \\
\hat{h}^{-}(\hat{x}) \quad \mbox{for } \hat{x}>\hat{x}_c
\end{cases} 
\end{equation}

A simple estimation of the final thickness can be obtained under the assumption that the system operates in optimal conditions (that is $\partial_{ \hat h}{\hat q}=0$), assuming that the maximum pressure gradient and shear stress act at the same location $\hat x^{*}$. From \eqref{Q_SIMPLE}, this gives:

\begin{equation}
\partial_{ \hat h}{\hat q}(\hat {h}^{*})=\hat {h}^{*2}\bigl(1-\partial_{\hat  x}\hat p_{g}^{*}\bigr)+\hat \tau_g^{*}\, \hat h^{*}-1=0\,,
\end{equation}

and thus the film thickness in this location is:

\begin{equation}
\label{0D}
\hat h^{*}=\frac{-\hat{\tau}_g^{*}+\sqrt{{\hat \tau_g^{*2}}+4\bigl(1-\partial_{ \hat x}\hat{p}_{g}^{*}\bigr)}}{2\bigl(1-\partial_{\hat x}\hat{p}_{g}^{*}\bigr)}\,.
\end{equation}

It is worth noticing that the downward orientation of the $\hat x$ axis in this work is opposite to the one used in the jet wiping literature \citep{Hocking2010,Tuck1984,Gosset2007}, but
in line with the falling liquid film literature \citep{Kalliadasis2012,Ruyer-Quil2014}. The wiping thus occurs in a region in which $\partial_{ \hat x}\hat p_{g}^{*}<0$ and $\hat\tau_g^{*}>0$.  Introducing this value of the film thickness in \eqref{Q_SIMPLE}, taking $\partial_{\hat x}\hat p_{g}=\partial_{\hat x}\hat p_{g}^{*}$ and $\hat \tau_g=\hat \tau_g^{*}$, allows for estimating the withdrawn flow rate $q(\hat h^{*})$.  The final coating thickness $\hat{h}_f$ can thus be estimated using again \eqref{Q_SIMPLE} in the far-field conditions (where $\hat \tau_g=\partial_{ \hat x}\hat p_{g}=0$). This approach, known as the 0D knife model, was proposed by \citet{Buchlin} and validated on several numerical and experimental works \citep{Gosset2007,Lacanette2006}. 

The polynomial in the far field condition is shown in figure \ref{PLOT_POLY} (see also \citealt{Snoeijer}). For a wiping condition yielding $\hat{q}({h}^*)=-0.1$, the corresponding final thickness ($\hat{h}_f$) and run back flow thickness ($\hat{h}_R$) are shown. At the lowest limit of the flow rate ($\hat{q}=-2/3$), only one solution is admissible for the film thickness (corresponding to $\hat{h}=1$ in the chosen scaling). This corresponds to the well-known limit for the drag-out problem \citep{B1964,Rio2017} in the gravity dominated regime.

\begin{figure}
\centering
\includegraphics[width=7.2cm]{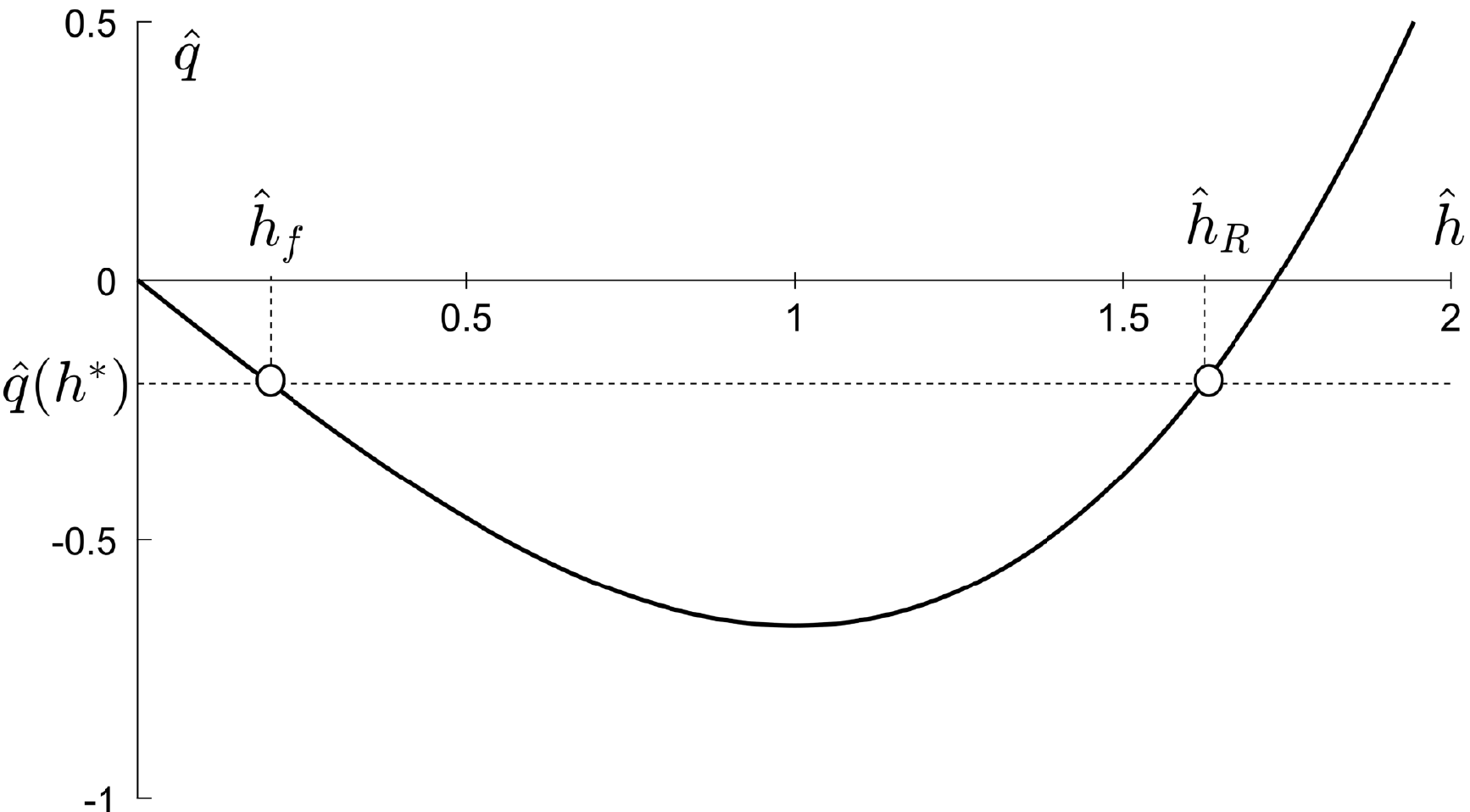}
\caption{Flow rate versus thickness relations for a film flowing along a vertical moving wall, assuming $\partial_{\hat x} \hat{p}_g=\hat{\tau}_g=0$ in \eqref{Q_SIMPLE} {and neglecting the surface $\partial_{\hat x\hat x\hat x}\hat{h}$}. For a given flow rate $\hat q(h^{*})<0$, the thin and the thick solutions are approximations of $\hat h_f=\lim_{\hat x \to +\infty} \hat{h}(\hat{x})$ and $\hat h_R=\lim_{\hat x \to -\infty} \hat{h}(\hat{x})$ respectively. Using $h^{*}$ from \eqref{0D} yields the 0D model formulation of the wiping process.}
\label{PLOT_POLY}
\end{figure}

Equation \eqref{Benney} allows for important considerations on the propagation of small disturbances on a flat film over an upward moving substrate. Neglecting the surface tension contribution, this equation simplifies to:

\begin{equation}
\label{WAVE_SIMP}
\partial_{\hat t} \hat h+\bigl( \hat h^2-1\bigr)\,\partial_{ \hat x}  \hat h=0\,,
\end{equation}

{\parindent0pt  that is } a standard \emph{kinematic wave equation} \citep{Whitham1999}: disturbances on a film $\hat h<1$ (that is in the final coat) propagate at a negative velocity (that is upward) while disturbances on a film $\hat h>1$ (that is in the runback flow) propagates at a positive velocity (that is downward).

\subsection{Integral Boundary Layer (IBL) Formulation}\label{KA}

As for the ZO model, this formulation considers only the first term in the velocity profile \eqref{uQS}, but it does not assume that the LHS of \eqref{I2} vanishes. Using \eqref{uQS} with the coefficient $a_0$ from \eqref{qQS}, the advection term $\mathcal{F}$ in \eqref{DEFs} yields

\begin{equation}
\label{ADV_eps}
\mathcal{F}=\frac{ \hat h^3\, \hat\tau_g^2}{120}+\frac{ \hat h\, \hat q\, \hat \tau_g}{20}+\frac{ \hat h^2 \hat \tau_g}{20}+\frac{6\, \hat q^2}{5\,  \hat h}+\frac{2\, \hat q}{5}+\frac{ \hat h}{5}\,.
\end{equation}

The shear stress term remains the one in \eqref{DeltaT}. The resulting model is an extension to the jet wiping process of the classical self-similar integral models proposed by \citet{KAPITZA} and \citet{Shkadov1971,Shkadov2017}. To the authors' knowledge, this integral model has never been used in the analysis of the jet wiping process.

\subsection{Weighted Integral Boundary-Layer Model (WIBL) }\label{WRM}

The full expansion in \eqref{Vel_PROF} is now considered, taking up to $N=4$ terms. This number can be derived based on the order of magnitude analysis (see \citet{Kalliadasis2012}, section 6.6).  
{The method of weighted residuals to derive the coefficients $\mathcal{A}=\{a_0,a_1,a_2,a_3,a_4\}$ in \eqref{Vel_PROF} for a falling liquid film was proposed by \citet{Ruyer-Quil2002}. Denoting the momentum equation \eqref{My} as an operator $\mathcal{M}(\hat{u})=0$, it is possible to construct the residuals $\mathcal{R}_j$ from the projections $\mathcal{R}_j=\langle w_j, \mathcal{M}(\sum_j a_j f_j(\hat{y}))\rangle$ with $\langle \cdot \,, \cdot \rangle$ denoting the continuous inner product over the space of possible solutions $u\in\mathcal{S}$ and $w_j$ a set of weight functions. Setting all residuals to $\mathcal{R}_j=0$ leads to a system of equations for the $a_j$ coefficients.}

{For the modelling of a film falling along a fixed wall and passive atmosphere, \cite{Kalliadasis2012} compare various weighted residuals methods, differing by choice of weight functions $w_j$. Their comparison shows that all methods converge, if four $w_j$'s are taken\footnote{\cite{Kalliadasis2012} also show that some methods converge `faster' than others: for instance, the Galerkin approach with $w_j=f_j$ converges with only one residual, while a collocation approach, with $w_j=\delta_j$, with $\delta_j$ a set of equally spaced Dirac functions, requires four residuals.}, to a model that can be derived following a polynomial matching procedure that involves neither weights nor residuals.}

{We here solely focus on this second approach, although this requires considerably more algebra, for two reasons. First, because this allows to derive an explicit equation for the coefficients $a_j$, while this is not needed in the weighted residual approach. Second, because this the approach does not depend on the choice of the $w_j$'s and allows for avoiding a discussion on convergence. Nevertheless, we still refer to the resulting model to as WIBL, in line with the literature on falling liquid films.}




Introducing the expansion of the velocity profile \eqref{Vel_PROF} in the momentum equation \eqref{Mx} yields a polynomial in the reduced coordinate $\overline{y}$:

\begin{equation}
\label{GRAND_POLY}
\mathcal{P}\Bigl(x,t,\overline{y}=\frac{y}{h}\Bigr)=\sum^{N}_{j=0}\,P_j\bigl (\mathcal{A},x,t\bigr )\,\overline{y}^j=0\,.
\end{equation}

Because this polynomial must be identically null, all the functions $P_j\bigl (\mathcal{A},x,t\bigr )$ must be null. Moreover one notices that the viscous term on the RHS of \eqref{Mx} \emph{decreases} the degree of the polynomials by two while the LHS \emph{increases} it by two. Therefore, only the $N=0$ should be introduced on the LHS, and up to $N=4$ terms should be introduced on the RHS, {recalling that $a_0\sim \mathcal{O}(1)$ and $a_i\sim \mathcal{O}(\varepsilon)$ $\forall i \in [1,4]$}. The resulting polynomial in \eqref{GRAND_POLY} is thus of order four.

Setting all of its coefficients to zero leads to a system of five equations of the form:

\begin{equation}
\label{System}
[P_j\bigl (\mathcal{A},x,t\bigr )]^{4}_{0}=0 \rightarrow \Gamma\,\mathcal{A}=\mathcal{G}\,,
\end{equation}

{\parindent0pt  where} the system matrix reads

\begin{equation}
\label{Gamma}
\Gamma=\frac{1}{h^2}
\begin{pmatrix}
1 & -2 &0 &0&0 \\
0 & 4 & -6 & 0 &0\\
0 &0 & 9 & -12&0 \\
0 & 0 & 0 & 16&-20\\
0 &0 & 0 & 0&25
\end{pmatrix}\,\,,
\end{equation}

{\parindent0pt  and} the vector $\mathcal{G}=\{G_0,G_1,G_2,G_3,G_4\}$ is shown in table \ref{Gs} of Appendix \ref{TABS}. The solution of the system leads to the full expression of the coefficient, given in table \ref{As} of Appendix \ref{TABS}. All the coefficients, except $a_0$, are then introduced in \eqref{Q} to identify the link between $a_0$ --thus the wall shear stress term in \eqref{D_Tau}-- and the flow rate. The resulting expression, following the classical notation from asymptotic expansions \citep{Howison2005}, is of the form:

\begin{equation}
\label{q_ASY}
\hat q=\hat q^{(0)} (a_0, \hat h)+\varepsilon Re \, q^{(1)}\bigl (\hat{h},a_0,\partial_{\hat t} a_0, \partial_{\hat x} a_0, \partial_{\hat x} \hat h, \partial_{\hat t} \hat h \bigr )\,,
\end{equation}

{\parindent0pt  having} considered only the functional dependency on the unknown variables. This expression can be used to compute $a_0$ following the same notation:

\begin{equation}
\label{a0_ASY}
a_0=a_0^{(0)} (\hat q, \hat h)+\varepsilon Re \,a_0^{(1)}(\hat{h},a_0,\partial_{\hat t} a_0, \partial_{\hat x} a_0, \partial_{\hat x} \hat h, \partial_{\hat t} \hat h )\,.
\end{equation}

Both expressions \eqref{q_ASY} and \eqref{a0_ASY} are shown in table \ref{Q_A} of appendix \ref{TABS} in their complete forms (\eqref{Q_0} and \eqref{A_0NS} respectively). As expected, the first order term recovers the coefficient from the leading-order model in \eqref{qQS}, while the other represents $\mathcal{O}(\varepsilon)$ corrections. At this stage, the coefficient $a_0$ is still implicitly defined. However, if $ Re\ll 1/\varepsilon$, the asymptotic expansion framework allows for substituting $a_0\approx a_0^{(0)}+\mathcal{O}(\varepsilon)$ in $a_0^{(1)}$ and neglect higher order terms, as done in \eqref{q_ASY} and \eqref{a0_ASY}.

The resulting shear stress is

\begin{equation}
\begin{split}
\label{LAST_E2}
\Delta \tau&=\frac{3}{2}\tau_g-\frac{3\,q}{h^2}-\frac{3}{h}+\varepsilon Re \Bigl (-\frac{19  {{h }^{3}}\, {{\tau}_g}   \partial_{x} {{\tau}_g}  }{3360}-\frac{17  h \, q \,  \partial_{x} {{\tau}_g}  }{560}- \frac{3  {{h }^{2}}\,\partial_{x} {{\tau}_g}}{560}-\frac{ {{h }^{2}}\,  \partial_{t} {{\tau}_g}  }{40}\\&-\frac{ h \, {{\tau}_g}   \partial_{x} q}{56}-\frac{18  q \,  \partial_{x} q  }{35 h }-\frac{4   \partial_{x} q  }{35}-\frac{  \partial_{t} q }{5}-\frac{ {{h }^{2}}\, {{{{\tau}_g}}^{2}}  \partial_{x} h}{112}-\frac{ q \, {{\tau}_g}   \partial_{x} h }{280}-\frac{3  h \, {{\tau}_g} \partial_{x} h  }{140}\\&+\frac{12  {{q }^{2}}\,  \partial_{x} h  }{35 {{h }^{2}}}+\frac{6  q \,  \partial_{x} h  }{35 h }+\frac{  \partial_{x} h  }{35} \Bigr)
\end{split}
\end{equation}

{The Weighted Integral Model (WIBL) for the jet wiping problem is obtained by introducing \eqref{LAST_E2} and \eqref{ADV_eps} in \eqref{I1} and \eqref{I2}.} Observe that the $\mathcal{O}(1)$ terms in \eqref{LAST_E2} are the ones in the IBL model. Among the fourteen $\mathcal{O}(\varepsilon)$ terms, it is worth noticing that one involving the partial time derivative of the flow rate (the eighth term). For computational purposes, it is convenient moving this term on the LHS of \eqref{I2}, resulting in a coefficient $\beta=6/5$ on the term $\partial_t \hat {q}$ (see section \ref{NUM}).

\section{The Transition and Turbulent Boundary Layer (TTBL)}\label{TBL}

A falling liquid film is laminar for Reynolds number below $\approx 100$, and it is turbulent above $\approx 400$ \citep{Alekseen1994,Ishigai,KARIMI19991305}. The intermediate-range is the transition region, and cannot be described solely in terms of Reynolds number.
Laminar integral boundary layer models have been proved successful (see \cite{Denner2018} for cases up to $Re\approx 80$) well above their theoretical range of validity (which sets $Re\sim \mathcal{O}(1)$), while much higher Reynolds numbers, like those considered in this work, might need a different treatment.

A classic theoretical formulation for turbulent liquid films is based on mixing length theory \citep{DRIEST1956,King1966,Geshev2014} and a Reynolds-averaged formulation of the velocity field in which the effect of turbulence is modeled by an additional eddy viscosity. This formulation is based on the statistically stationary assumption and is of difficult extension to the integral formulation of interest to this work. A different approach is commonly encountered in the literature of shallow water flows (see \citealp{James2019}), in which the most celebrated empiricism consists in introducing a correlation for the wall shear stress (see also \citealp{Katopodes}), and a shape factor for the velocity profile. As these allow for keeping the integral nature of the model formulation, a similar approach is pursued in this work.

The Transition and Turbulent Boundary Layer (TTBL) model for the jet wiping problem proposed in this work makes no pretension of completeness; on the contrary, {it offers a first attempt to analyze the possible impact of turbulence on the response of the coating thickness. Following the same self-similarity argument supporting the IBL model, the proposed model extends the IBL model to a turbulent liquid film. A similar extension of the WIBL for a falling liquid film, using the mixing-length formulation, is presented by \cite{Mukhopadhyay}.} The proposed closures for the wall shear stress and the advection terms are described in sections \ref{Tau_CLO} and \ref{F_CLO}, respectively.

\subsection{Closure for the Wall Shear Stress}\label{Tau_CLO}
Following the shallow water literature, the wall friction is modeled in terms of a friction coefficient $C_f$ that is function of the (local) Reynolds number.  In the jet wiping problem, where the liquid streamwise velocity component is composed of the terms in \eqref{U_Three}, one must first establish which of the Reynolds numbers in \eqref{Reynoldss} controls the transition to turbulence and the flow regime. Moreover, since the thickness of the liquid film varies significantly between the final coating region ($\hat{h}\ll1$) and the run-back flow region ($\hat{h}\sim 1$), different regimes (laminar, transition, fully turbulent) can be expected in different regions.
{We begin by introducing the skin friction coefficient in the laminar models. From \eqref{uQS} and \eqref{qQS}, the wall shear stress reads}

\begin{equation}
\label{Shear}
{
\partial_{\hat{y}}\hat{u}\Bigl |_{\hat {y}=0}=\frac{3 \hat{q}_F}{\hat{h}^2}+\hat{\tau}_g=\hat{\tau}_{wF}(\hat{q}_F)+\hat{\tau}_g}\,.
\end{equation}

{Introducing the skin friction coefficient based on the mean velocity $\hat{q}_F/\hat{h}$, one gets the following dimensional and dimensionless friction terms $\hat{\tau}_{wF}$}

\begin{equation}
    \label{Tau_WF}
    {
   {\tau}_{wF}=\frac{1}{2}\rho \frac{q_F|q_F|}{h^2} C_f\Longleftrightarrow \hat{\tau}_{wF}= \frac{1}{2} Re \frac{\hat{q}_F |\hat{q}_F|}{\widehat{h}^2} C_f\,,}
\end{equation}{ having used the reference quantities in table \ref{Scaling_Table}. Comparing \eqref{Shear} and \eqref{Tau_WF}, the skin friction coefficient in laminar conditions is $C_f=6/Re_F$.}

{To extend the model to turbulent films, while allowing for recovering \eqref{Shear} in laminar ones, two constraints should be considered. The first concerns the sign of ${\tau}_{wF}$, which is given by $\hat{q}_F$ in laminar conditions. It is worth noticing that in the run-back flow region, as later discussed in the example in figure \ref{PROFS_LAST}, this quantity is usually negative (i.e. directed upward) since the dominant role of the (positive) shear stress term yields}

\begin{equation}
\label{qF}
{\hat{q}_F=\hat{q}-\frac{1}{2} \hat{\tau}_g \hat{h}^2 +\hat{h}<0 \quad \mbox{since} \quad \frac{1}{2} \hat{\tau}_g \hat{h}^2>\hat{q}+\hat{h} }
\end{equation}
{In order to avoid discontinuities in the shear stress in the transition regions, we postulate that the sign of $\hat{\tau}_{wF}$ remains dictated by $\hat{q}_F$ also in turbulent conditions.}

{The second constraint is to ensure a smooth transition in ${\tau}_{wF}$ as the flow passes a certain critical Reynolds number. This can be easily ensured if both the mean velocity in the definition of $C_f$ and the correlations for $C_f$ are solely functions of $\hat{q}_F$.  }

{The TTBL proposed in this work is based on the simplest modeling solution respecting these two constraints. First, we keep \eqref{Tau_WF} also in turbulent conditions. Second, we adapt the skin friction in presence of turbulence to a correlation of the form $C_f\approx aRe_F^{b}$, with $b\approx -1/4$, as encountered in seminal works on turbulent lubrication \citep{Hirs1973,Elrod1967} and turbulent boundary layer theory \citep{BLT}.}
Assuming that the transition occurs at $Re_*$, the correlation allowing for a continuous transition is

\begin{equation}
\label{Cf}
C_f=
\begin{cases}
{6}/{Re_F}& Re_F<Re_*\\
(6 {Re^{-3/4}_*} )Re_F^{-0.25}  & Re_F>Re_*\\
\end{cases}
\end{equation}

where the critical Reynolds is here taken as $Re_*=100$. Equations \eqref{Tau_WF} and \eqref{Cf} can be used for computing the shear stress term $\Delta \hat{\tau}$ in \eqref{I2}.

{An alternative formulation could consist in defining the friction coefficient from a reference velocity $\hat{q}_T/h$, with $\hat{q}_T=\hat{q}_F+1/2\hat{\tau}_g\hat{h}^2 $, i.e. including both the first two flow rate contributions in \eqref{Reynoldss}. In this case, ensuring the aforementioned constraints becomes more challenging and requires the introduction of appropriate blending functions. }


\subsection{Closures for the Advection Term}\label{F_CLO}

As for the laminar case, the closure of the advection term in \eqref{DEFs} requires some assumptions on the velocity profile within the film. Self-similar profiles such as the one-seventh-power law have been borrowed from boundary layer theory in the liquid film literature and have shown reasonably good agreement in the prediction of film thickness \citep{Alekseen1994}. More sophisticated eddy viscosity models have been used to analyze phenomena such as gas absorption or heat transfer \citep{Mudawwar1986,Riazi1996} or the impact of the interface shear stress \citep{Geshev2014}.

Turbulence increases the momentum diffusion, flattening the velocity profile with respect to the laminar one. This has an impact on the advection term, and possibly on the response of the film thickness. To analyze this impact, we here consider the velocity profile of the falling film portion $\hat{u}_F$ (see \ref{U_Three}) as composed of two contributions:



\begin{equation}
\label{u_T}
\hat u_F(\hat x,\hat y,\hat t)= \hat{u}_L+\hat{u}_T= a_L(\hat x,\hat t)\,\Bigl[\Bigl(\frac{\hat y}{\hat h}\Bigr)-\frac{1}{2}\Bigl( \frac{\hat y}{\hat h}\Bigr)^2\Bigr]+ a_T(\hat x,\hat t)\Bigl [\Bigl(\frac{\hat y}{\hat h}-1 \Bigr)^{n_T}+1\Bigr],
\end{equation} where $n_T$ is an integer and odd number. By definition, the turbulent contribution, weighed by the coefficient $a_T$, satisfies the boundary conditions for the $\hat{u}_F$ term and reproduces boundary layers of different thicknesses with a flat profile above it. The combination of the two terms in \eqref{u_T} allows for representing various kinds of departure from the parabolic assumption. The coefficients $a_L$ and $a_T$ are constrained by the wall shear stress (obtained from \eqref{Tau_WF} and \eqref{Cf}) and the mass conservation:


\begin{equation}
\label{Coeff_BC}
\begin{dcases}
	\hat{\tau}_{wF}=\partial_{\hat y} \hat u_F\bigl|_{\hat{y}=0}\rightarrow \hat{\tau}_{wF}{\hat h}= a_L + n_T a_T\\
\hat{q}_F=\int_0^{\hat{h}} \hat u_F d \hat y \rightarrow {\hat{q}_F}=\frac{1}{3} a_L \,{\hat h} + \Bigl(\frac{n_T}{n_T+1}\Bigr)a_T\,{\hat h}\,.
\end{dcases} 
\end{equation}

The solution of the resulting linear system of equations gives:

\begin{equation}
\label{SoL_A}
\begin{dcases}
a_L=\frac{\hat h \hat{\tau}_{wF}n_T}{(n_T+1) c_T}- \frac{n_T\hat{q}_F}{\hat h c_T}\\
a_T=-\frac{\hat h \hat{\tau}_{wF}}{3 c_T}+\frac{ \hat{q}_F}{\hat h c_T}\,,
\end{dcases} 
\end{equation} where $c_T=(2n_T-n^2_T)/(3n_T+3)$. 
The model is closed for a given $n_T$. Regardless of the choice of $n_T$, this model recovers the IBL formulation in laminar condition, for which $\hat{\tau}_{wF}=3\hat q_F/\hat h^2$, as this leads to $a_T=0$. 

While the link between film thickness and flow rate requires the complete numerical analysis of the TTBL model, a first estimation of the range of validity of the model can be inferred from a simple physical constraint: the maximum velocity in \eqref{u_T} should be reached at the interface and no other extremes should occur within the liquid film:

\begin{equation}
\hat{h} \partial_{\hat{y}}\hat{u}_F =a_L\Bigl(1-\frac{\hat y}{\hat h}\Bigr)+a_T n_T \Bigl(\frac{\hat y}{\hat h}-1 \Bigr)^{n_T-1}\geq 0 \quad \forall \hat{y}\leq \hat{h}
\end{equation}

Solving the inequality for any given set of coefficients $a_L, a_T$ allows for computing the range of a validity of the model for a given $n_T$. This range is shown in figure \ref{Range_21} for $n_T=21$. For each combination of coefficient, it is possible to compute the shape factor of the corresponding profile, defined as:

\begin{equation}
\label{Up}
\Upsilon=\frac{\hat{h}}{\hat{q}}\int_0^{\hat{h}}\hat{u}^2d\hat{y}
\end{equation}

This parameter ranges from $1.2$ in laminar conditions to $\approx 1$ in case of a flat velocity profile. Using \eqref{Cf} and \eqref{SoL_A}, it is then possible to analyze how the shape factor changes as a function of $Re_F$ for a given $n_T$, as well as the maximum $Re_F$ tolerated by the model within its range of validity. This is shown in figure \ref{Beta} for $n_T=[7,15,21]$.

\begin{figure}
	\centering
	\begin{subfigure}[b]{0.42\textwidth}
		\centering
		\includegraphics[width=\textwidth]{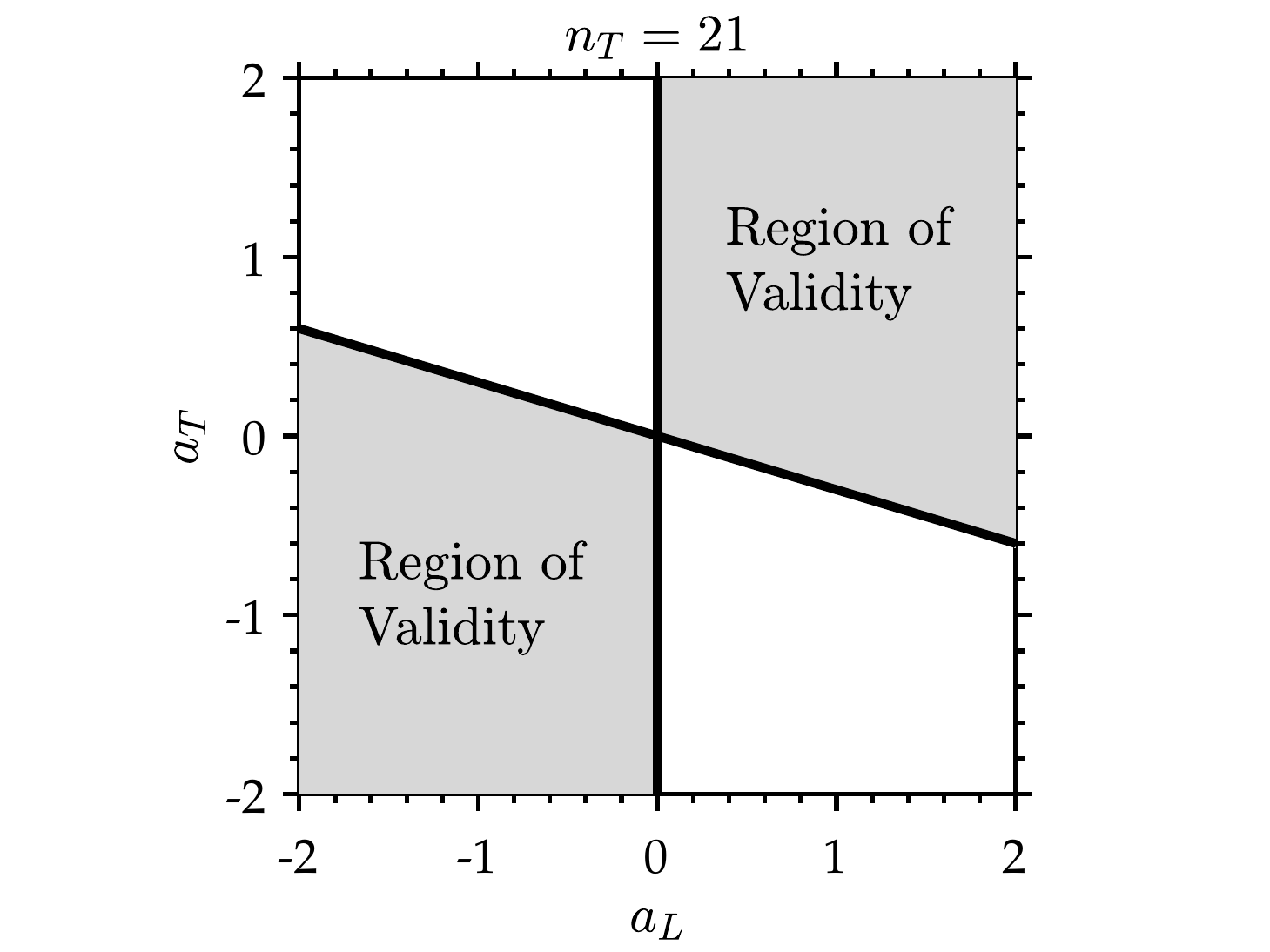}
		\caption{}
		\label{Range_21}
	\end{subfigure}
	\hspace{2mm}
	\begin{subfigure}[b]{0.53\textwidth}
		\centering
		\includegraphics[width=\textwidth]{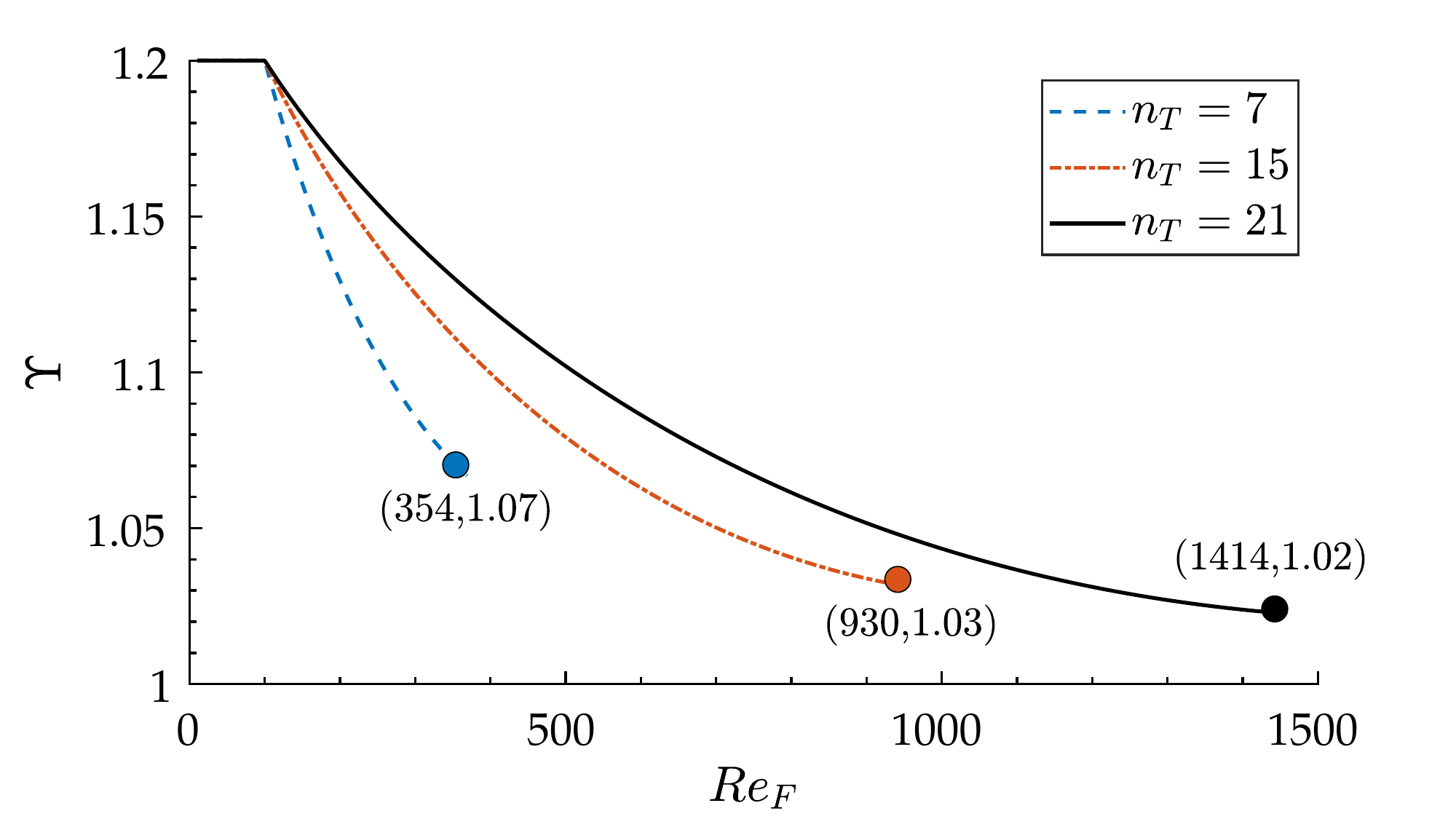}
		\caption{}
		\label{Beta}
	\end{subfigure}\\
		\begin{subfigure}[b]{0.32\textwidth}
		\centering
		\includegraphics[width=\textwidth]{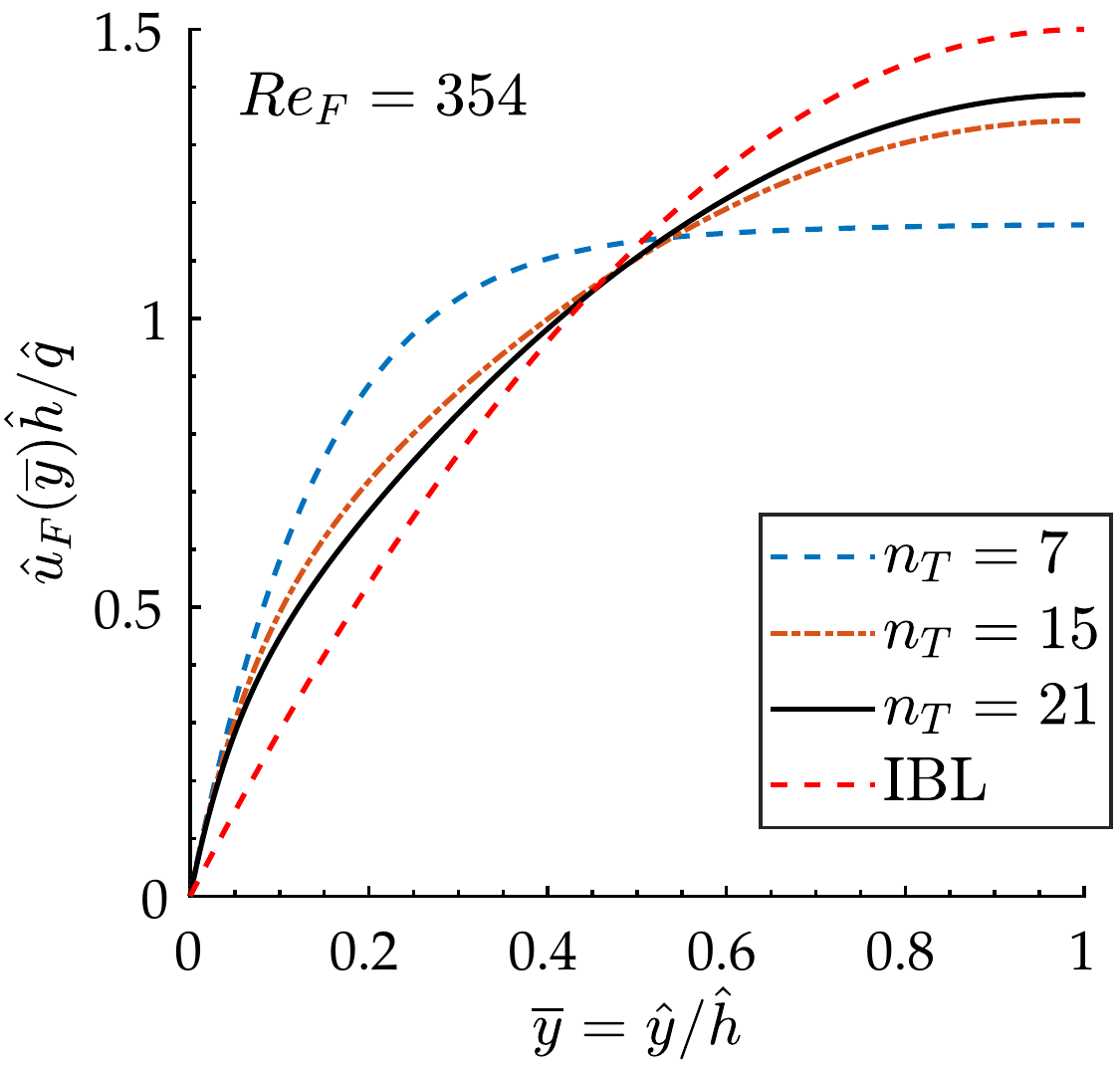}
		\caption{}
		\label{Profiles_n_T1}
	\end{subfigure}
			\begin{subfigure}[b]{0.32\textwidth}
		\centering
		\includegraphics[width=\textwidth]{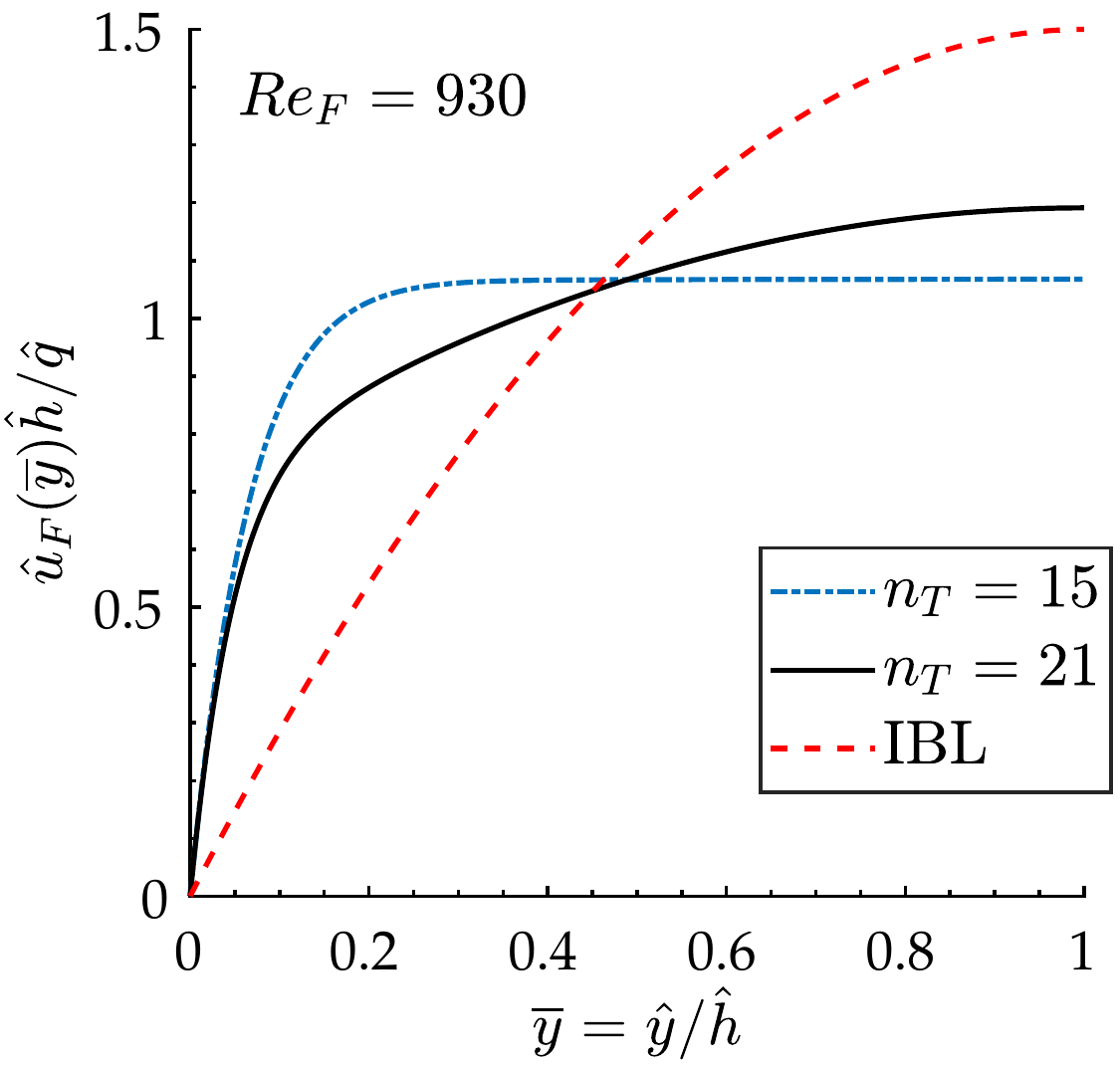}
		\caption{}
		\label{Profiles_n_T2}
	\end{subfigure}
	\begin{subfigure}[b]{0.32\textwidth}
		\centering
		\includegraphics[width=\textwidth]{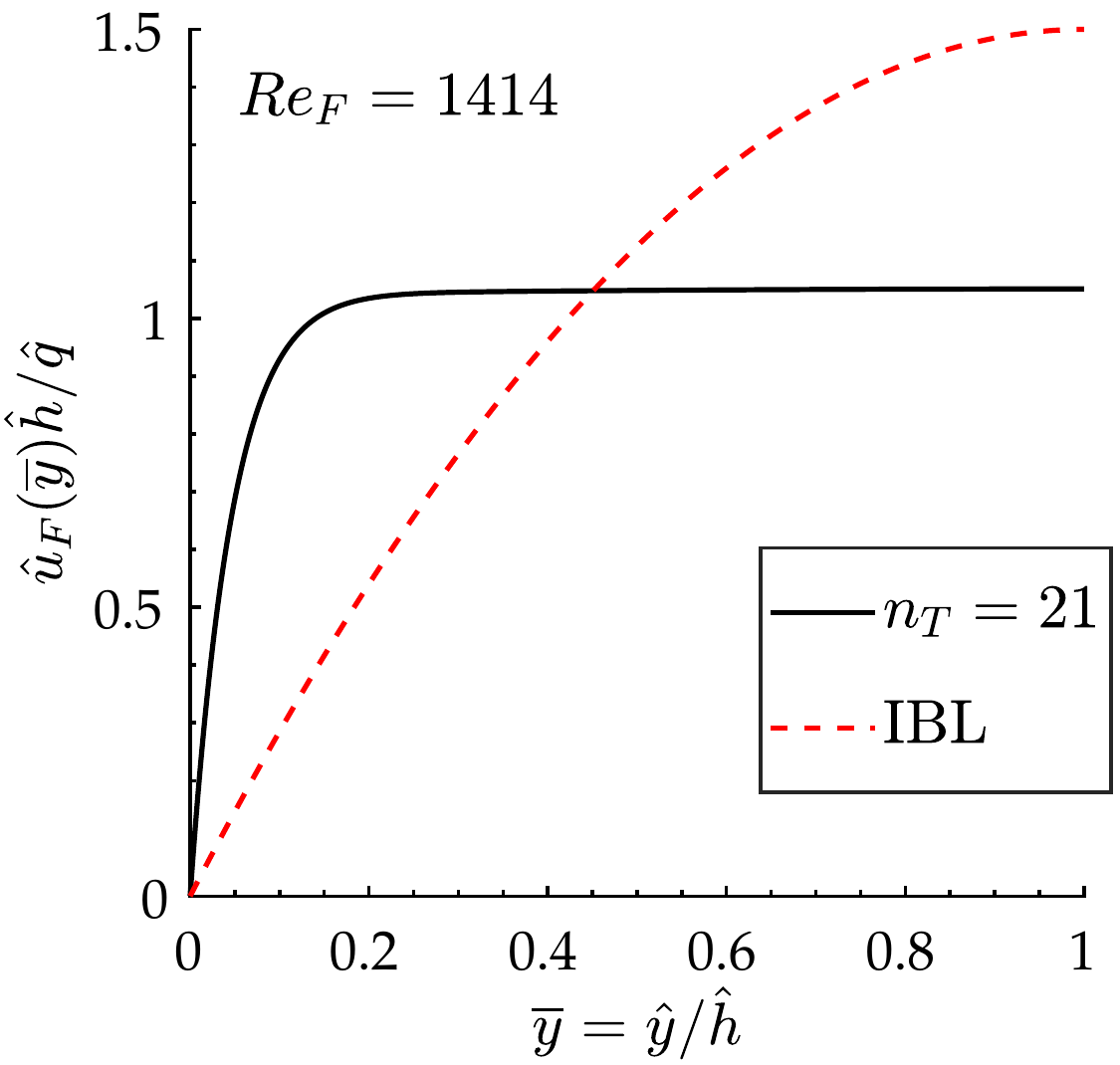}
		\caption{}
		\label{Profiles_n_T3}
	\end{subfigure}
	\caption{figures (a): range of admissible pairs $a_L$, $a_T$ for a velocity profile in \eqref{u_T}. figure (b): shape factor \eqref{Up} of the velocity profile in \eqref{u_T} as a function of the \emph{local} Reynolds number $Re_F$ for $n_T=7,15,21$. For each coefficient, the maximum admissible $Re_F$ is reported. figures (c)-(e)-(d) shows example of normalized velocity profiles for the three models at the $Re_F$ indicated in each plot, together with the parabolic profile assumed in the IBL model.}
	\label{Fig3}
\end{figure}

Figure \ref{Profiles_n_T1} compares the velocity profiles (normalized to have unitary mean) for the $n_T=[7,15,21]$ with the parabolic assumption at $Re_F=354$, i.e. the maximum admissible value for $n_T=7$. The same comparison is shown in figure \ref{Profiles_n_T2} at $Re_F=930$, the maximum admissible value for $n_T=15$, and in figure \ref{Profiles_n_T3} at $Re_F=1414$, the maximum admissible value for $n_T=21$.

For the purposes of this work, we limit our analysis of the impact of $n_T$ on the liquid film modeling to the definition of the upper limit within which such a model is valid. In what follows, a value of $n_T=21$ is considered. At the maximum Reynolds number, the resulting velocity profile features a boundary layer of approximately $\hat{h}/5$. Compared to the measurements in turbulent falling films (e.g., \citealt{Mudawar1993}), this estimation appears extreme, but well in line with the primary purpose of testing the impact of high turbulence in some portions of the falling film. Nevertheless, it is worth noticing that such extreme is not reached in any of the investigated test cases, among which the highest falling film Reynolds, in the run-back flow region, is $Re_F\approx 600$.

The advection term for the TTBL model, considering $n_T=21$, is computed by introducing \eqref{u_T} in \eqref{U_Three} and in the definition \eqref{DEFs}:

\begin{equation}
\label{F_TTBL}
\begin{split}
\hat{\mathcal{F}}=\int^{\hat{h}}_{0} \hat{u}^{2} d\hat y =\frac{1}{3} \hat h^3 \hat{\tau}^2_g
+\frac{252}{253} a_T \hat{h}^2 \hat{\tau}_g+\frac{5}{12} a_L\hat{h}^2\hat{\tau}_g -\hat{h}^2\hat{\tau}_g+\frac{441}{473} a^2_T \hat{h}\\ + \frac{175}{264} a_L a_T \hat{h}-\frac{21}{11} a_T \hat{h}+\frac{2}{15} a^2_L\hat{h}-\frac{2}{3}a_L \hat{h}+\hat{h}\,.
\end{split}
\end{equation} 

This terms is closed using \eqref{SoL_A} with \eqref{qF} and \eqref{Tau_WF}.

\section{The Wiping Actuators}\label{JETU} 

Following classical modeling strategies of the jet wiping process, the action of the gas jet is modeled via the pressure gradient and the shear stress produced on the liquid film. These two quantities, referred to as wiping actuators, are modeled via experimental correlations for gas jet impinging on a flat (dry) plate \citep{Beltaos1976,Tu1996,Elsaadawy,AnneThesis}, with minor adaptations to account for their time dependency, and under the assumption that the dynamics of the liquid film has no influence on their evolution.

Using the reference scales in table \ref{Scaling_Table}, the wiping actuators are of the form:

\begin{subequations}
\label{ACT_FORM}
\begin{align}
\partial_{\hat x} \hat p_{g}=\partial_{\hat x} \bigl[{P}_g(t)\,f_p\bigl(\tilde{x}(t)\bigr)/(\rho_l \,g)\bigr ]\label{P_C}\\
\hat \tau_{g}={T}_g(t)\,f_\tau(\tilde{x}(t))/(\sqrt{\mu_l\,\rho_l\,g\,U_p})\label{T_C}
\end{align}
\end{subequations}{where $\tilde{x}$ denotes a time dependent axis accounting for the possible oscillation of the jet, as described at the end of this section. } 

The functions $f_p$ and $f_\tau$ have range $\in [-1,1]$ so that the maximum values from these quantities are defined by the scalars ${P}_g(t)$ and $T_g(t)$.
Following the empirical correlation by \cite{Tu1996}, the pressure distribution for a gas jet impinging on a flat wall is:

\begin{equation}
\label{f_P}
f_p(\xi)=\,\exp\bigl(-0.693 \xi^2\bigr)+\frac{0.01895\, |\xi|}{1+(\xi-1.67489)^2}
\end{equation} where $\xi=x/b$ is a dimensionless coordinate, with the parameter $b$ controlling the spreading of the distribution. {A qualitative plot of the pressure distribution is shown in Figure \ref{Fig1} on the right, with a red dashed line.}

As proposed by \citet{Beltaos1976}, for a stand-off distance $Z/d>5$, this parameter can be computed as $b=0.125\,Z$. The maximum pressure ${P}_g$, for a statistically stationary impinging jet, can be computed as $P_g=6.5 {P_d\,d}/{Z}$, where $P_d=C_d\,\Delta P_N$ is the dynamic pressure at the nozzle outlet and $C_d$ is the discharge coefficient taking into account the losses due to friction and separation phenomena in the nozzle chamber. This parameter depends on the nozzle design and is taken as $C_d=0.8$ in this work. Considering the reference quantities in table \ref{Scaling_Table}, the role of the pressure gradient in the wiping capabilities of the jet is well described by the dimensionless group

\begin{equation}
\Pi_g=\frac{P_d\,d}{\rho_l\,g\,Z^2}=C_d\,\frac{\Delta P_N\,d}{\rho_l\,g\,Z^2}\,,
\end{equation} hereinafter referred to as the \emph{wiping number}. This number compares the maximum pressure gradient produced by the gas $(\sim C_d\,\Delta P_N d/Z^2)$ with the reference (hydrostatic) pressure gradient in the liquid film ($\rho_l g$).

The distribution of shear stress at the gas-liquid interface is computed following the numerical correlation proposed by \citet{Elsaadawy2007}. For $\xi\geq0$, this reads:

\begin{equation}
\label{T1}
f_{\tau}=
\begin{cases}
 \text{erf}\bigl(0.41\xi\bigr)+0.54\xi \exp(-0.22\xi^3)&\xi \leq1.73\\
  1.115-0.24\, \ln(\xi) &\xi >1.73\\
\end{cases}
\end{equation}

For $\xi<0$ this distribution is mirrored such that $f_{\tau}(\xi)=-f_{\tau}(-\xi)$. 
{A qualitative plot of the shear stress distribution is shown in Figure \ref{Fig1}, with a dash-dotted blue line.}

For a sufficiently high Reynolds number in the jet flow, like those considered in this work, the maximum shear stress is computed as ${T}_g=C_\tau P_d \,d/Z$, with $C_\tau=0.067$ \citep{Tu1996}. Considering the reference shear stress in table \ref{Scaling_Table}, the role of the shear stress in the wiping capabilities of the jet is measured by the dimensionless group

\begin{equation}
\mathcal{T}_g=\frac{T_g}{Z\,\sqrt{\rho_l\,g\,\mu_l\,U_p}}=C_d\,C_\tau\frac{\Delta P_N\,d}{Z\,\sqrt{\rho_l\,g\,\mu_l\,U_p}}\,, 
\end{equation}hereinafter referred to as the \emph{shear number}. This number compares the maximum shear stress produced by the gas flow ($\sim C_{\tau} P_d d/Z$) with the reference shear stress in the liquid film ($\mu_l U_p/[h]$).

It is worth noticing that in the case of wiping of very viscous liquids (such as, e.g., mineral oils or paint) the importance of this number decreases considerably. More information on the scaling laws of the jet wiping process and the typical operating conditions encountered in various industrial processes is presented in \citet{Gosset2019}.

The use of correlations for gas jet impinging on dry surfaces has been validated in various studies (see \citet{Lacanette2006}). While the validity of such simplification in time dependent conditions is certainly questionable, it is important to recall that the modeling of the shear stress produced by an impinging jet on a dry surface is still subject to extensive investigation, and large discrepancies exist in the correlation proposed by various authors. For more details, the reader is referred to \citet{Ritcey2017}. 

Finally, concerning the time dependency of the actuators, two possibilities are considered in this work: pulsations and oscillations. In the case of pulsations, the amplitude of the actuators $P_g(t)$ and $T_g(t)$ are set as harmonics with mean value equal to the correlations in steady-state conditions and amplitude of $30\%$. In the case of oscillations, the amplitudes are left stationary and equal to the correlations previously proposed, while the streamwise variable is taken as $\tilde{x}(t)=\hat{x}-Z\,\tan (W(\theta(t)))$, where { $\theta(t)$ is the angle of the oscillation with respect to the horizontal, taken as $\theta>0$ for a jet deflected upstream (on the runback flow side), and $W(\theta(t))$ is a possible waveform of the oscillation. }

{Three waveforms are considered. The first is a harmonic oscillation $W(\theta (t))=\theta_A\sin(2\pi f_h t)$.} The others are non-harmonic oscillations biased upstream or downstream. These are constructed by smoothing a square wave signal, which is symmetric around the mean but has a different duration of the positive/negative cycles. In the investigated test cases, a biased oscillation spends $80\%$ of its period upward or downward. The spatio-temporal evolution of the pressure gradient for an example of each of these oscillatory modes is shown in figures \ref{Example_WAVES}. Figures \ref{3a} and \ref{3b} shows a harmonic and a upward-biased oscillation, while figure \ref{3c} shows a pulsating perturbation.

\begin{figure}
     \centering
     \begin{subfigure}[b]{0.32\textwidth}
         \centering
         \includegraphics[width=\textwidth]{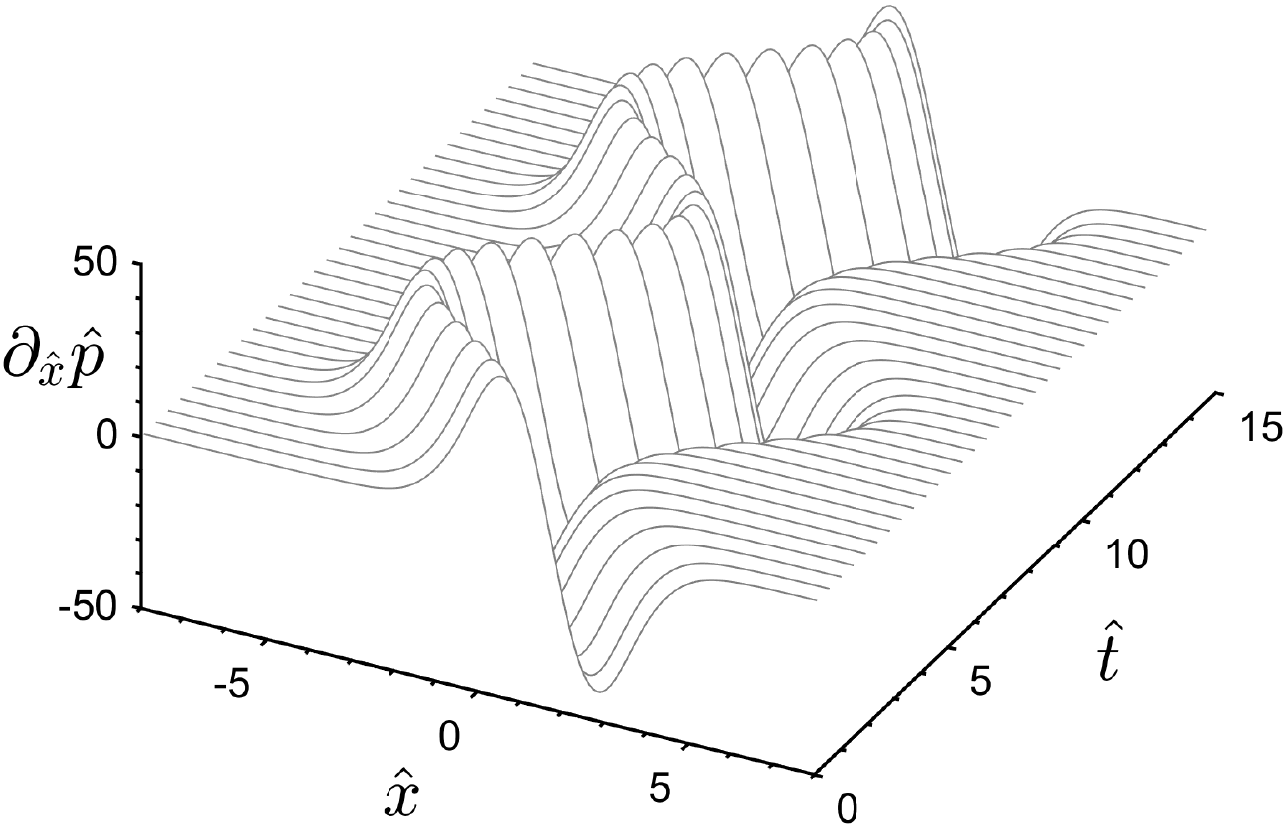}
         \caption{$y=x$}
         \label{3a}
     \end{subfigure}
     \hfill
     \begin{subfigure}[b]{0.32\textwidth}
         \centering
         \includegraphics[width=\textwidth]{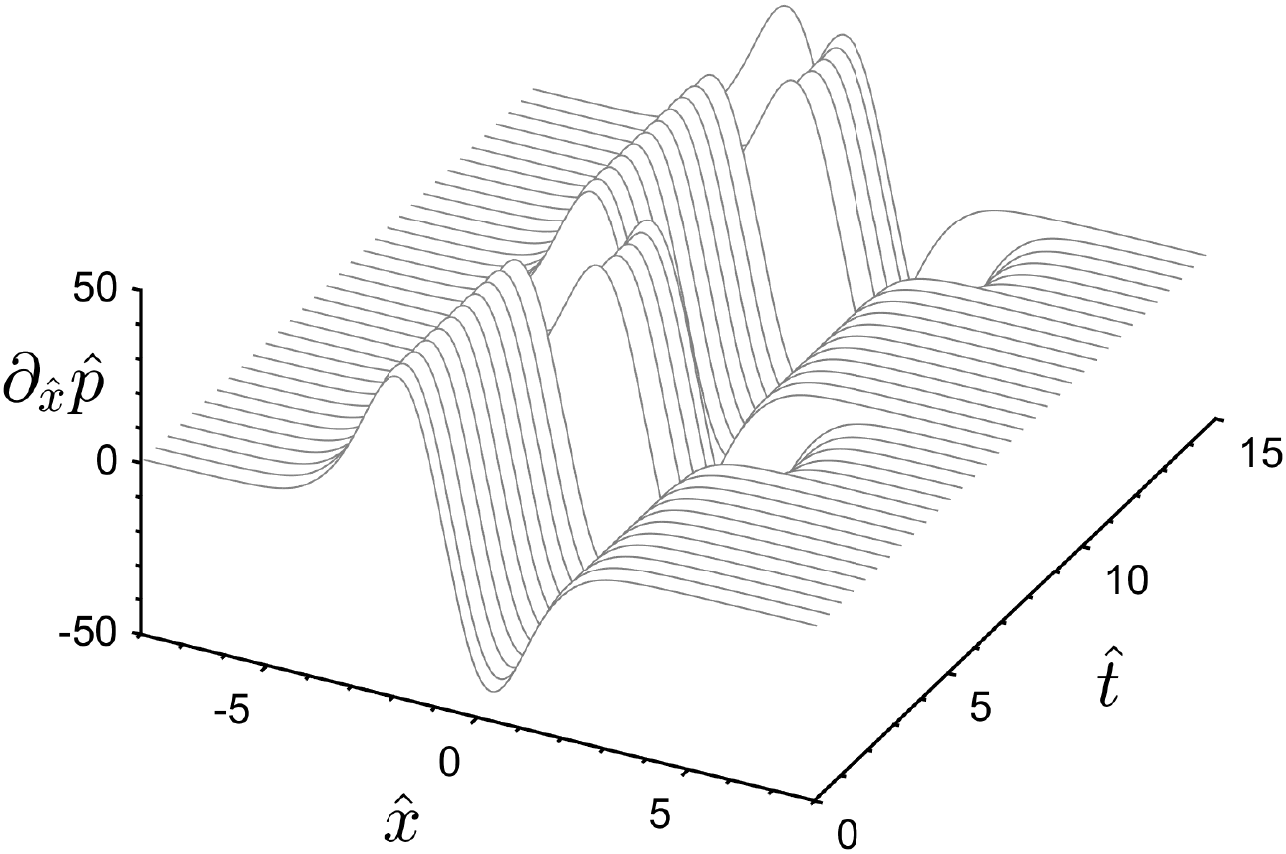}
         \caption{}
         \label{3b}
     \end{subfigure}
     \hfill
     \begin{subfigure}[b]{0.32\textwidth}
         \centering
         \includegraphics[width=\textwidth]{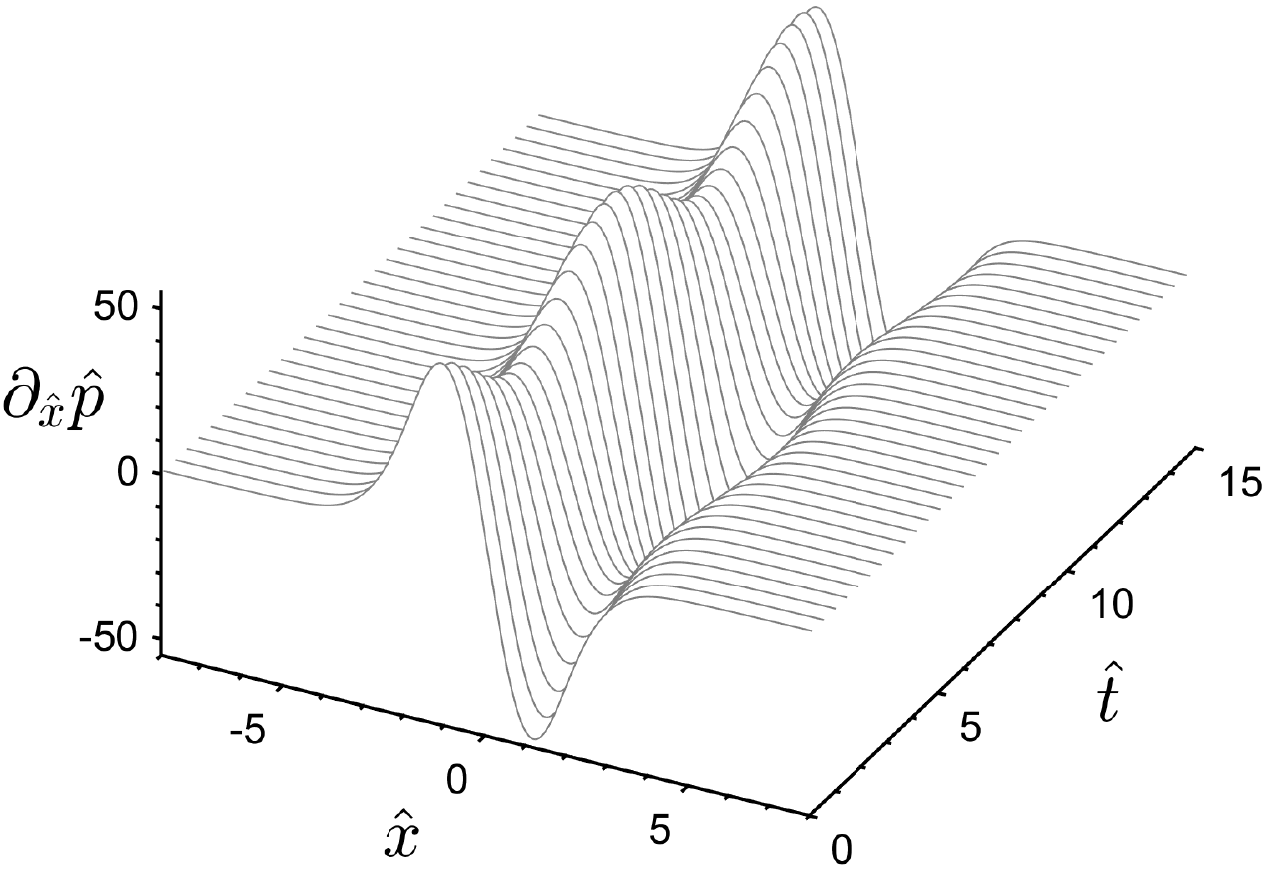}
         \caption{}
         \label{3c}
     \end{subfigure}
        \caption{Example of two oscillatory wave forms for the jet perturbation: (a) harmonic oscillation (b) upward biased ($\theta<0$) oscillation. The oscillation biased downward ($\theta>0$) is simply flipped along the $\hat x$ axis. The third case (c) is that of a jet pulsation.}
        \label{Example_WAVES}
\end{figure}

{These time dependent jet perturbations were designed to mimic different oscillatory modes in the impinging jet flow. Jet flow oscillations in the wiping process are experimentally investigated in \cite{Mendez2019} while a similar oscillatory mechanism was analyzed in \cite{Mendez2018} on a deformable interface reproducing both fixed and moving surfaces. While it is now known that jet oscillations are coupled to the interface instability, this work aimed at analyzing the response of the liquid film to possible jet disturbances, such as oscillations and pulsations, disregarding coupling effects.}

\section{Numerical Methods}\label{NUM}

\subsection{1D Solver for Integral Models}\label{SOLV1}

We here introduce the numerical methods to solve the set of equations \eqref{INT_GEN} . This is a system of hyperbolic PDEs that can be written in the general form 

\begin{equation}
\partial_t\,\mathbf{V}(x,t)+\partial_{x}\,\mathbf{F}(x,\mathbf{V})=\mathbf{S}(x,t,\mathbf{V})
\label{GENERAL_NUM}
\end{equation} with $\mathbf{V}$ the state vector of the problem, $\mathbf{F}$ the conservative flux and $\mathbf{S}$ the source term. In both the IBL and the WIBL model, the state vector is $\mathbf{V}=[h \,,\, q]^T$, the flux term is $\mathbf{F}=[q \,,\,\mathcal{F} \beta]$ and the source term is

\begin{equation}
 \mathbf{S}=\begin{bmatrix}0  \\ \Bigl(\hat h+ \hat h \partial_{ \hat x \hat x\hat x} \hat h- \hat h \partial_{\hat x}\hat p_g\Bigr)+\Delta {\hat \tau} \end{bmatrix} \frac{\beta}{\varepsilon\,Re}\,,
\end{equation} where the coefficient $\beta=6/5$ is introduced only for the WIBL and is $\beta=1$ otherwise. This coefficient is introduced to account for the contribution $\partial_t \hat q/5$ which appears in the shear stress term  $\Delta \tau$ (see \eqref{LAST_E} in Appendix \ref{TABS}). 

The system of PDEs in \eqref{GENERAL_NUM} has been extensively treated in the literature of non-homogeneous Shallow Water (SW) equations (in which the source term typically accounts for bed topography) and a wide range of suitable Finite Volume (FV) schemes for their numerical analysis is described in various textbooks  \citep{Toro2001,LeVeque2002}. Among these, two major classes can be distinguished in the literature: methods based on the (approximated) solutions of the Rieman problem (arising from Godunov's scheme) and methods based on centered fluxes (arising from the Lax-Friedrich scheme). The first class of methods is better suited to handle strong gradients, such as hydraulic jumps, while the second has the advantage of a much lower computational cost \citep{Kurganov2012,Hernandez-Duenas2016}. Because the investigated simulations do not produce shocks within the space and time domain of interest, this work focused on the second class of methods.

Centered schemes are usually used with a certain amount of artificial viscosity (see \cite{Mattsson2014,Ginting2018} and references therein), which can be introduced by a suitable combination of low order schemes and high order schemes. This combination is achieved using flux limiters \citep{LeVeque2002} to blend a high order scheme (e.g., Lax Wendroff) in regions where the solution is sufficiently smooth with a low order scheme (e.g., Upwind or Lax Friedrich) in regions of strong gradients. This approach combines the advantages of the two options: first-order schemes prevent numerical oscillations (dispersion) at the cost of excessively smoothing the solution, while the reverse is true for high order methods.

A standard Finite Volume (FV) formulation using explicit methods with three-point stencil in conservative form discretizes  \eqref{GENERAL_NUM} as:

\begin{equation}
 \mathbf V_{i}^{k+1} =  \mathbf V_{i}^{k}-\frac{\Delta t}{\Delta x}\Bigl[ \mathbf{F}^{+}-\mathbf{F}^{-}\Bigr]+ \Delta  t\, \mathbf{S}^{k}_{i}
\end{equation} where the $\mathbf{F}^{+}=\mathbf{F}(\mathbf V_{i}^{k},\mathbf V_{i+1}^{k})$ and $\mathbf{F}^{-}=\mathbf{F}(\mathbf V_{i}^{k},\mathbf V_{i-1}^{k})$ are the fluxes on the right and the left boundaries of each cell. In a flux limiting scheme, these are

\begin{equation}
   \mathbf{F}_i=\mathbf{F}^H_i+\Bigl(\mathbf{F}^L_i-\mathbf{F}^H_i\Bigr)\phi_i\,,
\end{equation} where $\phi_i$ is the flux limiting function, $\mathbf{F}^H$ is the flux calculated from an high order scheme and $\mathbf{F}^L$ is the flux calculated from a low order scheme.

In this work, we select the two steps Lax Friedrich Scheme (LxF in \cite{Shampine2005}) as a low order flux $\mathbf{F}^L$ and Richtmyer’s two-step variant of the Lax-Wendroff (LxW) method as high order flux $\mathbf{F}^H$. An efficient implementation of both schemes in \textit{Matlab} is provided by \citet{Shampine2005a}, and this work proposed minor modification to combine the two. These schemes are described in Appendix \ref{A_N}.

\begin{figure}
\centering
\includegraphics[width=3.8cm]{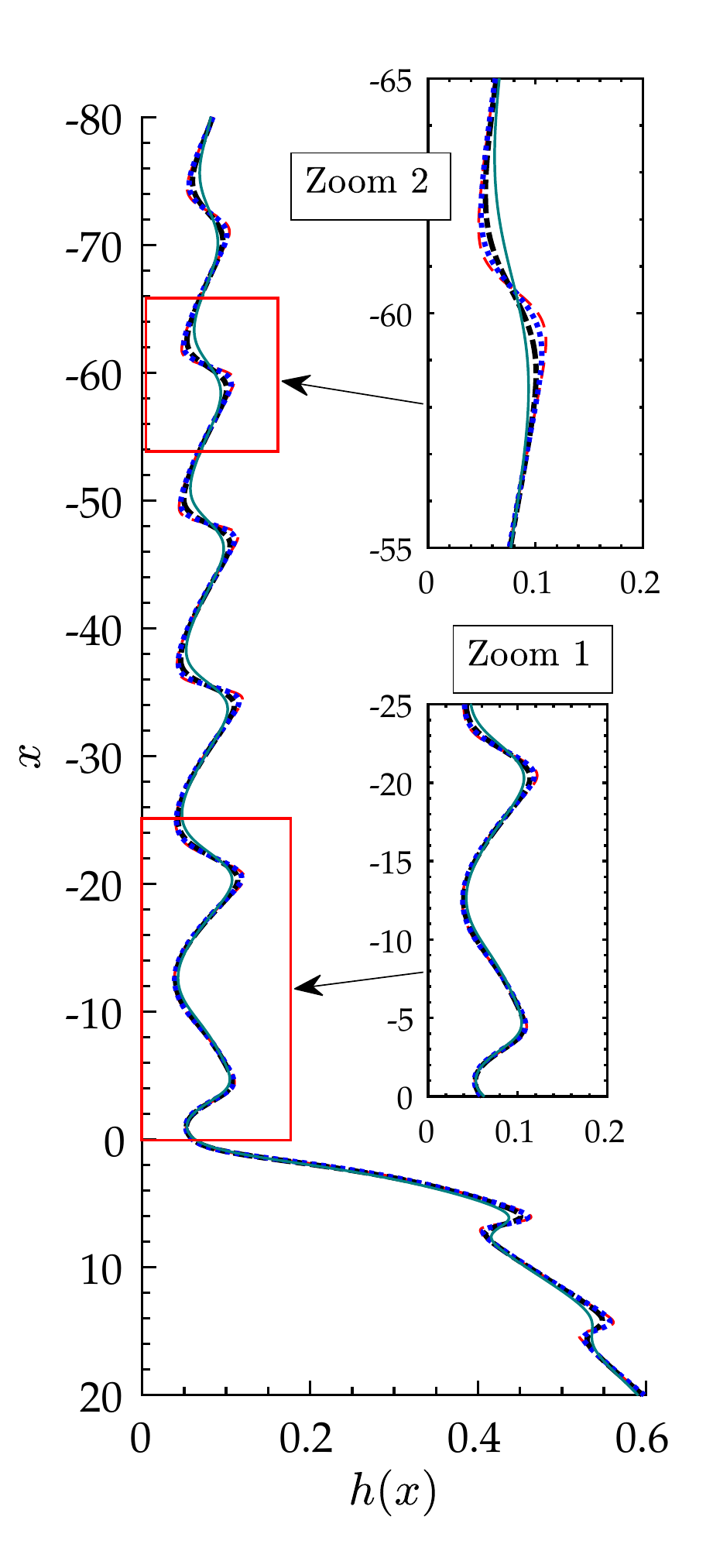}
\includegraphics[width=3.8cm]{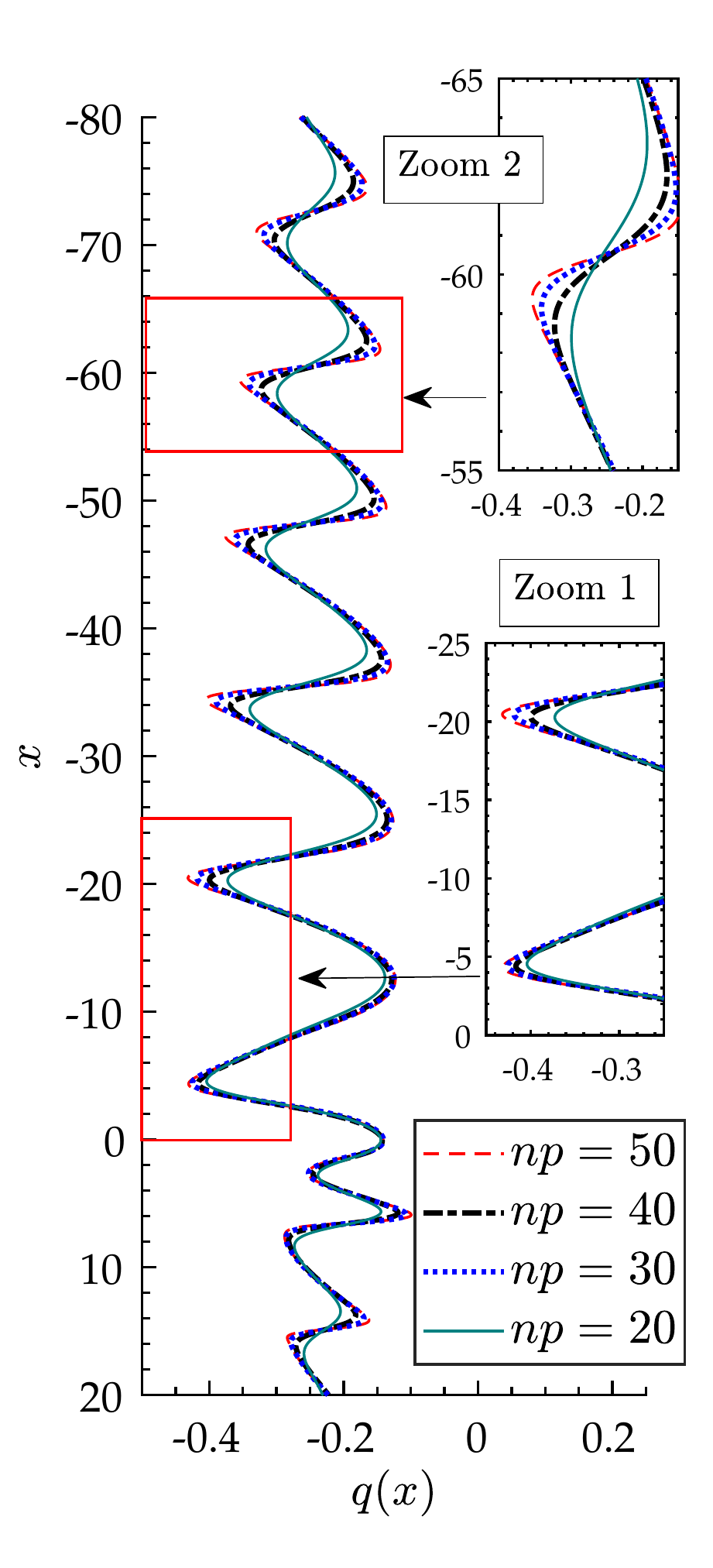}
\caption{Grid dependency analysis on the film thickness (left) and the flow rate (right). In the zoomed region far downstream the wiping point (Zoom 2), the effect of numerical diffusion makes the convergence {harder than in the wiping region (see Zoom 1). The simulations are carried out with $\Delta P_N=30 kPa$, $\theta_A=30^o$, $Z=15 mm$, $d=1.5mm$, $U_p=3 m/s$, $\nu_g=1.5e-5m^2/s$, using zinc as working fluid}.
}
\label{GRID_IND}
\end{figure}

The chosen limiter function is the classical min-mod:

\begin{equation}
\label{limiter}
\phi_i=\max\bigl(0,\min(1,\theta_i)\bigr)
\end{equation} where $\theta_i=(h_{i}-h_{i-1})/(h_{i+1}-h_{i})$ is the smoothness parameter based on the liquid film thickness as it is customary also in SW problems (e.g, \cite{Zhou2001}).

The proposed strategy allows for avoiding the calculation of the Jacobian and its eigendecomposition and is therefore computationally cost-effective. On the other hand, the numerical diffusion added by the low order scheme results in the smoothing of the waves in the liquid film. This smoothing becomes more evident as the waves move away from the wiping point from which they originate.

An example of a mesh independency study is shown in figure \ref{GRID_IND} for an instantaneous thickness and flow rate profile and four different meshes with $n_x=[1940, 2909, 3879, 4848]$ mesh points respectively. These are computed by setting the number of mesh points $n_P$ within the half-width $b$ of the Gaussian pressure distribution at the wall from \eqref{f_P}, so that $\Delta \hat x=b/([x] n_P)$ in dimensionless form. The simulations shown are computed with $n_P=[20,30,40,50]$, respectively, hence ensuring a reasonable accuracy in the calculation of the pressure gradient.
 
 The effect of numerical diffusion, as the number of mesh points is reduced, is evident in figure \ref{GRID_IND}. {However, as the focus of this work is placed on the response of the liquid film within a relatively short distance from the introduced perturbation, this is not considered as a limitation. }

In all the simulations of this work, {the boundary conditions are set as non-reflecting open boundaries while the initial solution is taken from the simplified 1D formulation. The simulation is run for about until a fully periodic response is produced in the film before the data for post processing is acquired.} The time step is taken by setting the $\Delta t=0.4 \Delta x$, i.e., assuming a $CFL=u_w\,\Delta t/\Delta x=0.8$ for waves traveling at $u_w\approx 2$, that is twice the substrate speed. Such estimation revealed to be rather conservative. 

\subsection{DNS Simulations and validation of the long-wave formulation}\label{OPEN}

{Before considering the jet wiping problem, we analyze the validity of long-wave formulations on a much simpler test case, namely the flow of a wavy liquid film over an upward moving substrate. This configuration is relevant to describe the dynamics of the final coating much downstream the wiping region, where the pressure gradient and the shear stress produced by the impinging gas jet vanish.}

{Although this test case is too simple for complete validation of the models, which is out of the scope of this work, we here focus on the validity of the long-wavelength formulation of the jet wiping problem at large Reynolds numbers. Moreover, we analyze the validity of the proposed Skhadov-like scaling by considering two liquids with largely different properties operating at the same dimensionless thickness $\hat{h}$ and re-scaled Reynolds number $\delta=\varepsilon Re$. The liquids considered are water and zinc.}

The validation has been carried out using Direct Numerical Simulations (DNS) of the two-phase flow using the VOF (Volume of Fluid) method, in which the liquid-gas interface is tracked on a fixed grid. Surface tension is accounted for through the Continuum Surface Force method \citep{Brackbill1992}, in which the surface force due to capillarity is converted into a volume force that acts across the interface thickness. The computational domain is shown in figure \ref{VALIDATION_GEO} along with the required boundary conditions. The liquid film has a mean (and initial) thickness $h_o$.

\begin{figure}
\centering
\begin{subfigure}[b]{0.5\textwidth}
\centering
\includegraphics[width=6.8cm]{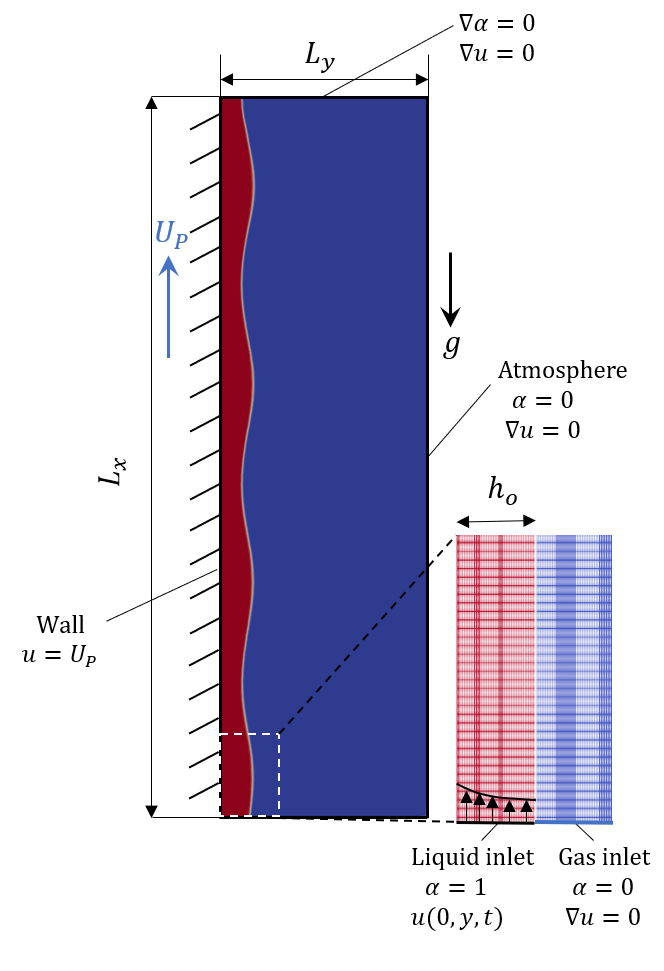}
\caption{}
\label{VALIDATION_GEO}
	\end{subfigure}
\begin{subfigure}[b]{0.32\textwidth}
\centering
\includegraphics[width=3.8cm]{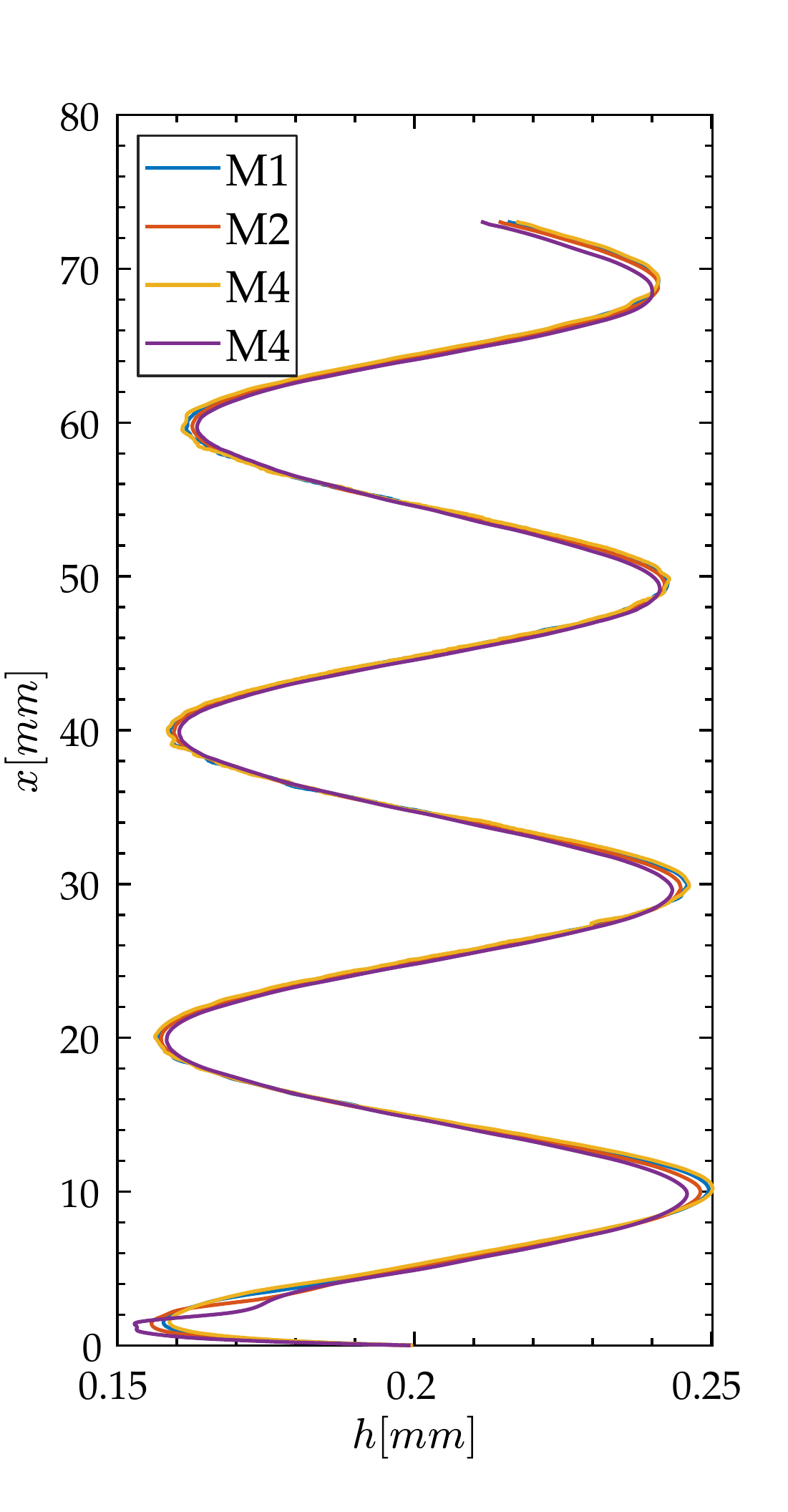}
\bigskip
\bigskip
\caption{}
\label{GRID_IND_ANNE}
	\end{subfigure}
\caption{Figure \ref{VALIDATION_GEO}: schematic of the flow configuration used for validation purposes using the OpenFoam solver \emph{InterFoam}: flow domain and zoom on the near-wall mesh. Fig \ref{GRID_IND_ANNE}: snapshot of the thickness evolution for the four meshes in table \ref{Meshes_density}.
} 
\end{figure}

The computational domain is rectangular, with a dimensionless length $L_x=8400 h_o$ for water and $L_x=12300 h_o$ for zinc in the streamwise direction, and $L_y=7.8 h_o$ for water and $L_y=7.5 h_o$ for zinc in the cross-stream direction. A perturbation of the film flow rate is introduced at the inlet of the domain via a pulsation of the film velocity (as in \cite{Doro2013}). At the liquid inlet (with thickness held constant and equal to $h_o$), the film velocity profile is prescribed as:

\begin{equation}
\label{PERT}
    \hat{u} (\hat{x}=0,\hat y)=\Biggl[\hat y \Biggl (\frac 1 2 \hat y - \hat h_o \Biggr)+1\Biggr]\Biggl[1+q_A\sin(2\pi\hat{f}\hat t)\Biggr]
\end{equation} where $\hat f=f [t]$ is the dimensionless oscillation frequency, $\hat t$ is the dimensionless time, and assuming that the $x$ axis points out in the same direction than the substrate velocity $U_P$. The corresponding flow rate per unit width is therefore:

\begin{equation}
    \hat{q}=\Bigl[\frac 1 3 \hat{h}^3-\hat h\Bigl]\Bigl[ 1+q_A\sin\bigl(2\pi\hat{f}\hat t\bigr) \Bigr]
\end{equation}

For these computations, the interFoam solver of the finite volume code OpenFoam is used. This solver has been extensively validated in the CFD literature \citep{Deshpande2012} and in various studies on falling liquid films (e.g. \cite{Gao2003,Doro2013,Dietze2014}) as well as co-current and counter-current gas-liquid flows \citep{Dietze2013}. The solver assumes incompressible and isothermal flow and features an interfacial compression flux term that activates at the interface to mitigate the effects of numerical smearing of the liquid-gas boundary. 

The liquid volume fraction $\alpha$ is fixed at the inlet, together with a Neumann condition for pressure. On the inlet boundary located in the gas phase ($ x _0=0$, $h_o \leq y \leq 7.8h_o$) and on the right hand side of the gas boundary ($y=7.8 h_o$, $0 \leq x \leq L_x$), $\alpha$ is fixed to 0, with a zero derivative for the velocity and a fixed total pressure. At the wall ($y=0$, $0 \leq x \leq L_x$), a no-slip condition is prescribed ($\hat u(\hat y=0)=1$), with a zero flux for $\alpha$ and a fixed pressure condition. At the outlet, a zero gradient is set for both velocity and liquid volume fraction. The flow field is initialized with the nominal thickness $h_o$. The velocity profile within the liquid is initially parabolic ($\hat{t}=0$ in \eqref{PERT}) while the velocity is set to 0 in the gas phase. In agreement with \eqref{WAVE_SIMP}, we consider cases with $\hat{h}\ll1$ such that the waves propagate upstream, which is in the direction of the substrate motion.

 \begin{table*}
\centering
\begin{tabular}{c|c|c}
Mesh Number & $\Delta x /h_o$ & $\Delta y /h_o$ \\
\hline
M1 & 0.110 & 0.022 \\
M2 & 0.156 & 0.031 \\
M3 & 0.235 & 0.047 \\
M4 & 0.391 & 0.078 \\
\end{tabular}
\caption{Mesh densities for the sensitivity study.}
\label{Meshes_density}
\end{table*}
 
 The simulations are carried out using a second-order backward Euler scheme in time for the transient term, and second-order discretization schemes for the convective (van Leer scheme for the $\alpha$ transport equation), diffusive and pressure terms. The coupling between pressure and velocity is solved using a standard PISO algorithm. The time step was set adaptatively, based on a maximum value of 0.3 for the global CFL number in the $\alpha$ equation. This leads to time steps of the order of $4.5 \times 10^{-6}$ s with water and $2.7 \times 10^{-6}$ s with zinc.
 
 In order to evaluate the influence of mesh density on the results, simulations are performed on four different grids with an increasing mesh density (see table \ref{Meshes_density}): the streamwise cell size $\Delta x$ is varied between $0.11 h_o$ and $0.39 h_o$, and the cross-stream cell size $\Delta y$ between $0.022$ and $0.078 h_o$. For these tests, water was used as the working fluid, and the length of the domain was reduced to save computational time ($L_x=1500 h_o$). The substrate speed is fixed to 1 \SI{}{m/s}. The results in figure \ref{GRID_IND_ANNE} show that the selected mesh densities are sufficient to capture the sinusoidal waves that form shortly after the inlet, and that beyond the density of mesh M2, the thickness profiles are almost insensitive to the size of the cells. The meshes used in the validation process have cell sizes of the order of $\Delta x=0.23 h_o$ and $\Delta y=0.046 h_o$, resulting in grids of 3.3 millions for the test case with water and 4.6 millions for the one with zinc.

\begin{table*}
\centering
\begin{tabular}{@{}cccccccccccc@{}}
\toprule
 & $\hat{h}_o$ & ${h}_o$ & $\hat{f}_o$ & $\rho_l$ & $\sigma$ & $\nu$ & $f$ & $U_p$ & $Re$ & $Ca$ & $\delta=\varepsilon Re$ \\ \midrule
 & $[-]$ & $[\mu m]$ & $[-]$ & $[kg/m^3]$ & $[N/m]$ & $\times 10^{-6} m^2/s$ & $[Hz]$ & $[m/s]$ & $[-]$ & $[-]$ & $[-]$ \\ \midrule
Water & $0.2$ & $63.9$ & $0.05$ & $998.2$ & $0.073$ & $1$ & $37.4$ & $1$ & $319$ & $0.0137$ & $76.3$ \\
Zinc & $0.2$ & $42.7$ & $0.05$ & $6500$ & $0.78$ & $0.45$ & $37.4$ & $1$ & $478$ & $0.0037$ & $73.9$  \\ \bottomrule
\end{tabular}
\caption{Physical parameters and operating conditions of the two test cases used for validation purposes}
\label{Table3}
\end{table*}

\section{Results}\label{Res}

\subsection{Validation test cases}\label{RES_A}

 The physical parameters and the operating conditions for the two simplified cases with water and zinc are recalled in table \ref{Table3}. Both test consider a dimensionless liquid thickness of $h_o/[h]=0.2$, perturbed at the inlet with a flow rate pulsation with amplitude $q_A=0.2$ and dimensionless frequency $\hat{f}=0.05$. The substrate velocity is taken as $U_p=1\, \SI{}{m/s}$ for both cases. These two liquids differ by one order of magnitude in surface tension and almost an order of magnitude in density and dynamic viscosity. However, their kinematic viscosity is comparable, resulting in similar time scales (cf. table \ref{Scaling_Table}), and similar rescaled Reynolds number $\delta=\varepsilon Re$.

\begin{figure}
    \centering
    \begin{subfigure}{0.4\textwidth}
        \centering
       \hspace{2mm} Zinc \hspace{19mm}  Water\\
 \includegraphics[width=2.4cm]{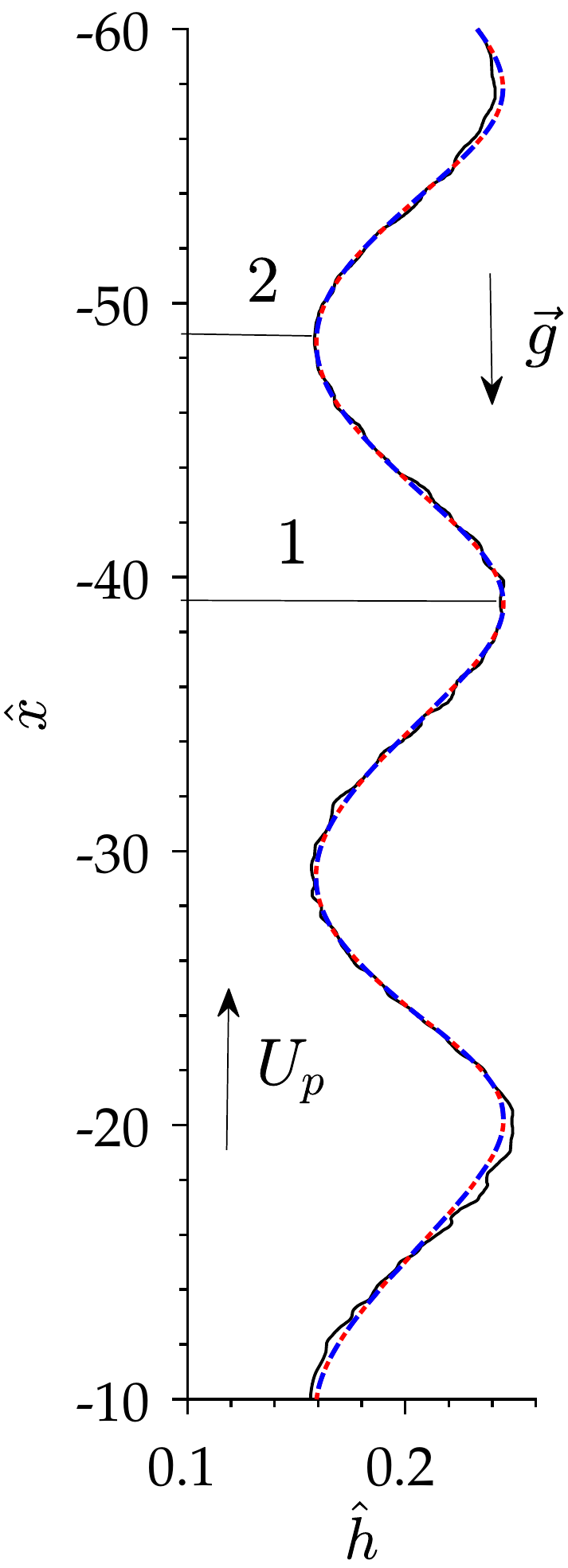}
 \hspace{3mm}
\includegraphics[width=2.3cm]{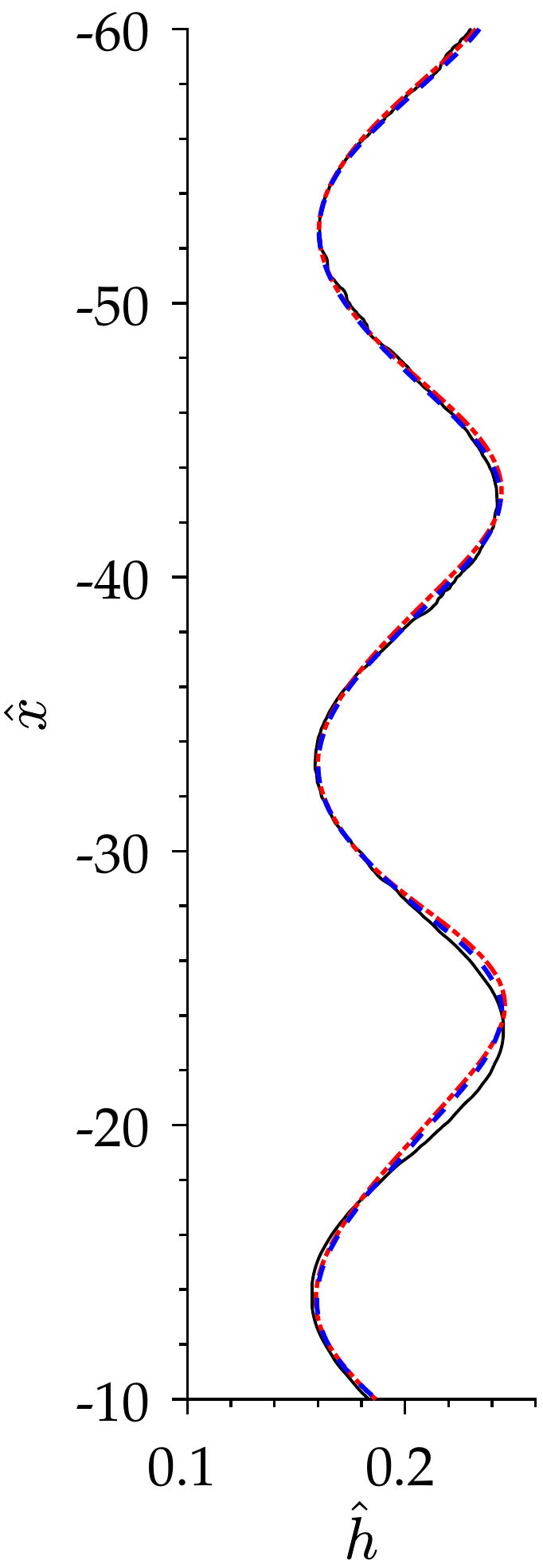}\\
\hspace{2mm}
\includegraphics[width=3.5cm]{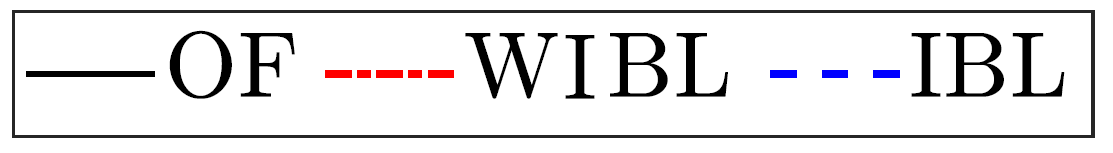}
        \caption{}
        \label{Validation_Profiles}
    \end{subfigure}
    \begin{subfigure}{0.45\textwidth}
        \centering
    \includegraphics[width=7cm]{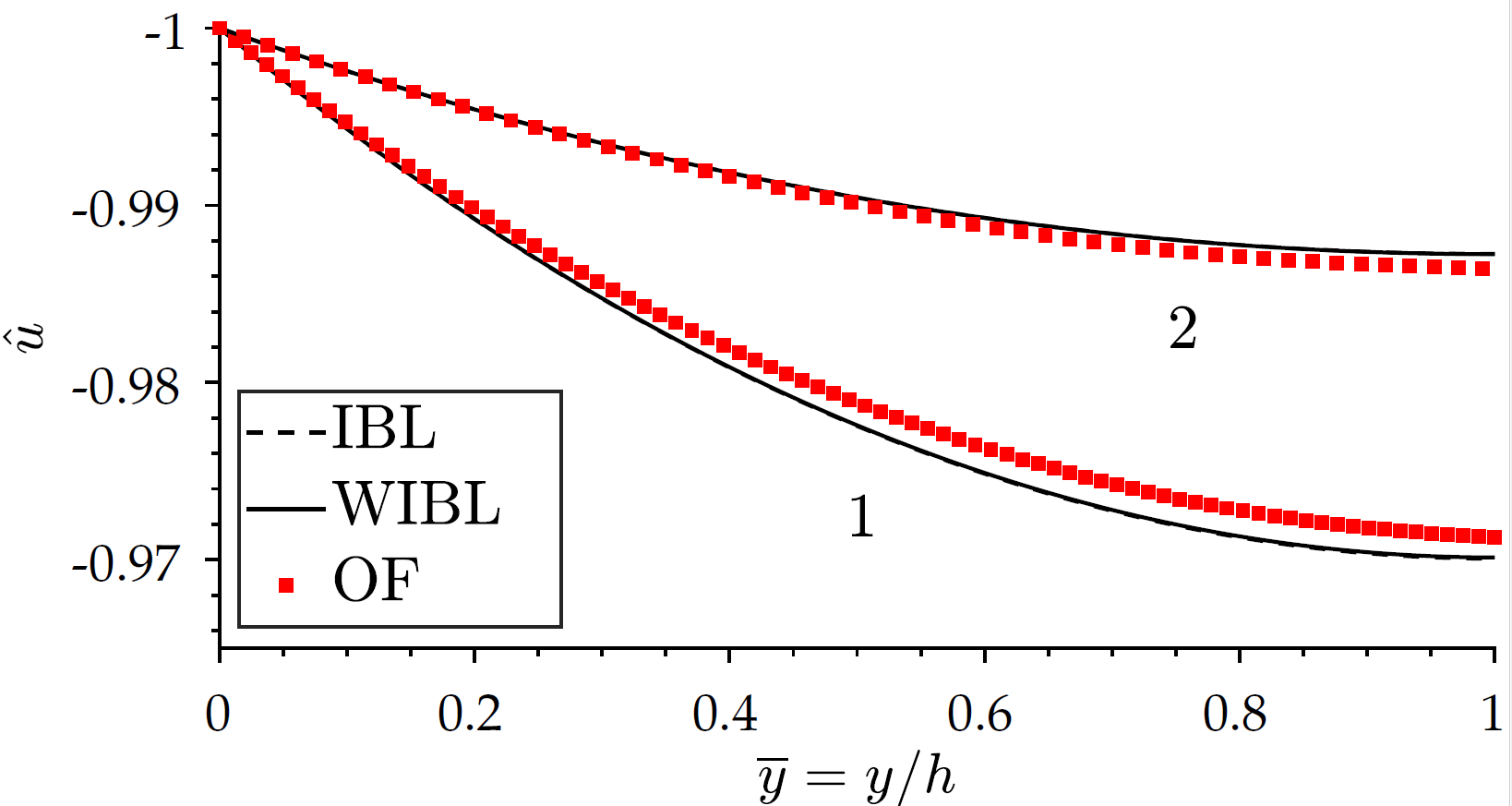}
        \caption{}
        \label{Vel_Profiles}
    \end{subfigure}
    \caption{ a) Instantaneous thickness profile for the liquid film thickness the liquid zinc (left) and the water (right) test cases. b) Comparison of the velocity profiles extracted from the zinc simulations in OpenFoam and IBL/WIBL simulations. The profiles are extracted from the wave maxima (location 1) and wave minima (location 2) in the instantaneous in figure a).}
\end{figure}

\begin{table*}
\centering
\begin{tabular}{@{}c|c|c|cc@{}}
\cmidrule(l){2-5}
 & \multicolumn{2}{c|}{Zinc ($\delta=74.2$, $\hat{f}=0.05$)} & \multicolumn{2}{c}{Water ($\delta=76.5$, $\hat{f}=0.05$)} \\ \cmidrule(l){2-5} 
 & IBL/WIBL & OF & \multicolumn{1}{c|}{IBL/WIBL} & \multicolumn{1}{c|}{OF} \\ \midrule
$\hat{q}_{\rm{min}}$ & -0.158 & -0.155 & \multicolumn{1}{c|}{-0.159} & \multicolumn{1}{c|}{-0.153} \\
$\hat{q}_{\rm{max}}$ & -0.240 & -0.238 & \multicolumn{1}{c|}{-0.239} & \multicolumn{1}{c|}{-0.234} \\
$\hat{h}_{\rm{min}}$ & 0.159 & 0.156 & \multicolumn{1}{c|}{0.160} & \multicolumn{1}{c|}{0.158} \\
$\hat{h}_{\rm{max}}$ & 0.245 & 0.244 & \multicolumn{1}{c|}{0.244} & \multicolumn{1}{c|}{0.242} \\
$\hat{\lambda}$ & 18.9 & 18.7 & \multicolumn{1}{c|}{18.8} & \multicolumn{1}{c|}{18.6} \\ \bottomrule
\end{tabular}
\caption{Summary of the results in terms of flow rate and thickness maxima/minima (subscript min-max) and wavelength ($\hat{\lambda}$) for the IBL/WIBL models and the Openfoam (OF) simulation.}
\label{TAB_VAL}
\end{table*}

An instantaneous thickness profile is shown in figure \ref{Validation_Profiles} for both cases, comparing the Integral Boundary Layer Model (IBL), the Weighted Integral Boundary Layer (WIBL) and the OpenFoam (OF) simulations. Since $Re_F\approx 95$ in zinc and $Re_F\approx 65$ in water, the TTBL recovers the IBL model and its results are not shown. The thickness profiles are perfectly overlapping, demonstrating the validity of the integral formulation and the numerical methods, as well as the capability of the long-wave formulation to model the flow.

Table \ref{TAB_VAL} collects all results in terms of flow rate and thickness maxima and minima, while figure \ref{Vel_Profiles} shows the velocity profile underneath a maximum (1) and a minimum (2) of the film waves in the simulations of zinc.
The location in which the profiles are re-computed is indicated in figure \eqref{Validation_Profiles}.
For both IBL and WIBL, these profiles are reconstructed from the results of the simulation ($\hat{h},\hat{q}$). In the IBL, this is a straightforward implementation of equation \eqref{uQS}; in the case of the WIBL, this involves all the relations in table \ref{As} with equations \ref{Vel_PROF} and \eqref{U_Three}. For the WIBL, reconstructing the velocity profile is an ill-posed problem, since the closure of the model (i.e. the expression for $\Delta \hat{\tau}$ in \eqref{LAST_E}) assumes that $\delta \ll1$ while in this case $\delta\sim 75$. Nevertheless, in these simple examples, the $\Delta \tau^{(1)}$ term is small enough to let the WIBL converge on the IBL despite the large $\delta$, and the reconstructed profile reflects this convergence. As later discussed in section \ref{RES_D}, this does not happen for the jet wiping configurations at higher Reynolds numbers, as the higher order corrections of the velocity profile becomes more important than the zero-th order.

Finally, it is worth highlighting excellent agreement between the integral models and the DNS calculations, which reveals parabolic velocity profiles. {This result shows the important effect of the substrate motion as compared to the classic falling film problem, where the departure from the parabolic profile occurs at a much lower Reynolds number than the values considered here (see \cite{Denner2018}).}

\subsection{The relative contribution of forces}\label{RES_B}

This section focuses on the relative importance of all the terms in the integral momentum formulation \eqref{I2} as the wiping strength (as measured by the wiping number $\Pi_g$) is increased. The WIBL and the TTBL models are considered. As in the previous section, both liquid zinc and water are analyzed as working fluids. 

The liquid properties are the same as the previous section, as listed in table \ref{Table3}. In the case of zinc, the wiping conditions are taken to be representative of an industrial galvanizing line, with a nozzle having opening $d=1.5$ \SI{}{mm} and stand-off distance $Z=15\,\rm{mm}$. The substrate velocities $U_p=[1,2,3]$ \SI{}{m/s}, corresponding to $Re=[478,1352,2483]$ and $\delta=[74,263,554]$ are considered and, for each of these, the pressure in the nozzle is varied in the range $\Delta P_N=[3,40]$\SI{}{kPa}. This leads to wiping numbers in the range $\Pi_g=[0.7,2.9]$ and a shear stress number in the range $\mathcal{T}_g=[5,70]$.

In the case of water, the nozzle parameters $Z$ and $d$ are the same as for the zinc cases. The substrate velocity is reduced to $U_p=[0.2,0.3,0.4]$ \SI{}{m/s}, corresponding to $Re=[28,52,80]$ $\delta=[4,8.4,14.3]$, and the nozzle pressure decreased down to the range $\Delta P_N=[0.7,1.5]$ \SI{}{kPa}. This leads to much lower wiping numbers, in the range $\Pi_g=[0.3,0.8]$, but comparable shear stress number, in the range $\mathcal{T}_g=[30,80]$. The simulated wiping conditions falls within the operational range of the Essor VKI facility \citep{Buchlin}.

\begin{figure*}
\centering
\begin{centering}
 Liquid Zinc \hspace{58mm} Water\\
\end{centering}
\includegraphics[trim={0.3cm 0 1.5cm 0},clip,width=6.7cm]{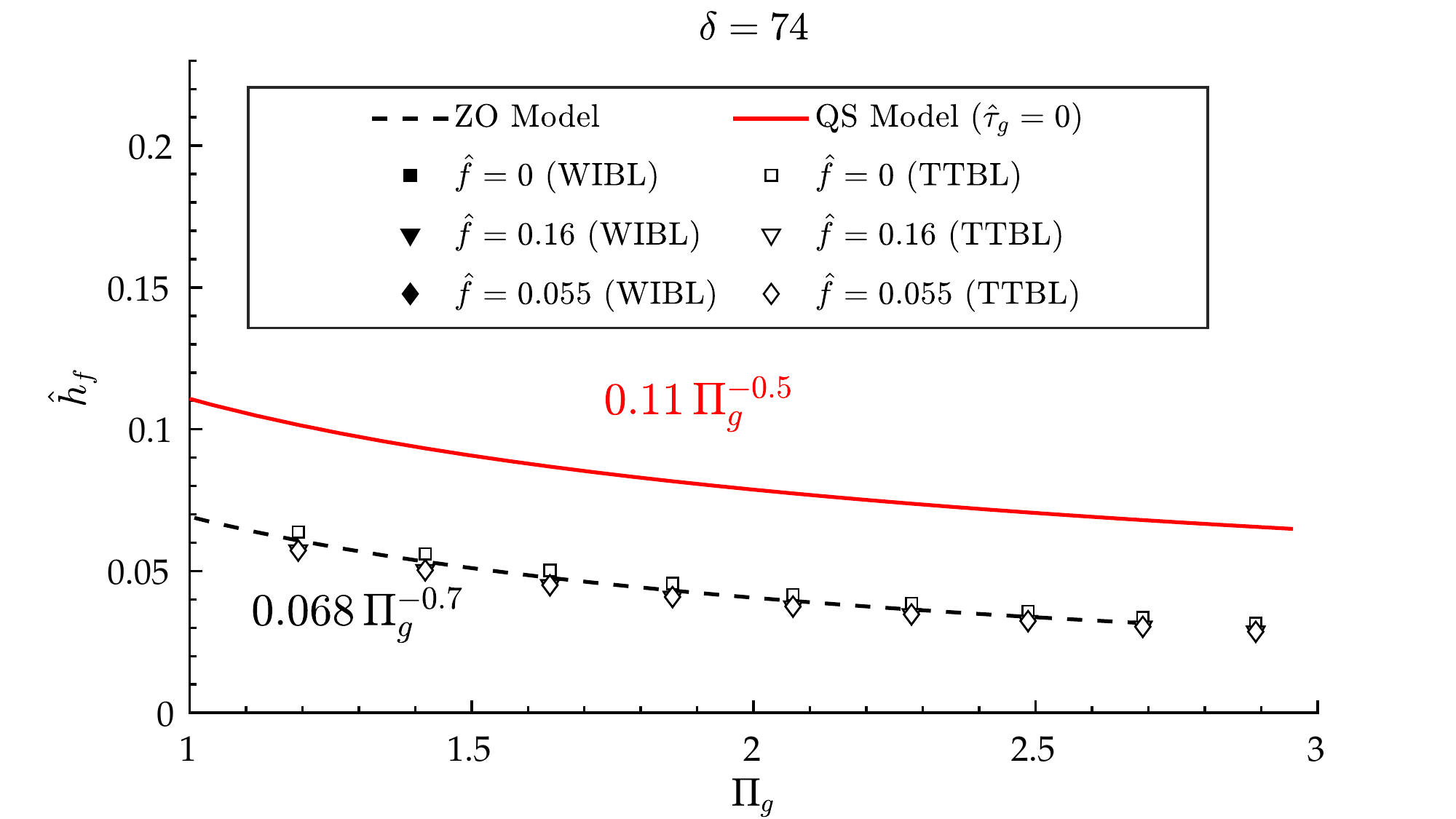}
\includegraphics[trim={0.3cm 0 1.5cm 0},clip,width=6.7cm]{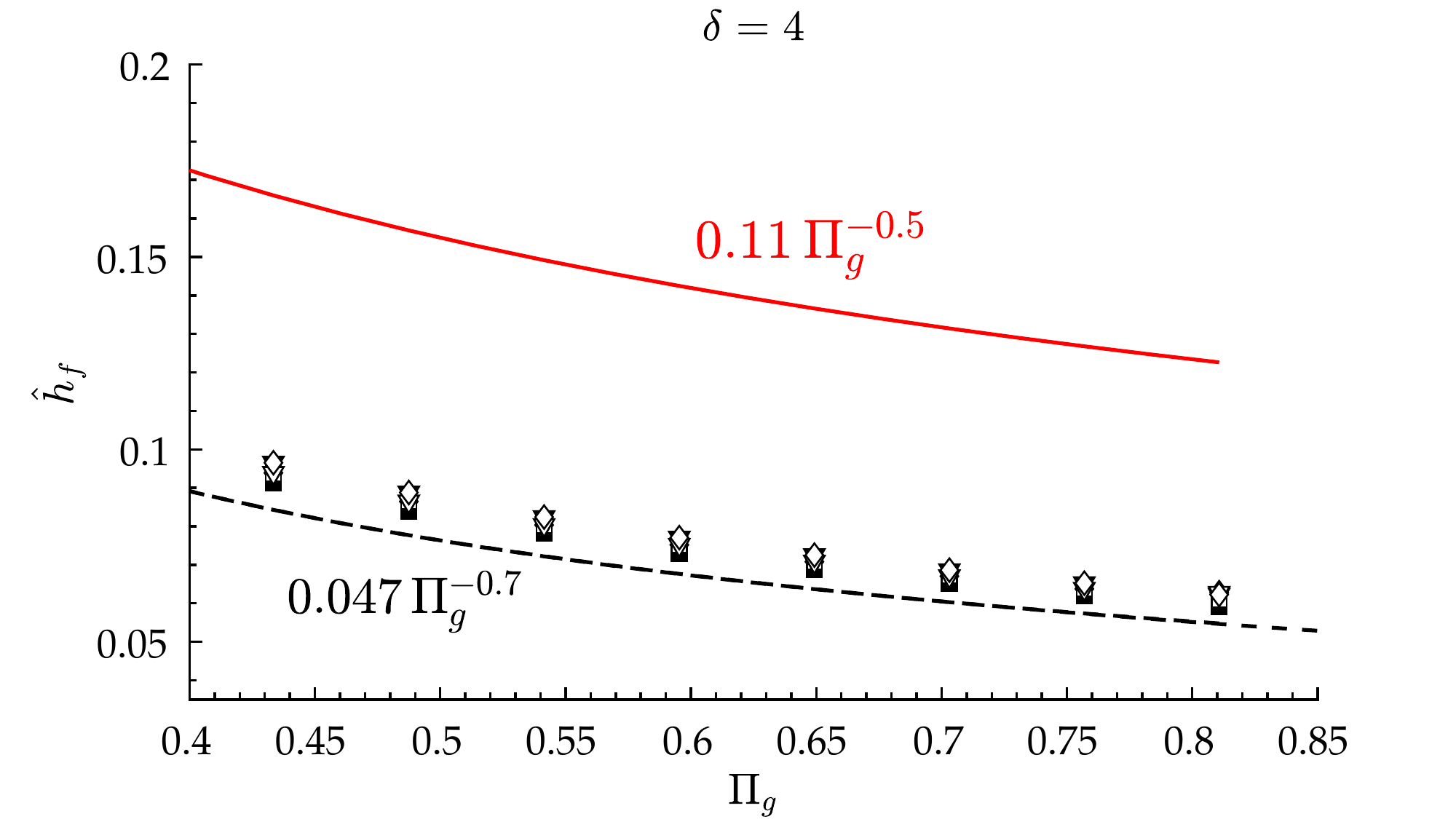}\\
\includegraphics[trim={0.3cm 0 1.5cm 0},clip,width=6.7cm]{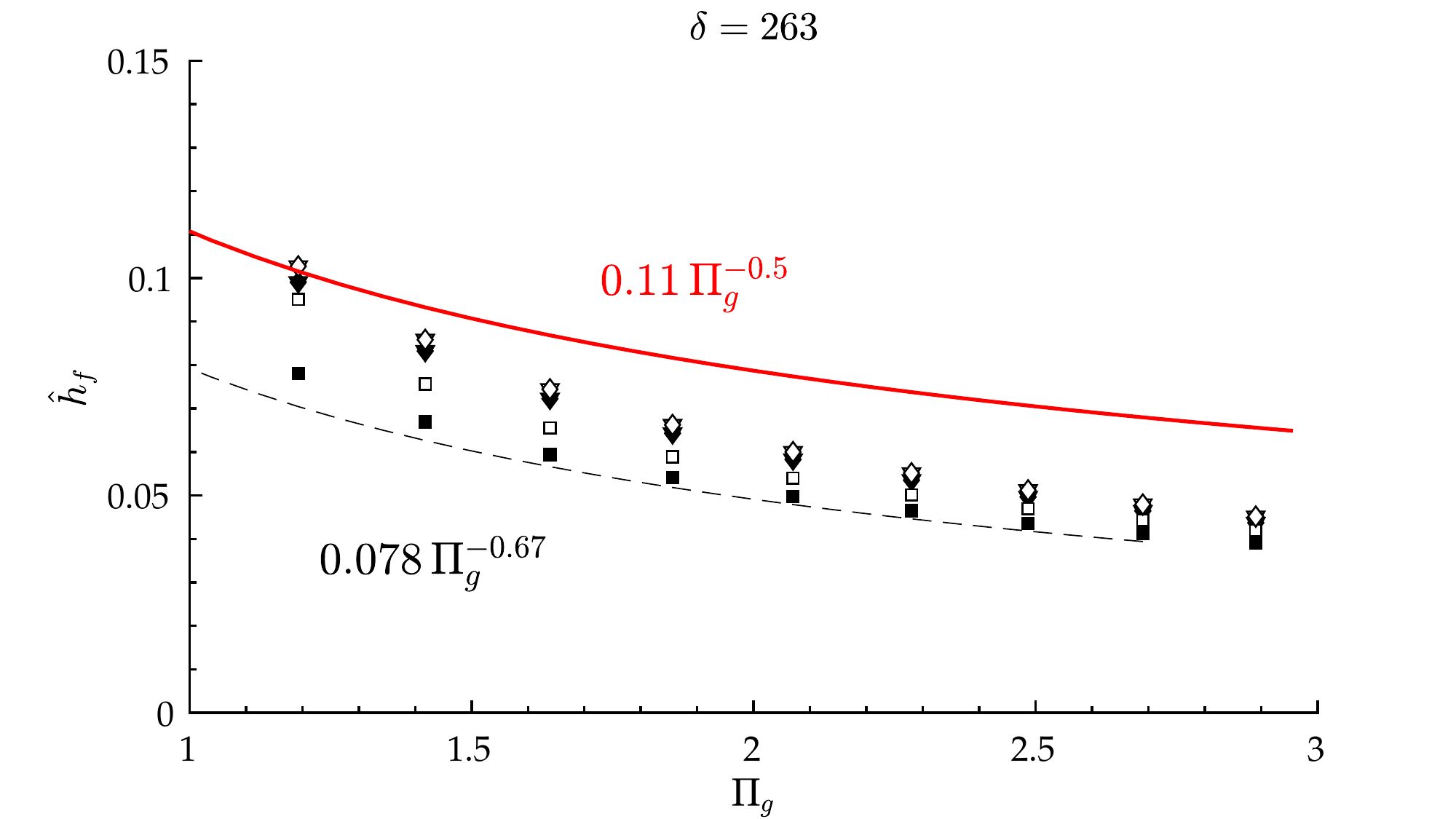}
\includegraphics[trim={0.3cm 0 1.5cm 0},clip,width=6.7cm]{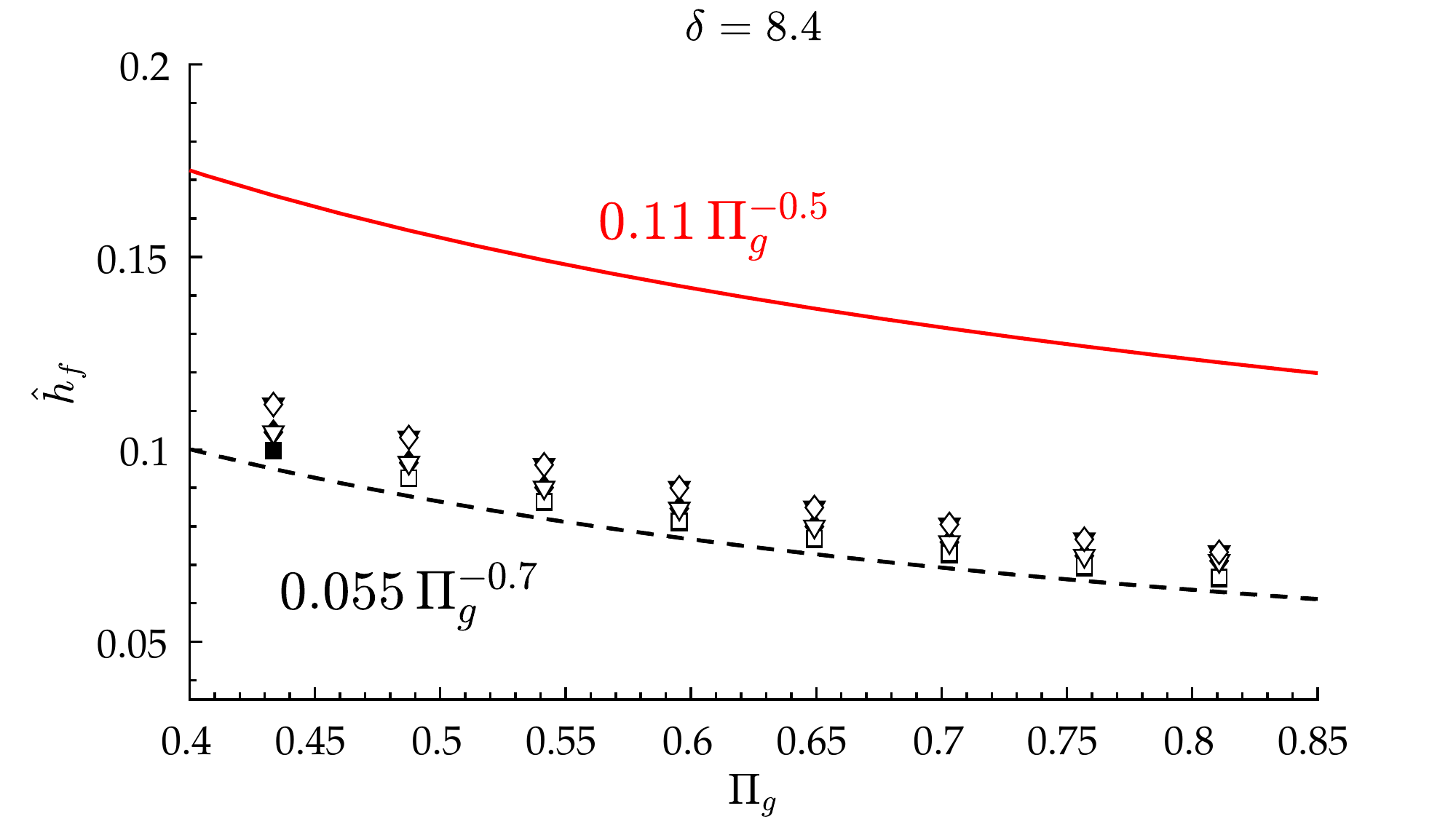}\\
\includegraphics[trim={0.3cm 0 1.5cm 0},clip,width=6.7cm]{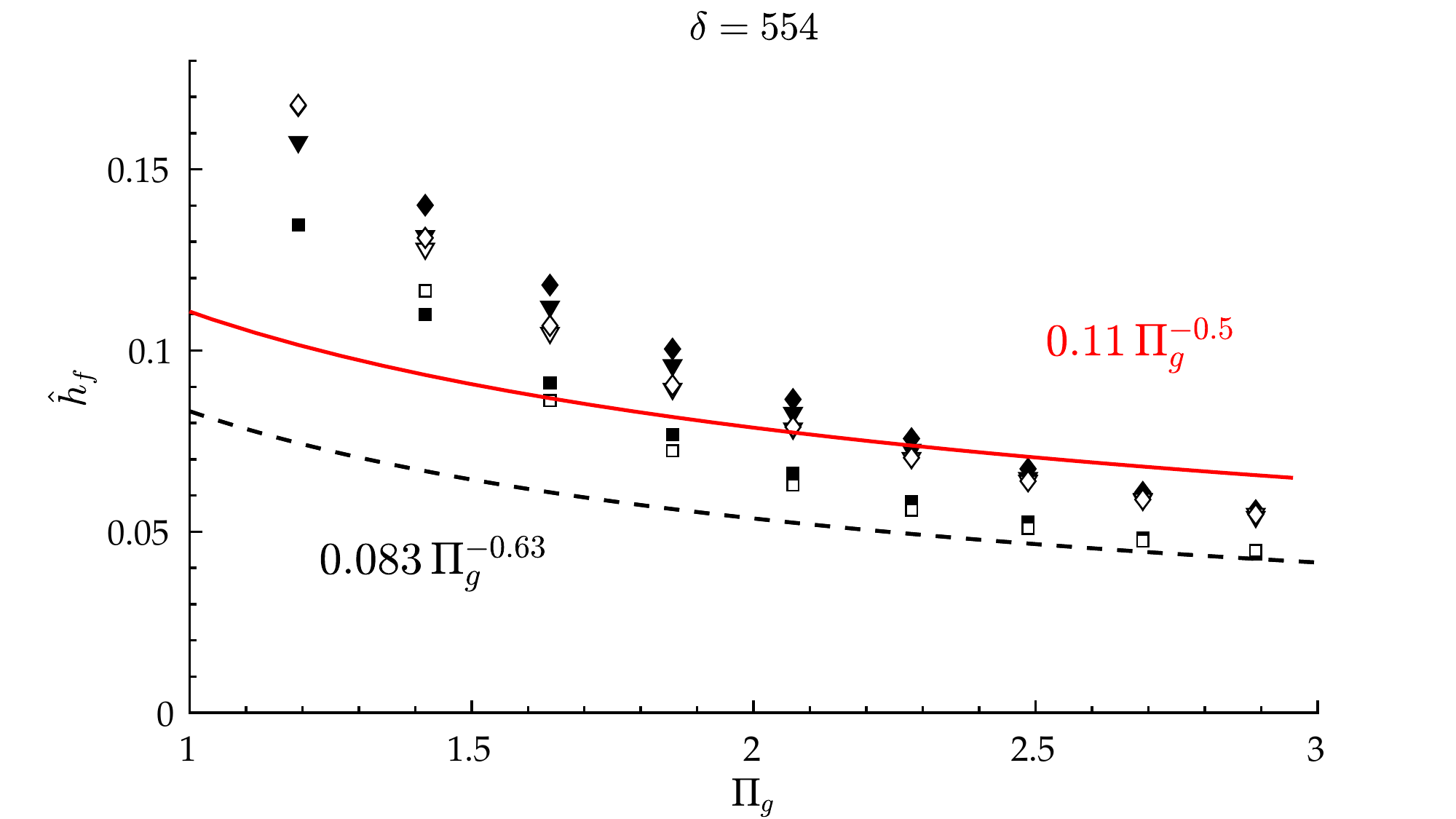}
\includegraphics[trim={0.3cm 0 1.5cm 0},clip,width=6.7cm]{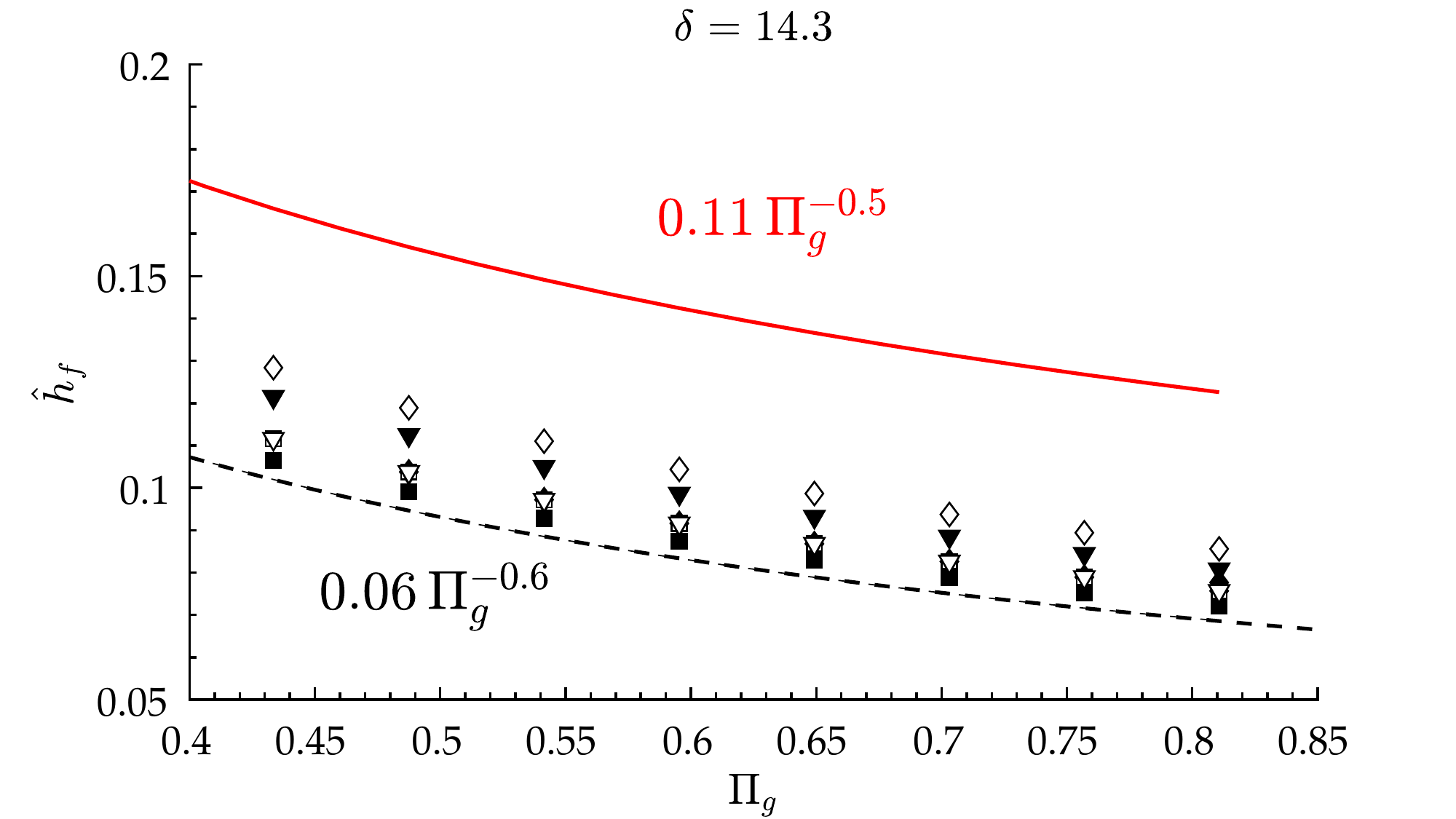}
\caption{Time averaged thickness in the final coating film for three rescaled Reynolds number (indicated in each plot) and various oscillation frequencies ($\hat{f}=[0,0.055,0.16]$). The black dashed line and the red continuous line correspond to the prediction from the Zero Order (ZO), the first taking into account the contribution of the shear stress, the second setting the shear stress to zero. Black marker refers to simulations using the WIBL model; white markers refers to simulations using the TTBL model. 
}
\label{RES_1}
\end{figure*}

In all the investigated configurations, the harmonic oscillations of the jet have an amplitude of $\theta_A=10^{o}$. It is worth noticing that in the Skhadov scaling, this leads to different amplitudes of the oscillations along the streamwise direction ($\hat{x}$) as the jet stand-off distance is the same while the streamwise length scale is not.

Three dimensionless frequencies are considered, i.e. $\hat{f}=[0,0.055,0.16]$. The zero-frequency simulates the process in steady-state conditions, yet accounting for the role of inertia and surface tension. Several preliminary tests in this configuration confirms that the flow is absolutely stable, meaning that initial disturbances in the film move away from the domain (towards $\hat{x}\rightarrow \infty$ if the disturbance is located in the run backflow; towards $\hat{x}\rightarrow -\infty$ otherwise) leaving the steady-state solution unvaried. The highest frequency $\hat{f}=0.16$, as further discussed in section \ref{RES_C}, {is damped by the liquid film and results in no appreciable undulation}. \textcolor{black}On the contrary, the frequency $\hat{f}=0.055$ is in the range of maximum receptivity of the liquid film and produces the largest waves.

The time-averaged coating thickness downstream of the wiping is shown in figure \ref{RES_1} for both liquids. The left column refers to the wiping conditions in zinc, and the right column to the wiping conditions in water. From top to bottom, the velocity of the substrate is increased, and the corresponding rescaled Reynolds number $\delta=\varepsilon Re$ is indicated.
Each figure compares the prediction of the zero order (ZO) model presented in section \ref{QS}. This comparison is made with (black dashed lines) and without (continuous red line) the shear stress in the model. The results of the ZO models are well described by power laws of the form $a \Pi^{b}_g$, reported in the figures.

In the presence of shear stress, the power correlation changes slightly with the substrate speed, while this remains unaltered if the shear stress is removed. This result is in remarkable agreement with the experimental correlations presented in previous experimental works \citep{Gosset2019,Mendez2019}. It is interesting to observe that these experimental works were carried out on a much more viscous mineral oil, producing similar wiping numbers (in the range $\Pi_g=[0.1,0.8]$) but much lower shear stress numbers (in the range $\mathcal{T}_g=[1,8]$). This highlights the role of the shear stress in the wiping process for liquids with low kinematic viscosity such as zinc or water and confirms the experimental observation that the wiping of highly viscous liquids is mostly governed by the dimensionless group $\Pi_g$.

Concerning the role of the substrate speed on the mean film, the ZO model predicts a negligible impact in all the test cases, while the WIBL and TTBL models reveal a discrepancy that becomes more important at lower wiping numbers and large Reynolds numbers. Overall, the WIBL and the TTBL models are in good agreement in all the simulations analyzed (at small Reynolds numbers, these are indistinguishable). The results from these models at $\hat{f}=0$ (trends with square markers) show that surface tension and advection produce a significant departure from the wiping curve obtained by the simplified model at $\Pi_g\rightarrow 0$ while the agreement is asymptotically reached at $\Pi_g\rightarrow \infty$.

It is worth noticing that the cases with jet oscillation yield a higher mean coating thickness regardless of the oscillation frequencies. This result is particularly interesting if one considers that the case at $\hat{f}=0.16$ yields no undulation in the final coat as later discussed in section \ref{RES_C}. Yet, the mean thickness increases, as if the oscillation spreads momentum and results in an effective distribution that is closer to the time-averaged profiles. A similar phenomenon is also observed by \citet{Lunz2018}, who have studied the response of a liquid film to an oscillatory pressure source and discussed the calculation of the effective pressure.

\begin{figure}
\centering
\begin{tabular}{@{}ccc@{}}
      & $Re=478$\quad $\delta=74$                   & $Re=2483$\quad $\delta=554$ \\ 
\rotatebox{90}{\hspace{22mm} $\hat{f}=0$}  &
\multicolumn{1}{c|}{
\includegraphics[width=2cm]{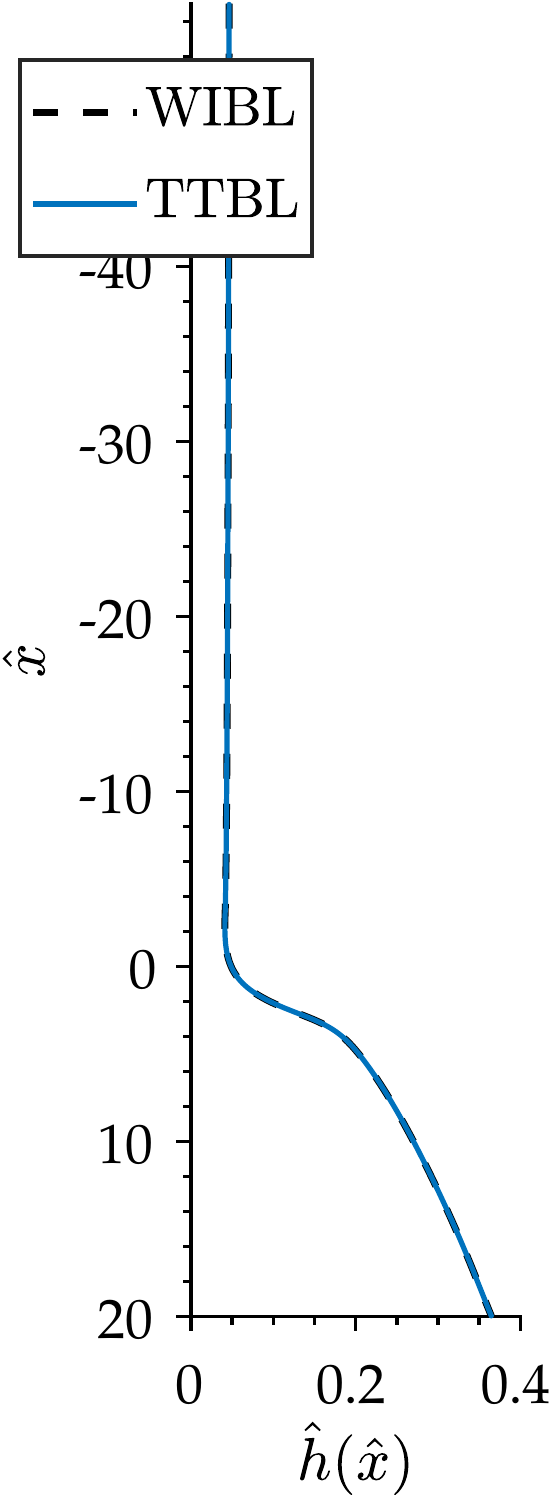}
\includegraphics[width=2cm]{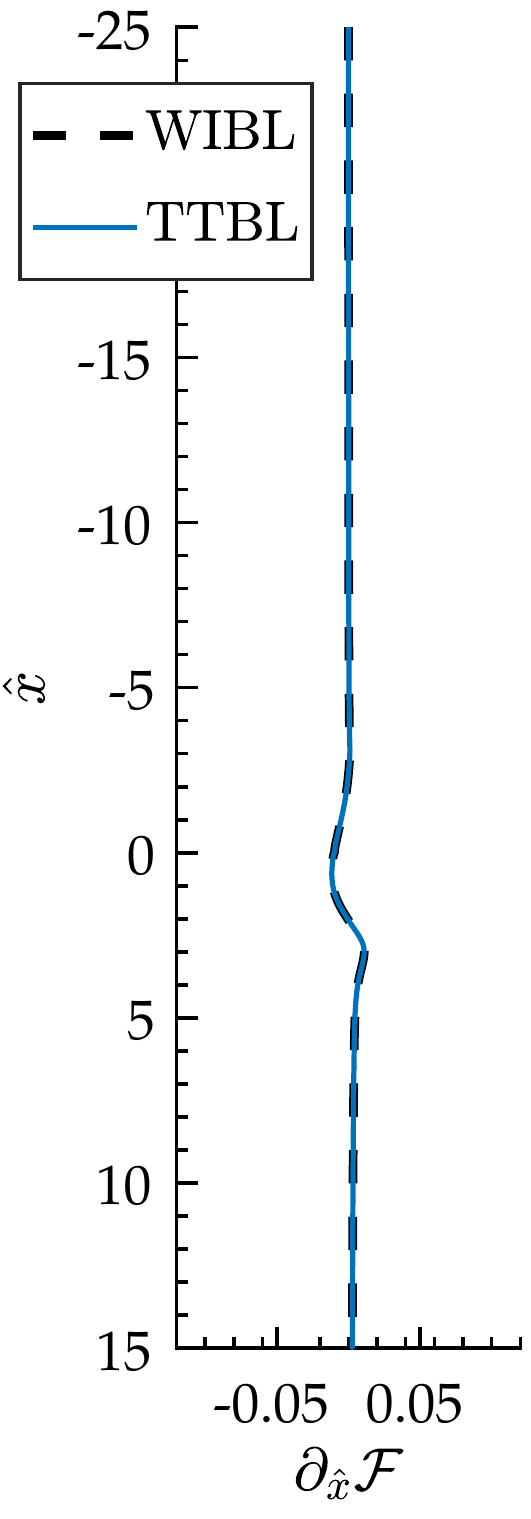}
\includegraphics[width=2cm]{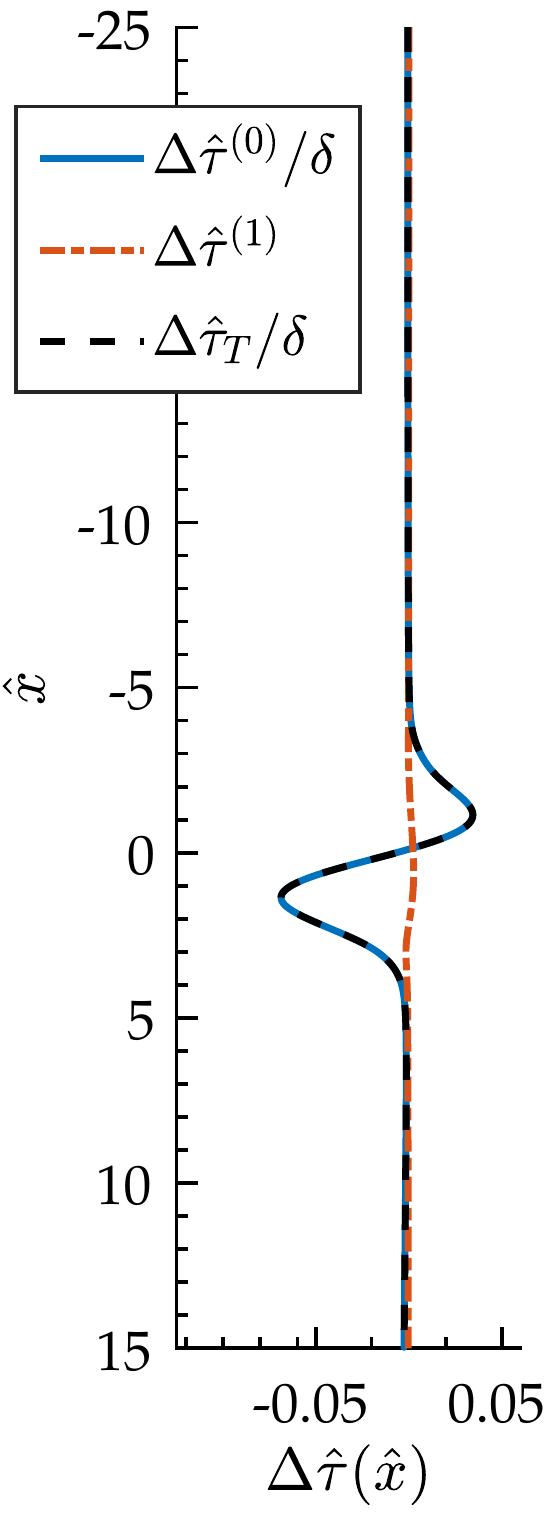}} & 
        \multicolumn{1}{c}{
\includegraphics[width=2cm]{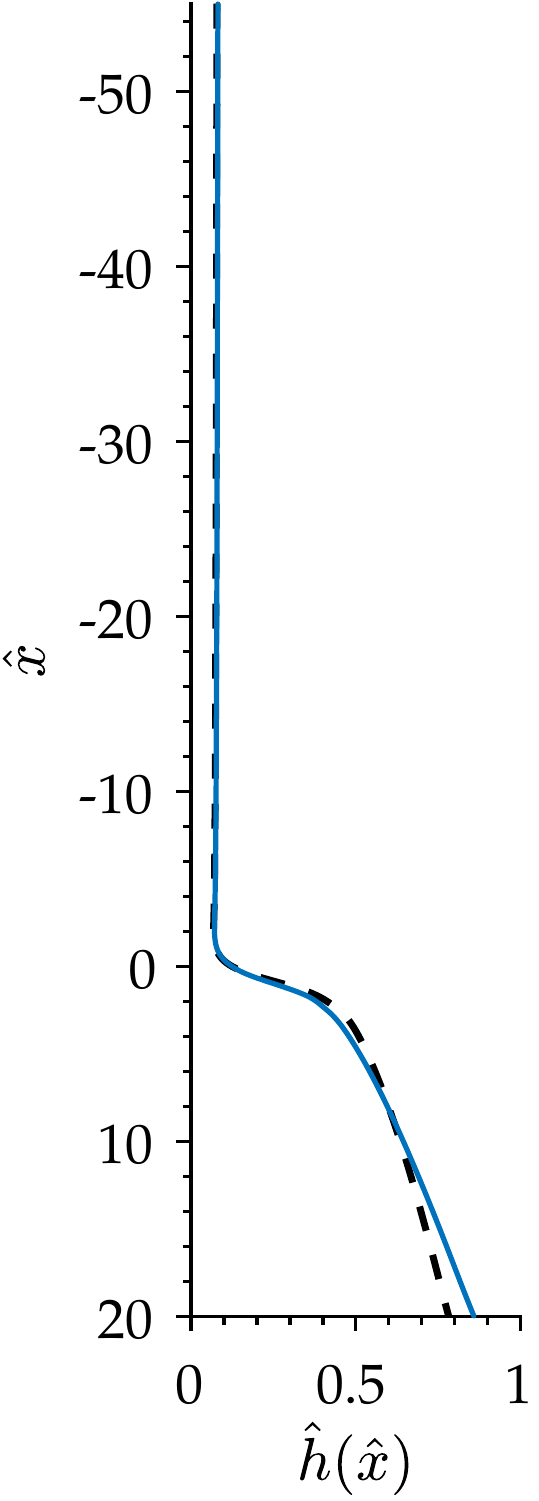}
\includegraphics[width=2cm]{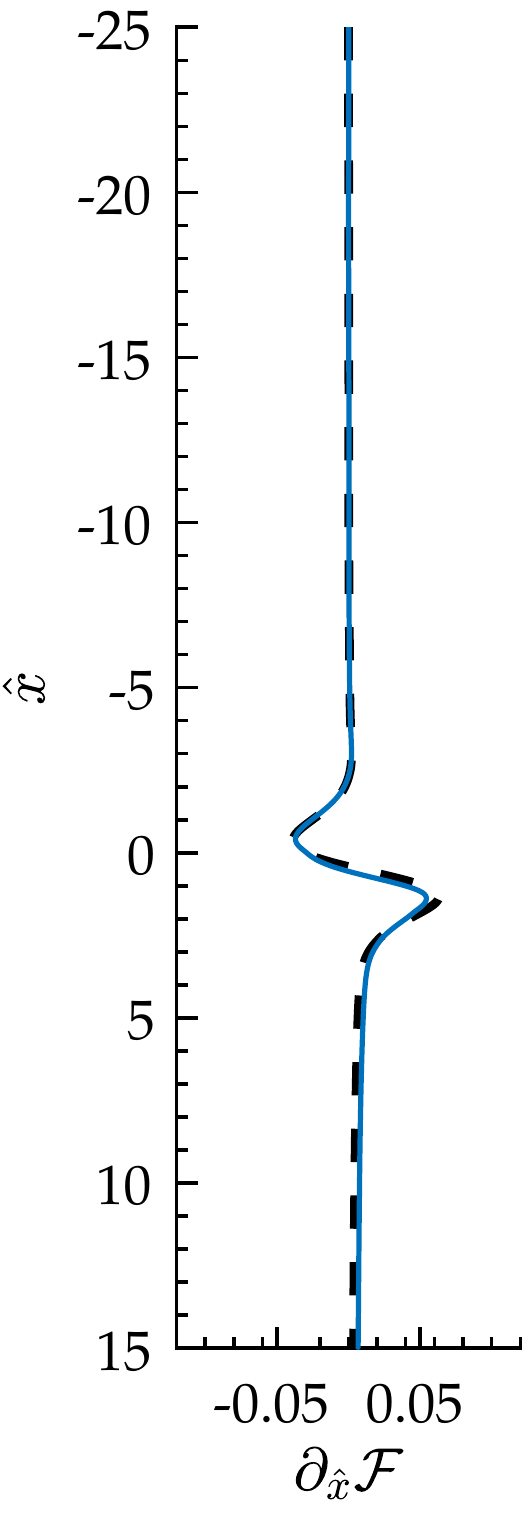}
\includegraphics[width=2cm]{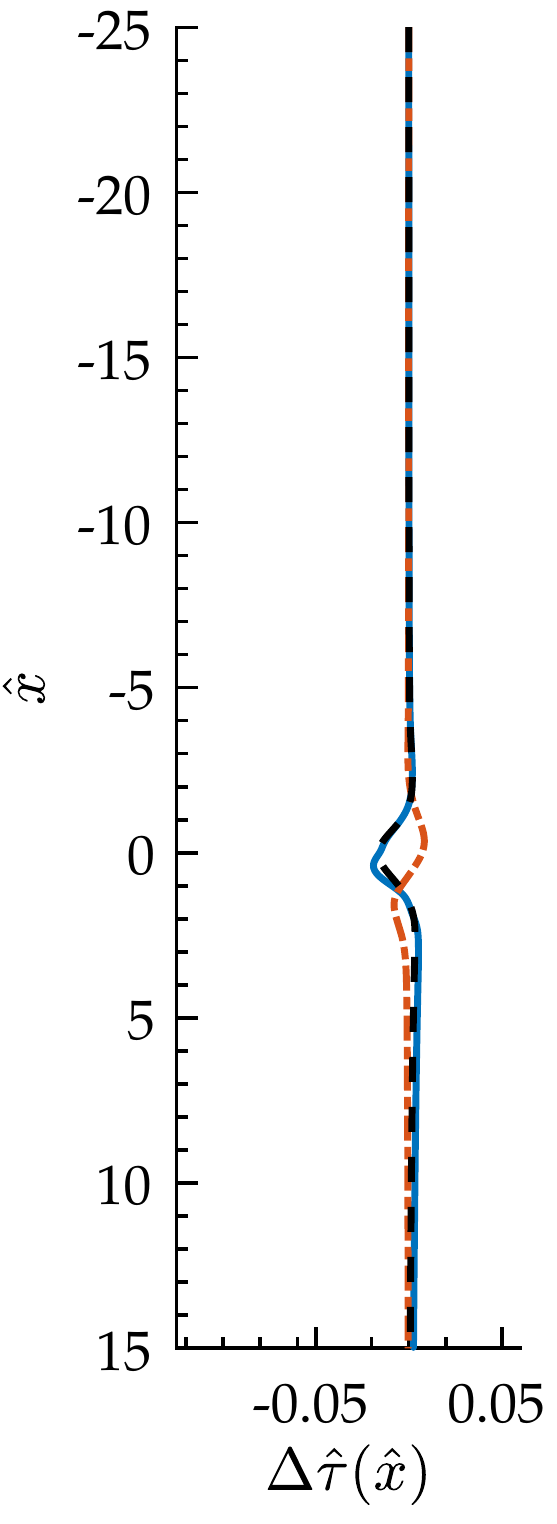}}\\
       
         \hline
        \rotatebox{90}{\hspace{19mm} $\hat{f}=0.055$}   &
\multicolumn{1}{c|}{
\includegraphics[width=2cm]{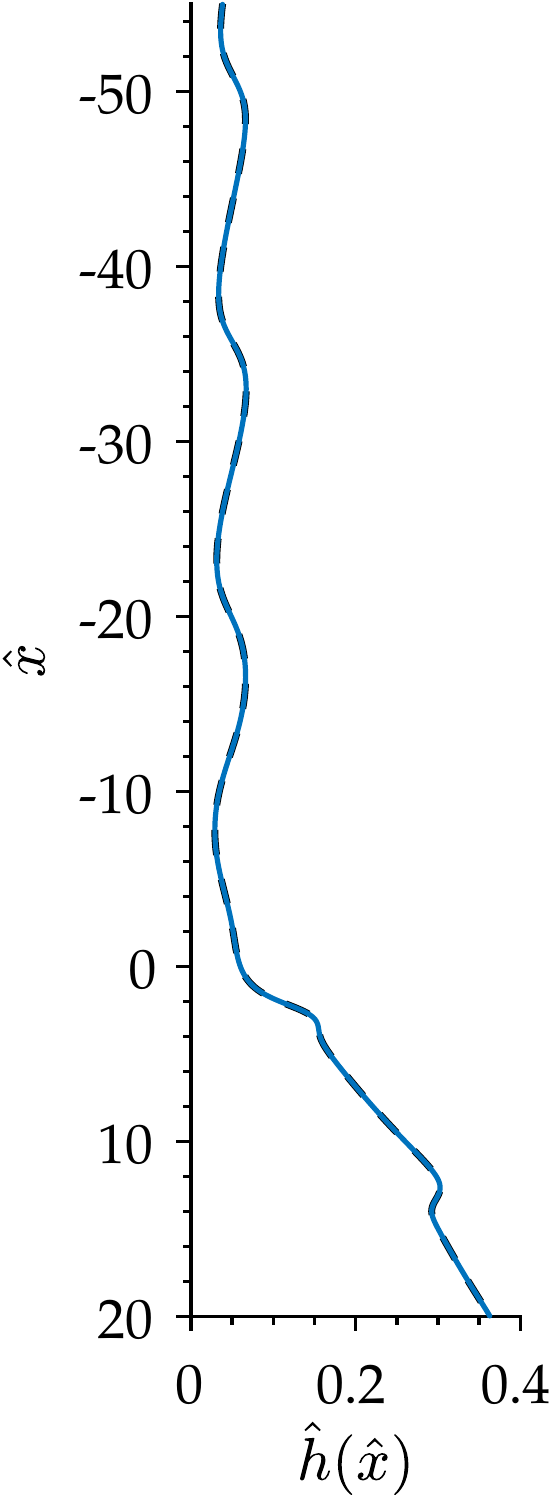}
\includegraphics[width=2cm]{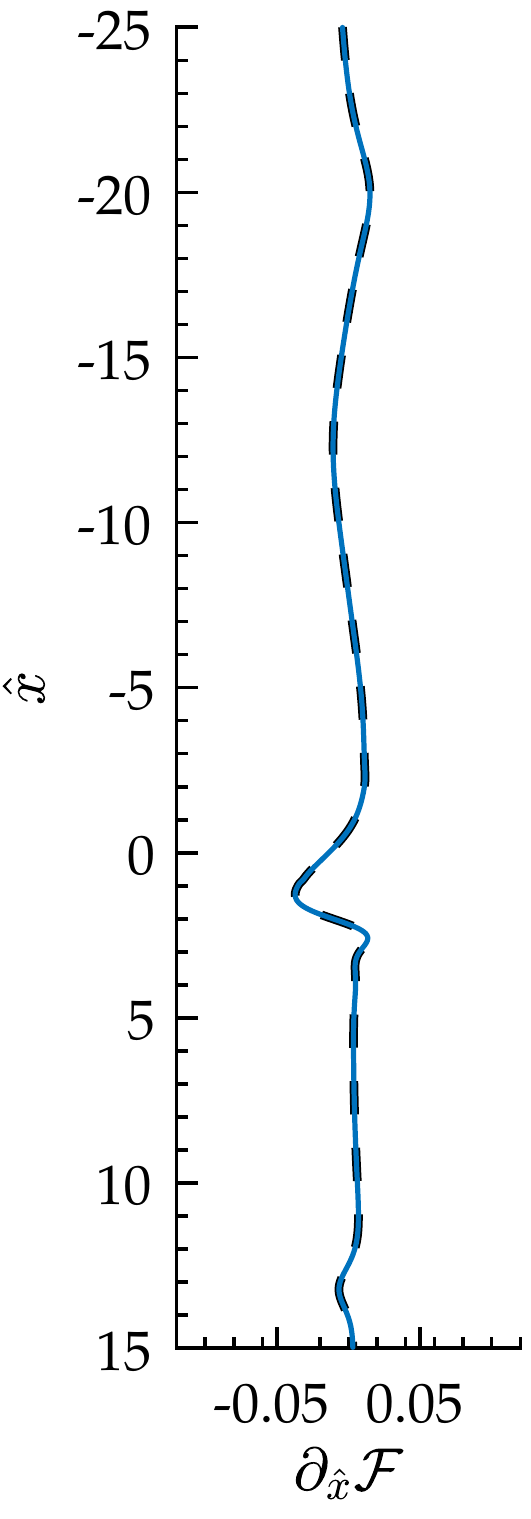}
\includegraphics[width=2cm]{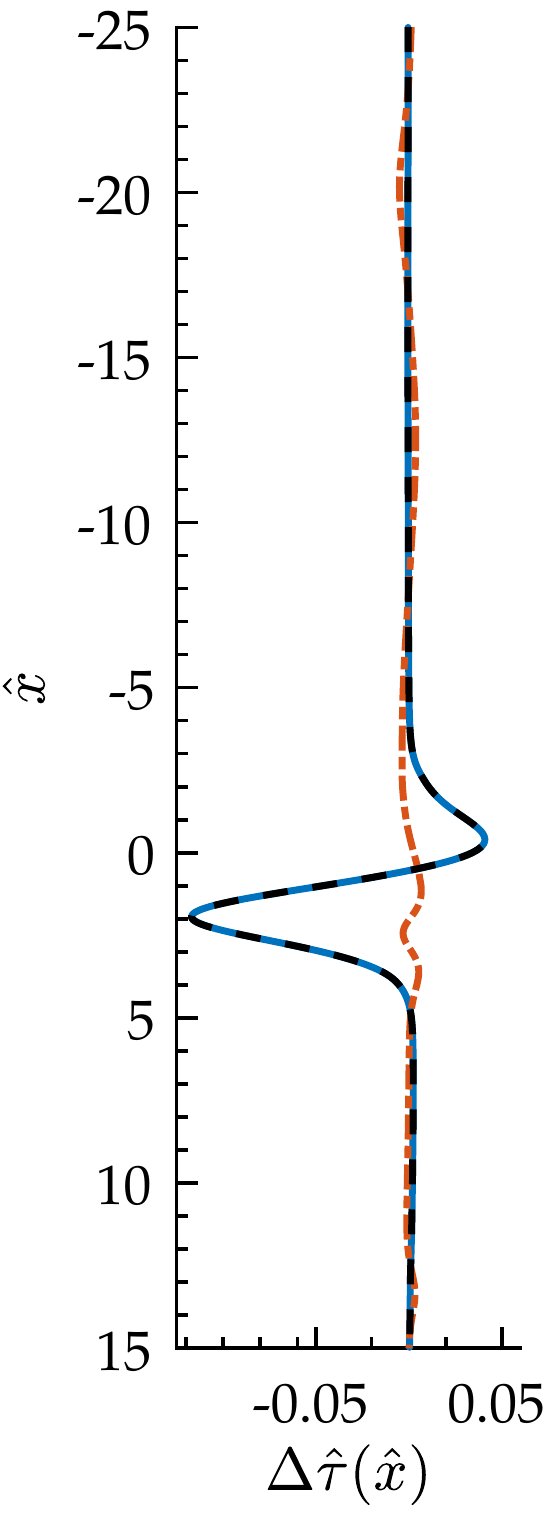}} \hspace{2mm} & 
       \multicolumn{1}{c}{
\includegraphics[width=2cm]{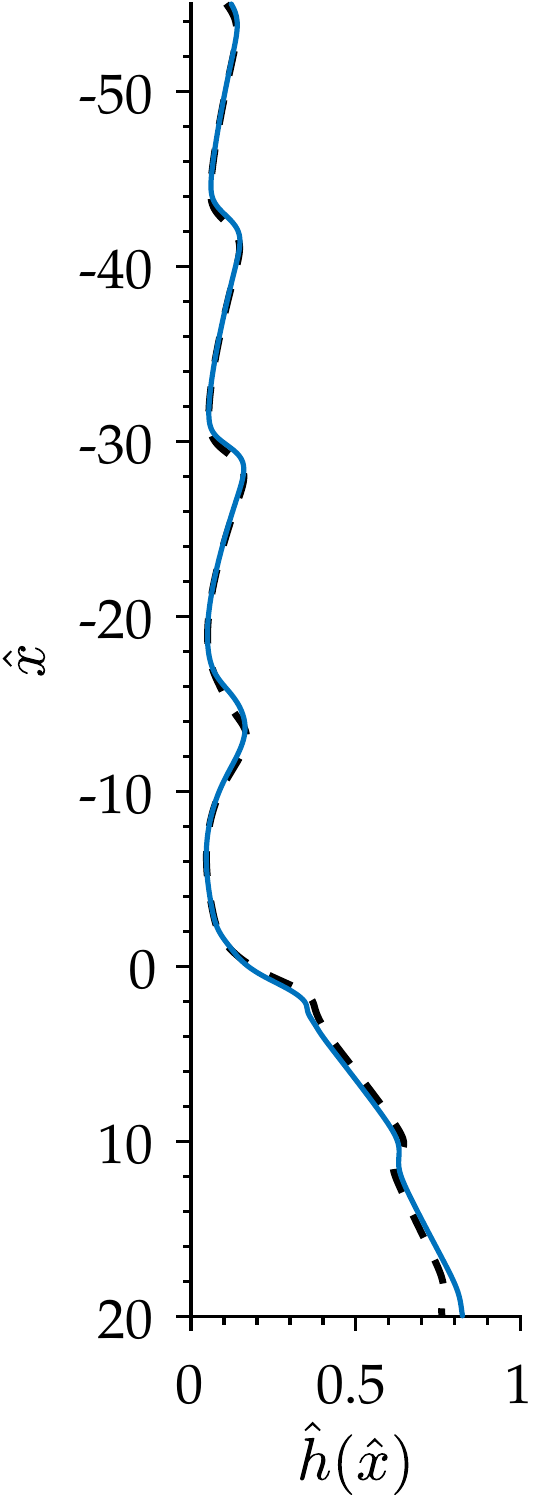}
\includegraphics[width=2cm]{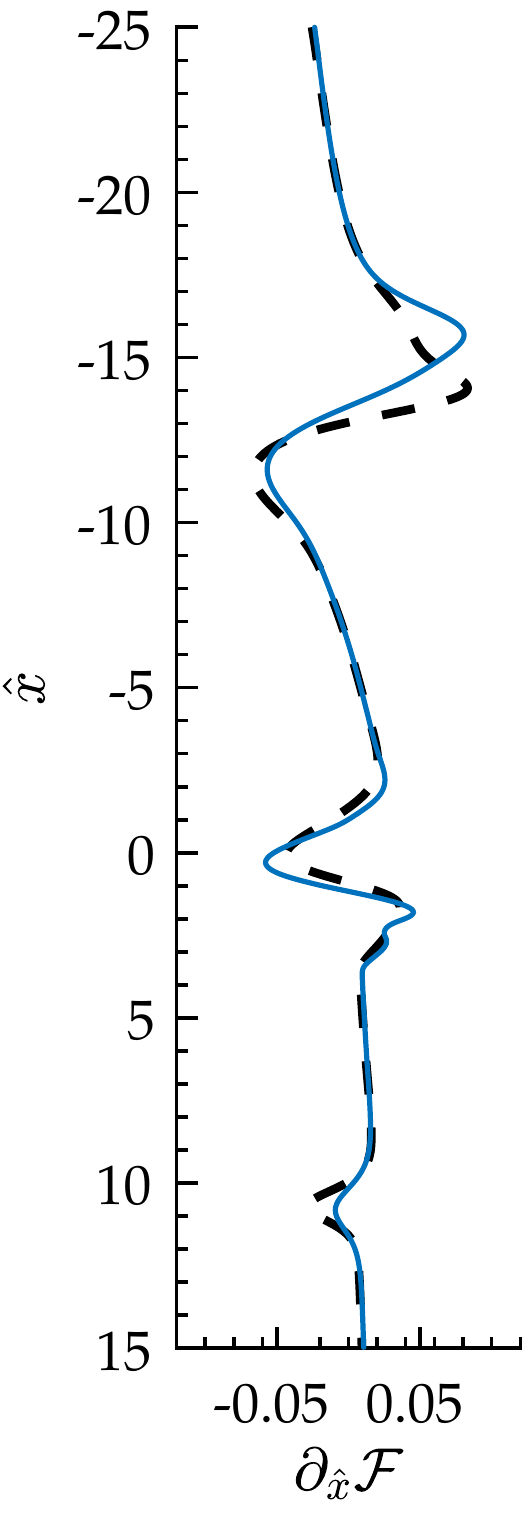}
\includegraphics[width=2cm]{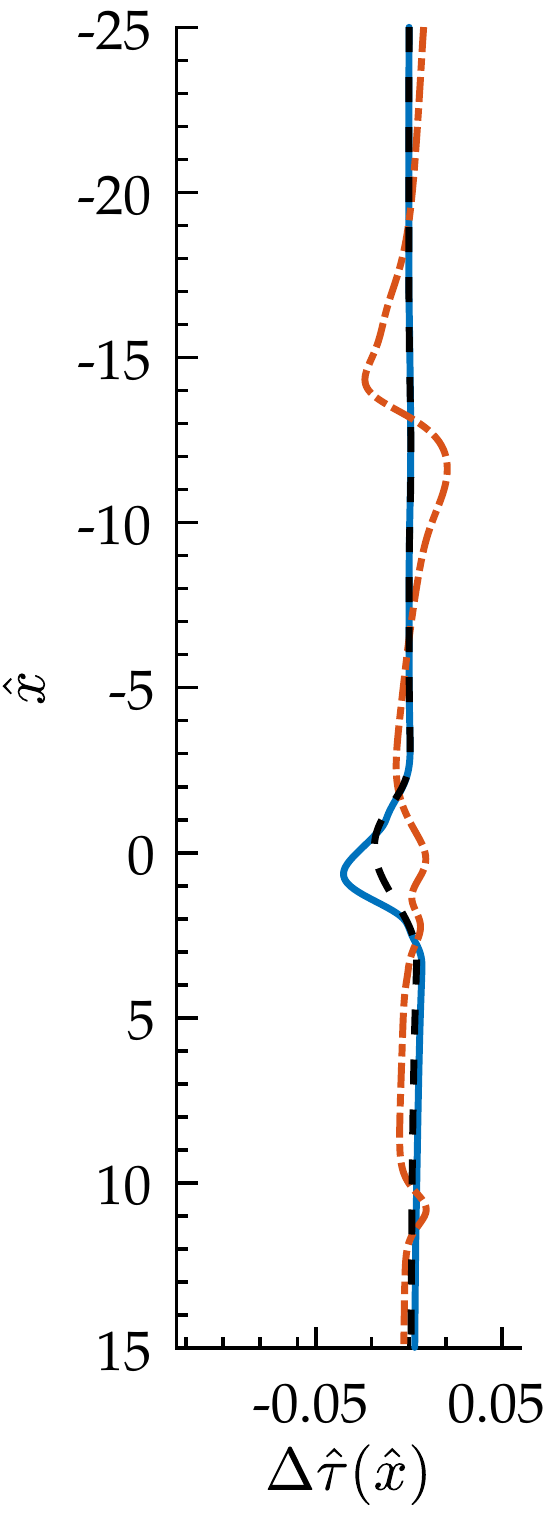}}\\        
                 \hline
    \rotatebox{90}{\hspace{20mm} $\hat{f}=0.16$}  &
\multicolumn{1}{c|}{
\includegraphics[width=2cm]{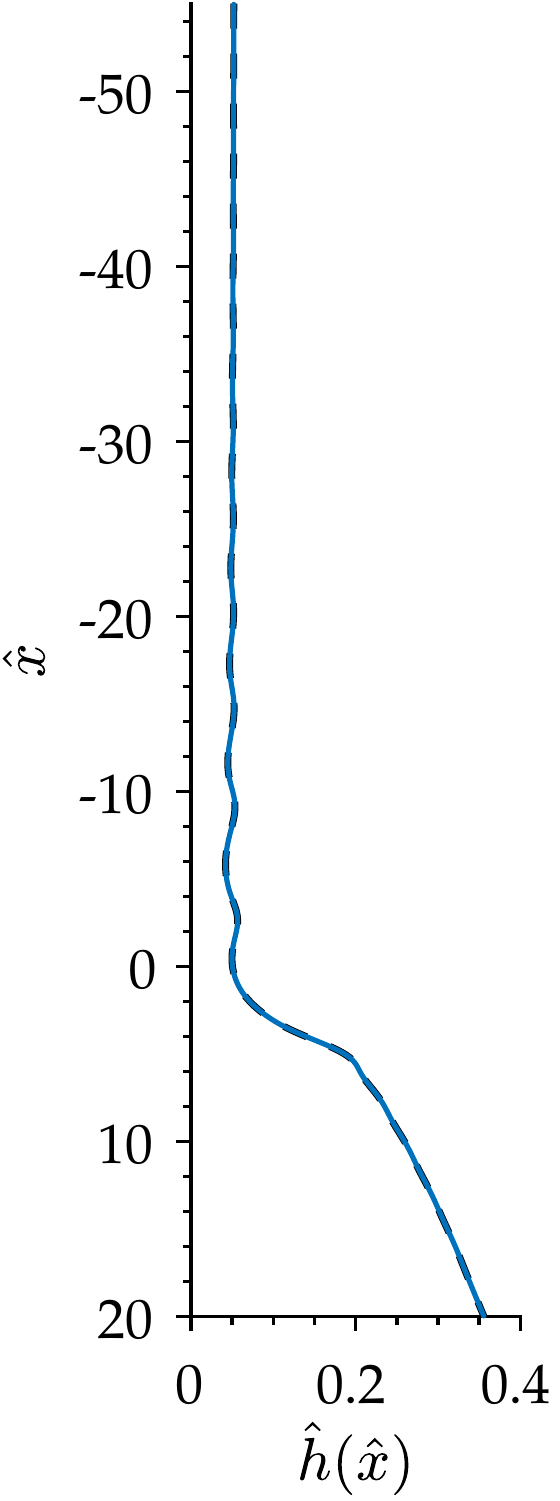}
\includegraphics[width=2cm]{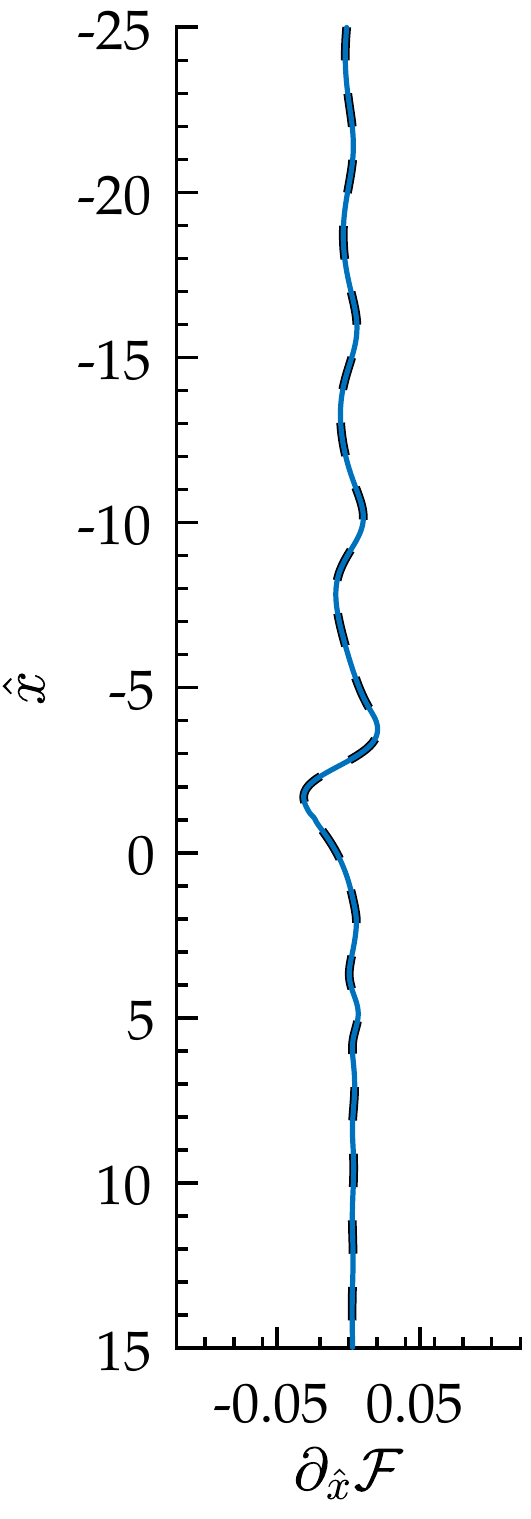}
\includegraphics[width=2cm]{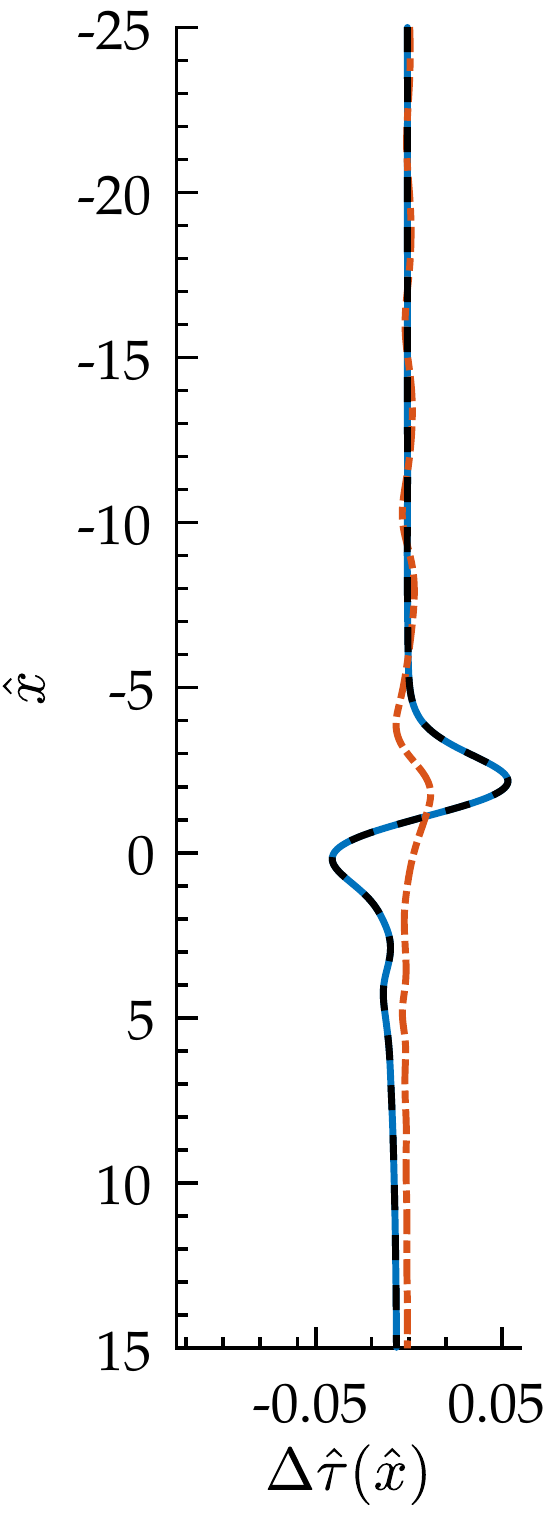}} \hspace{2mm} & 
        \multicolumn{1}{c}{
\includegraphics[width=2cm]{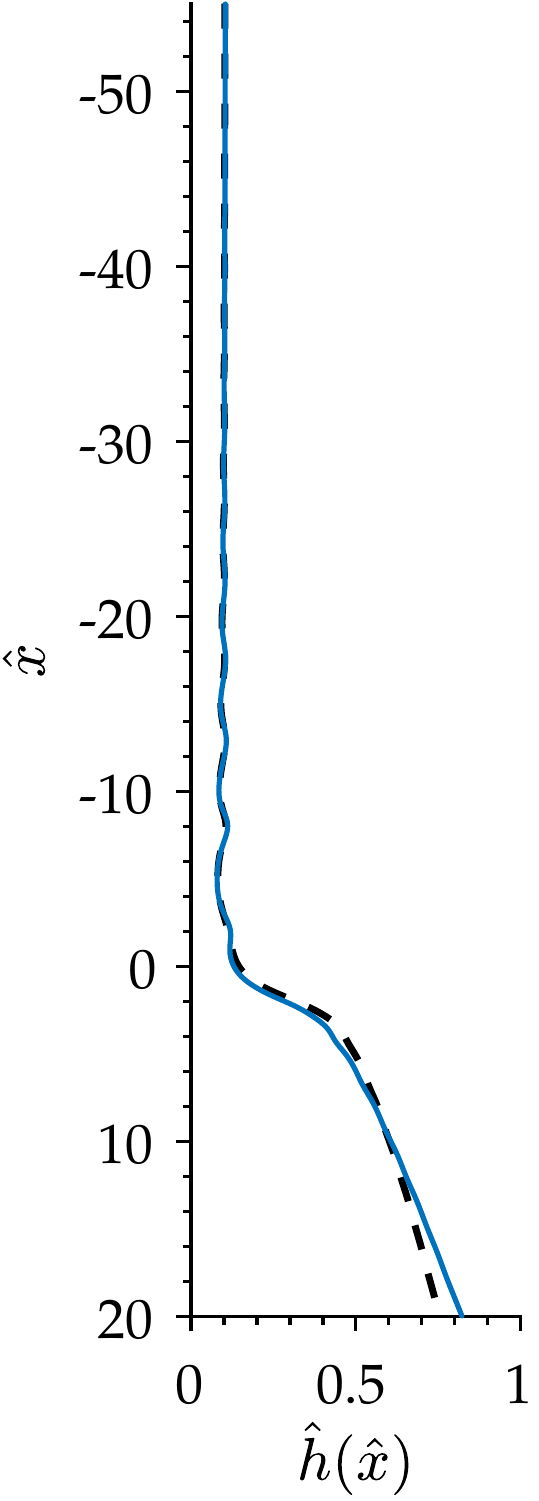}
\includegraphics[width=2cm]{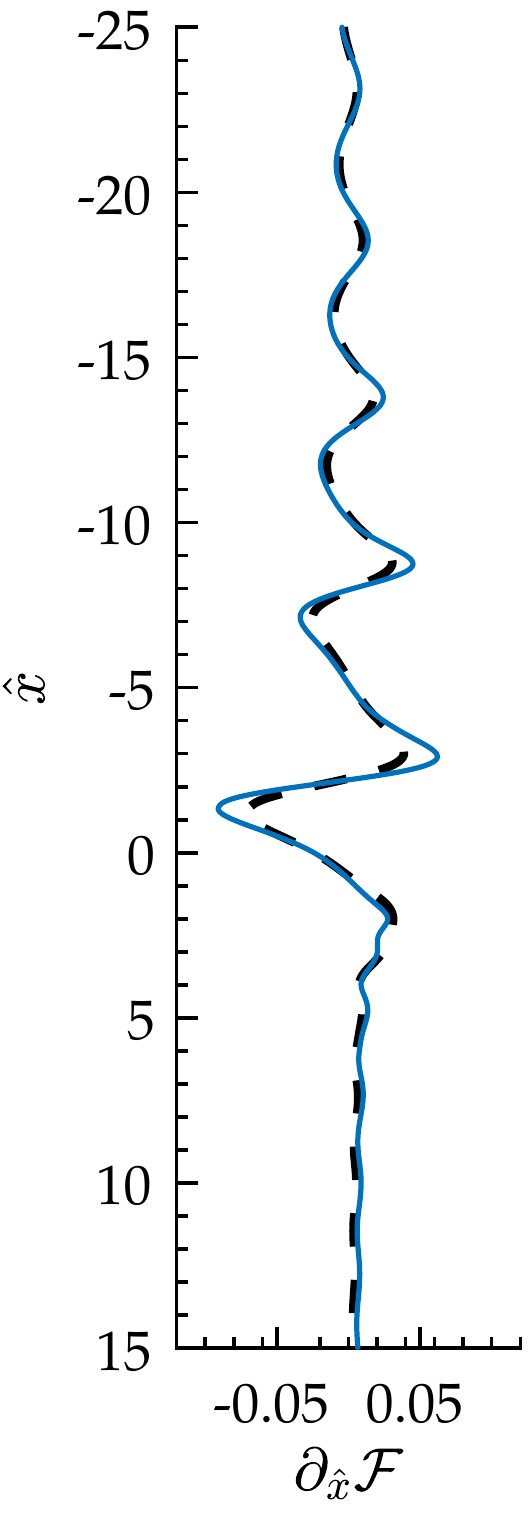}
\includegraphics[width=2cm]{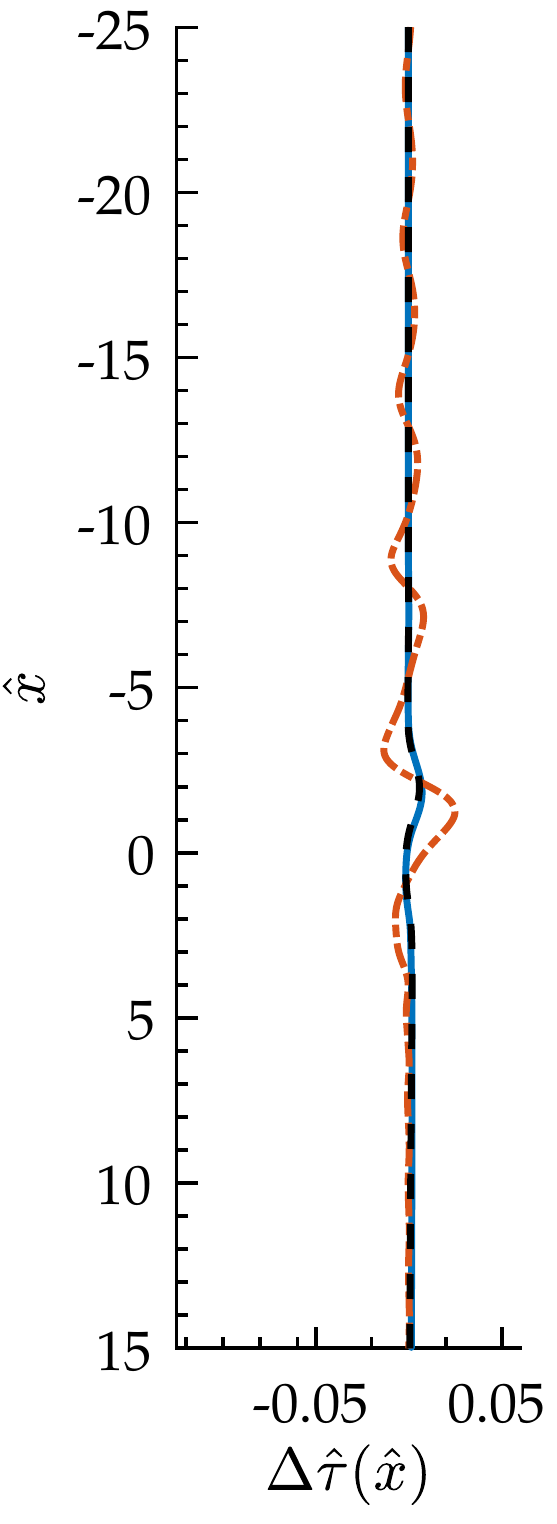}}\\ 
\end{tabular}
\caption{ Results from the simulations of the jet wiping {in galvanizing conditions} using the WIBL and the TTBL models for $\Pi_g=1.856$, two Reynolds numbers ($Re=478$ on the left column and $Re=2483$ on the right) and three oscillation frequencies ($\hat{f}=[0,0.055,0.16])$, along the rows). In each panel the first plot from the left shows an instantaneous film thickness, the second and the third the corresponding advection and shear stress terms respectively.}
\label{Table_Zinc_FORCES_1}
\end{figure}

Focusing on the wiping cases with zinc, figure \ref{Table_Zinc_FORCES_1} presents several instantaneous profiles for six representative test cases taken from the results in figure \ref{RES_1}. These include the lowest ($Re=478$) and the largest ($Re=2483$) Reynolds numbers and the three perturbation frequencies, keeping the wiping number at $\Pi_g=1.856$. In each of the six panels, the first plot compares the instantaneous dimensionless film thickness profiles for WIBL and TTBL models. The second plot compares the advection terms $\partial_x \mathcal{F}$ profiles for both models while the third collects the contributions to the wall shear stress. The term $\Delta \hat{\tau}^{(0)}$ is the zero-order term from the WIBL, which is the one in the IBL model. The term $\Delta \hat{\tau}^{(1)}$ is the first-order contribution that distinguishes the WIBL from the IBL. Finally, the term $\Delta \hat{\tau}_T$ is the wall shear stress term from the TTBL model.

As expected, no differences are observed between the three models at the lowest $Re$, except for the cases at the highest perturbation frequency $\hat{f}=0.16$. At $Re=478$, the wall shear stress terms have a larger contribution to the liquid film dynamics, especially in the proximity of the wiping region. In this condition, the first-order term $\Delta \tau^{(1)}$ appears to have a negligible contribution; hence the WIBL model corresponds almost everywhere to the IBL model (not shown). In the case at $Re=2483$, the contribution of this term increases but remains less important than the advection term, which mostly dominates the dynamics of the liquid waves. The TTBL departs from the laminar models in the run-back flow region, while no appreciable difference is observed in the final coating region as this is characterized by $Re_F<100$. This further highlight the convective nature of the problem with two opposite characteristic lines: the dynamics in the final coating film appears to be insensitive to the dynamics of the run back flow region. Finally, in terms of the frequency response of the liquid coat, the three models reveal that the perturbation frequency of $\hat{f}=0.16$ is too high to generate any appreciable wave in the final coat. The influence of the modeling strategy on the harmonic response of the flow is discussed in section \ref{RES_D}; the next section focuses on the harmonic response of the film considering only the WIBL model.

\subsection{The frequency response of the liquid film}\label{RES_C}

This section focuses on the frequency response of the liquid film subject to different kinds of jet perturbation, producing pressure gradient evolutions of the form described in figure \ref{Example_WAVES}. For a wiping number $\Pi_g=1.2$ and rescaled Reynolds number $\delta=554$, figure \ref{OSCI_ZINC} shows {several contour maps of the mean-centered thickness $\tilde{h}=\hat{h}(\hat{x},\hat{t})-\overline{h}(\hat{x})$, obtained by subtracting the temporal average $\overline{h}(\hat{x})=\frac{1}{T}\int^T_0 \hat{h}(\hat{x},\hat{t})d\hat t$, with $T$ the dominant wave period}. Four jet perturbations are considered, namely three oscillations (harmonic, upward biased, and downward biased) and a harmonic jet pulsation (cf figure \ref{Example_WAVES}). For each of these, the frequencies considered are $\hat{f}=[0.02,0.05,0.08]$.  For the jet oscillation test cases, a white dashed line indicates the evolution of the impingement point, i.e., the region of maximum pressure and zero gas shear stress at the interface.

\begin{figure}
\vspace{3mm}
\centering
\hspace{3mm}Harmonic Oscillation \\
\vspace{2mm}
\begin{tabular}{@{}cccc@{}}
       $\hat{f}=0.02$  & $\hat{f}=0.05$   & $\hat{f}=0.08$\\ 
\includegraphics[trim={0.4cm 0 2cm 0},clip,width=4.4cm]{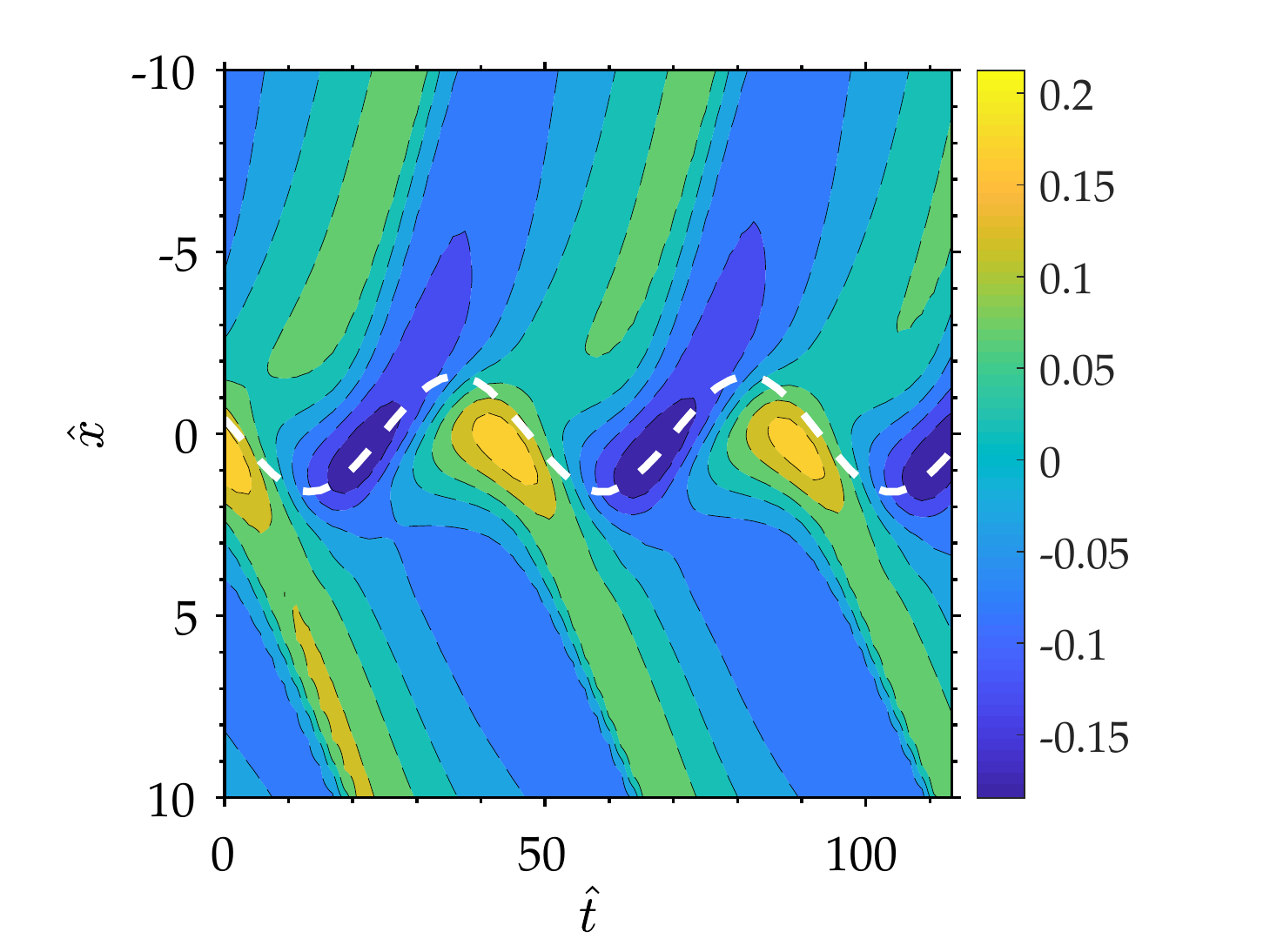} &  
\includegraphics[trim={0.4cm 0 2cm 0},clip,width=4.4cm]{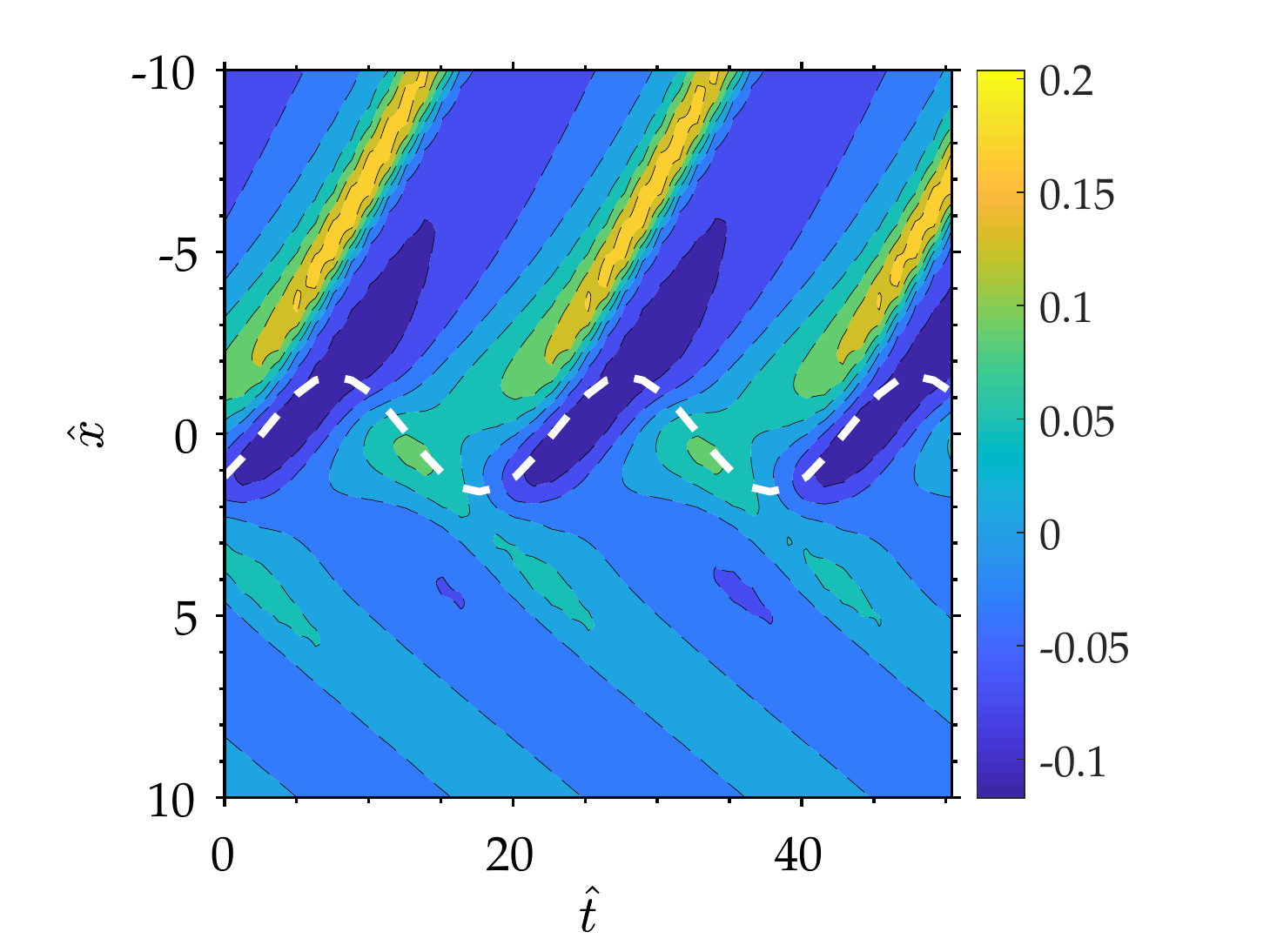} &  
\includegraphics[trim={0.4cm 0 2cm 0},clip,width=4.4cm]{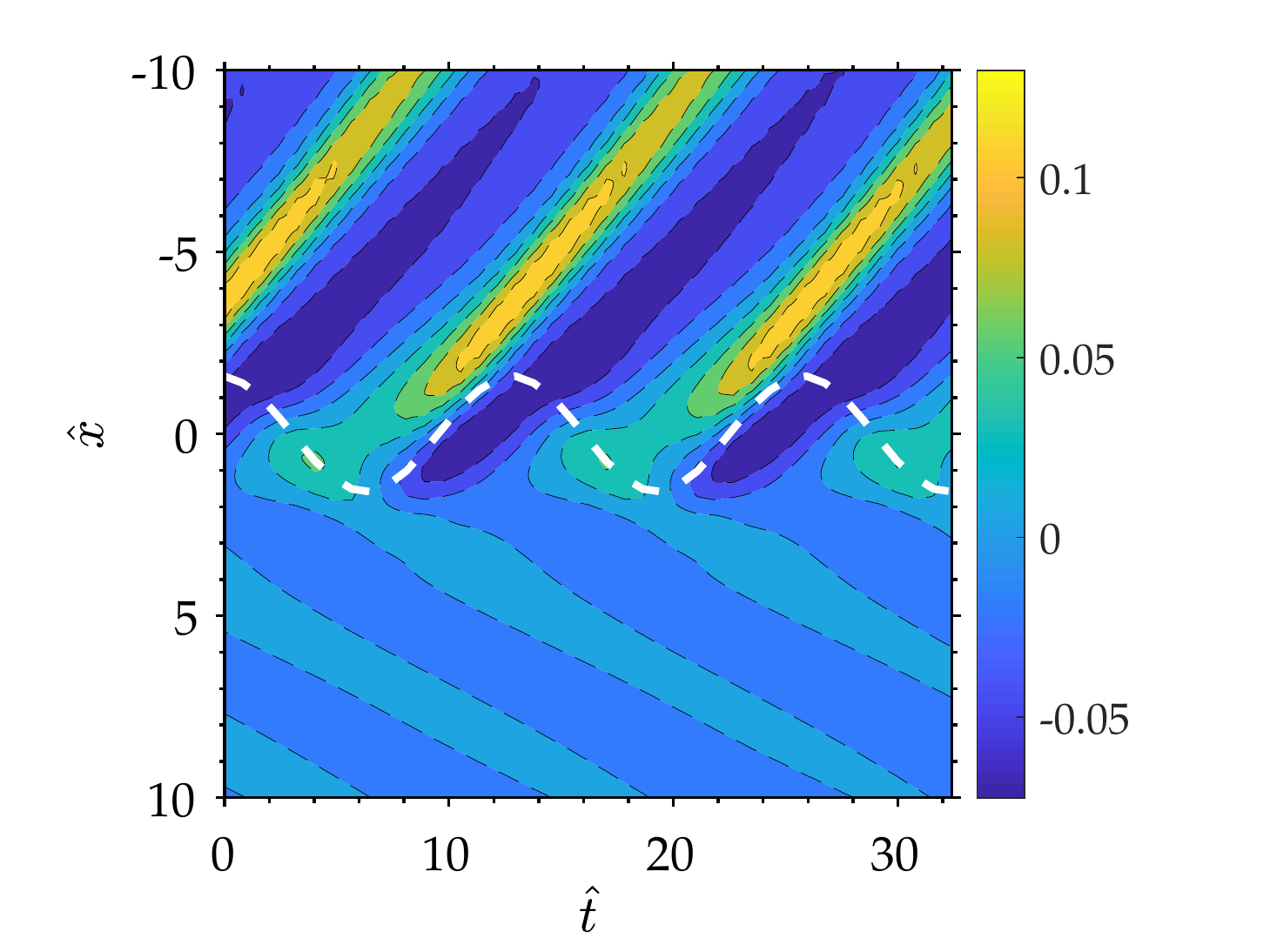} \\
\end{tabular}
\vspace{1mm}
\hspace{2mm}Upward-Biased Oscillation \\
\begin{tabular}{@{}cccc@{}}
       

\includegraphics[trim={0.4cm 0 2cm 0},clip,width=4.4cm]{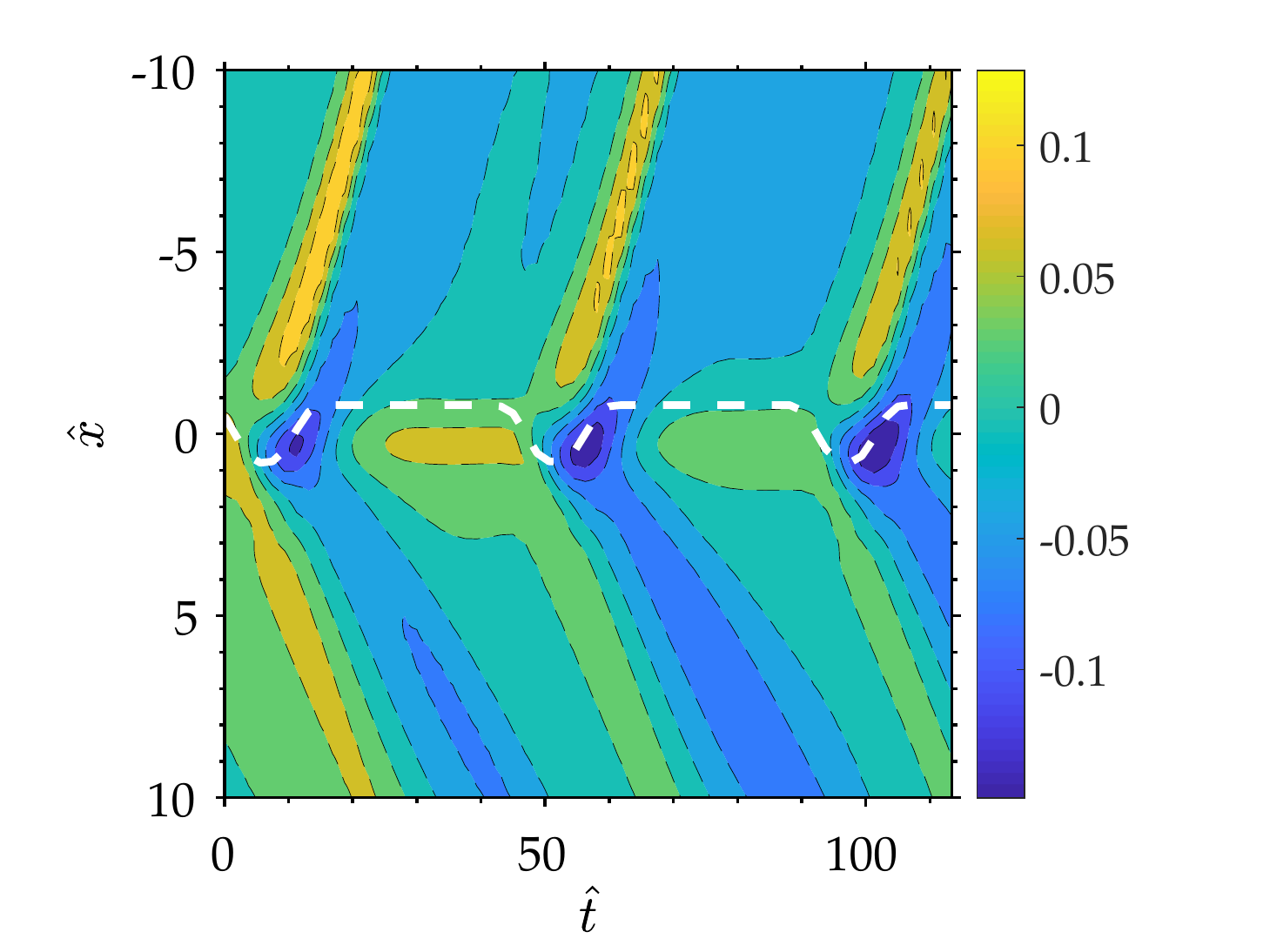} &  
\includegraphics[trim={0.4cm 0 2cm 0},clip,width=4.4cm]{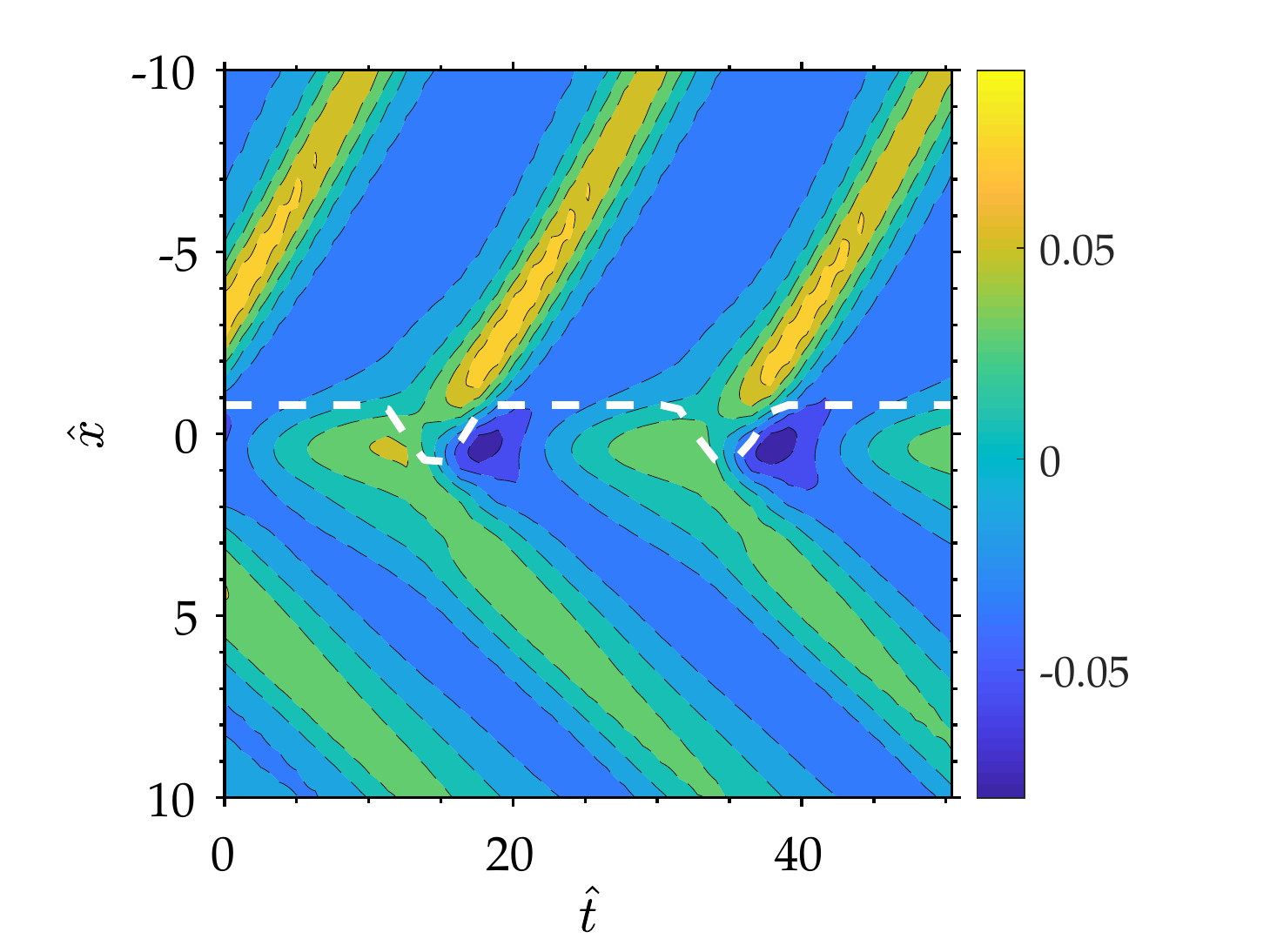} &  
\includegraphics[trim={0.4cm 0 2cm 0},clip,width=4.4cm]{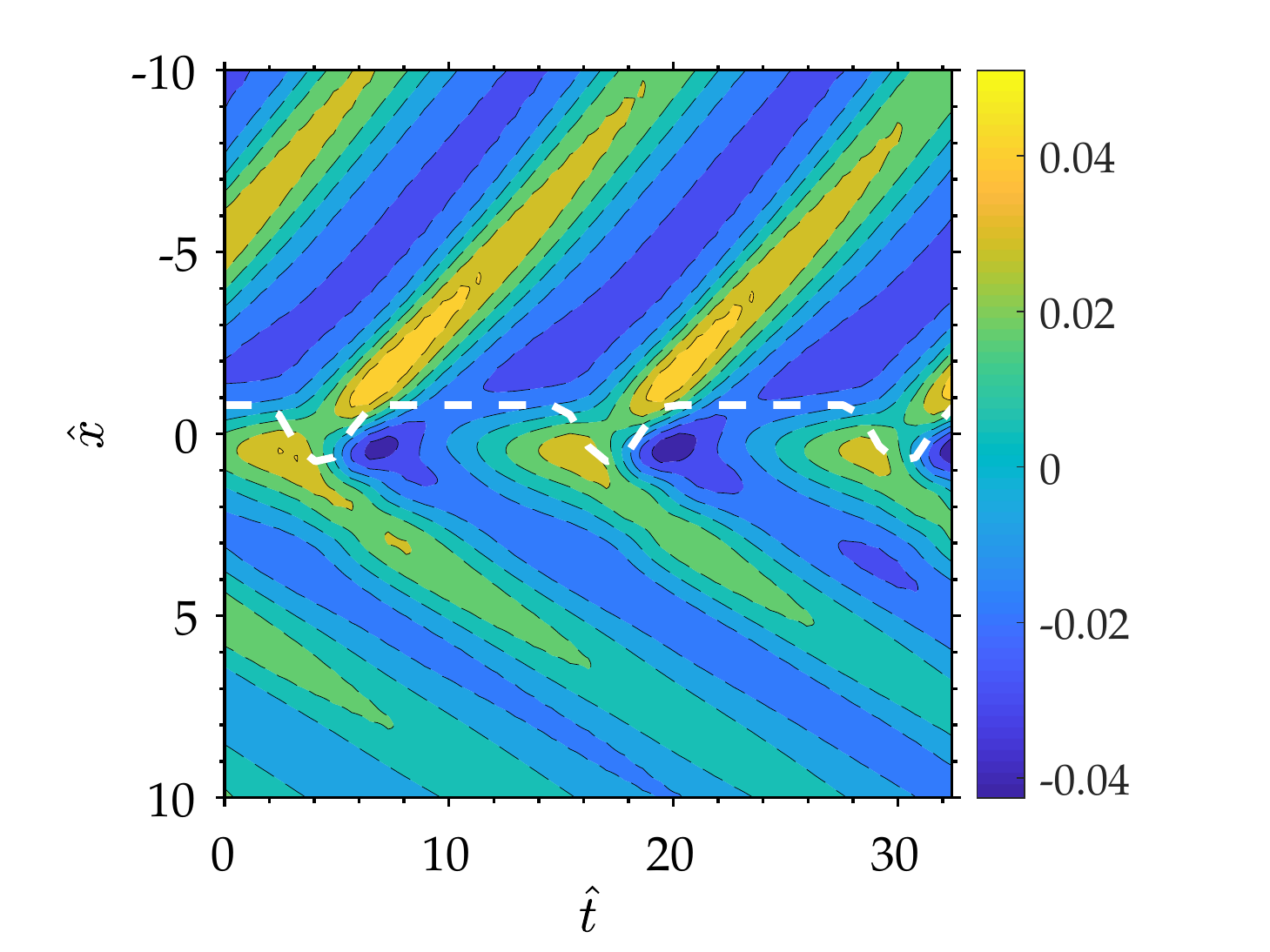} \\
\end{tabular}
\vspace{1mm}
\hspace{2mm}Downward-Biased Oscillation \\
                 \begin{tabular}{@{}cccc@{}}
\includegraphics[trim={0.4cm 0 2cm 0},clip,width=4.4cm]{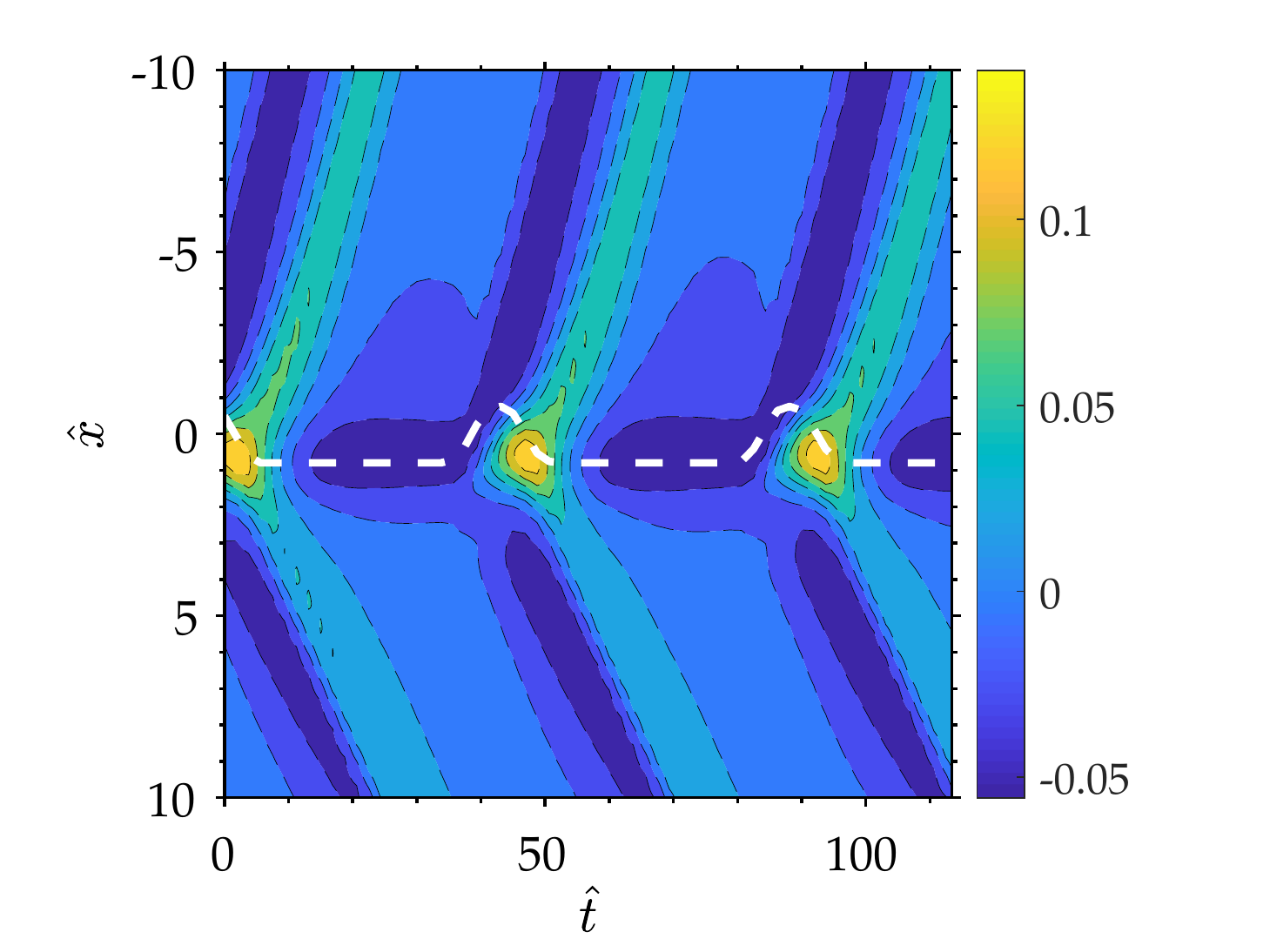} &  
\includegraphics[trim={0.4cm 0 2cm 0},clip,width=4.4cm]{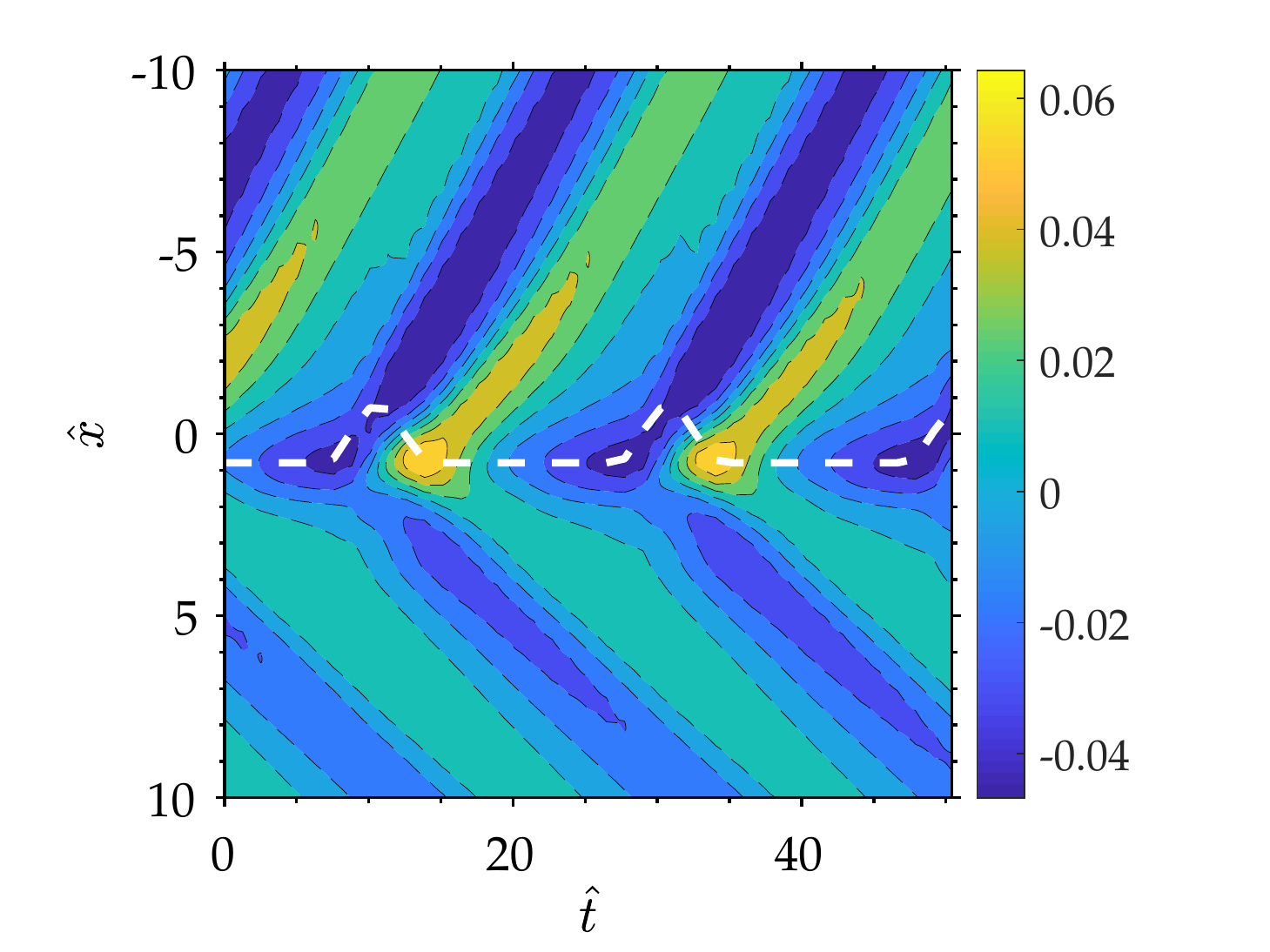} &  
\includegraphics[trim={0.4cm 0 2cm 0},clip,width=4.4cm]{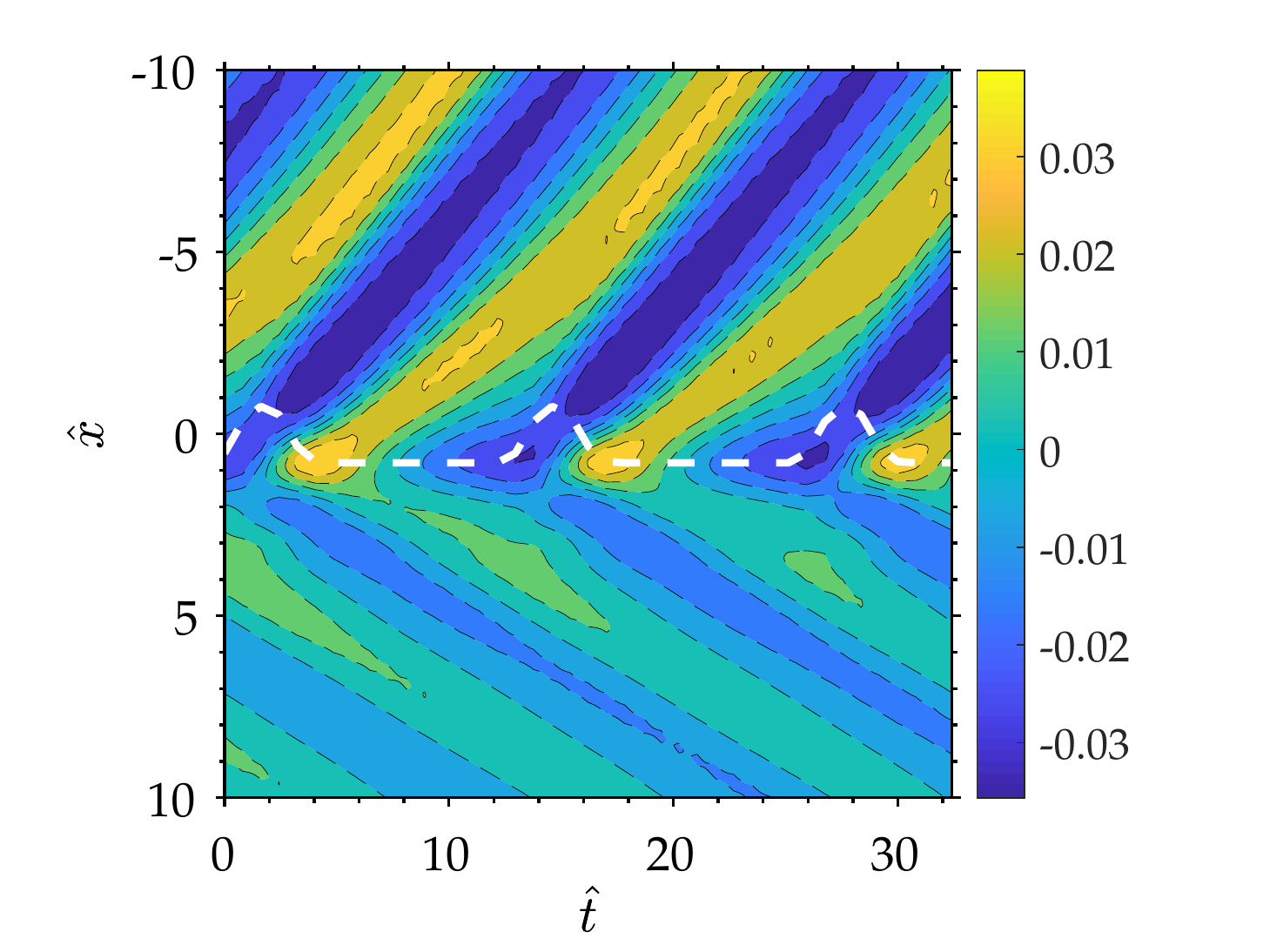} \\
\end{tabular}\\
 \hspace{2mm}Pulsating Jet Flow \\
                 \begin{tabular}{@{}cccc@{}}
\includegraphics[trim={0.4cm 0 2cm 0},clip,width=4.4cm]{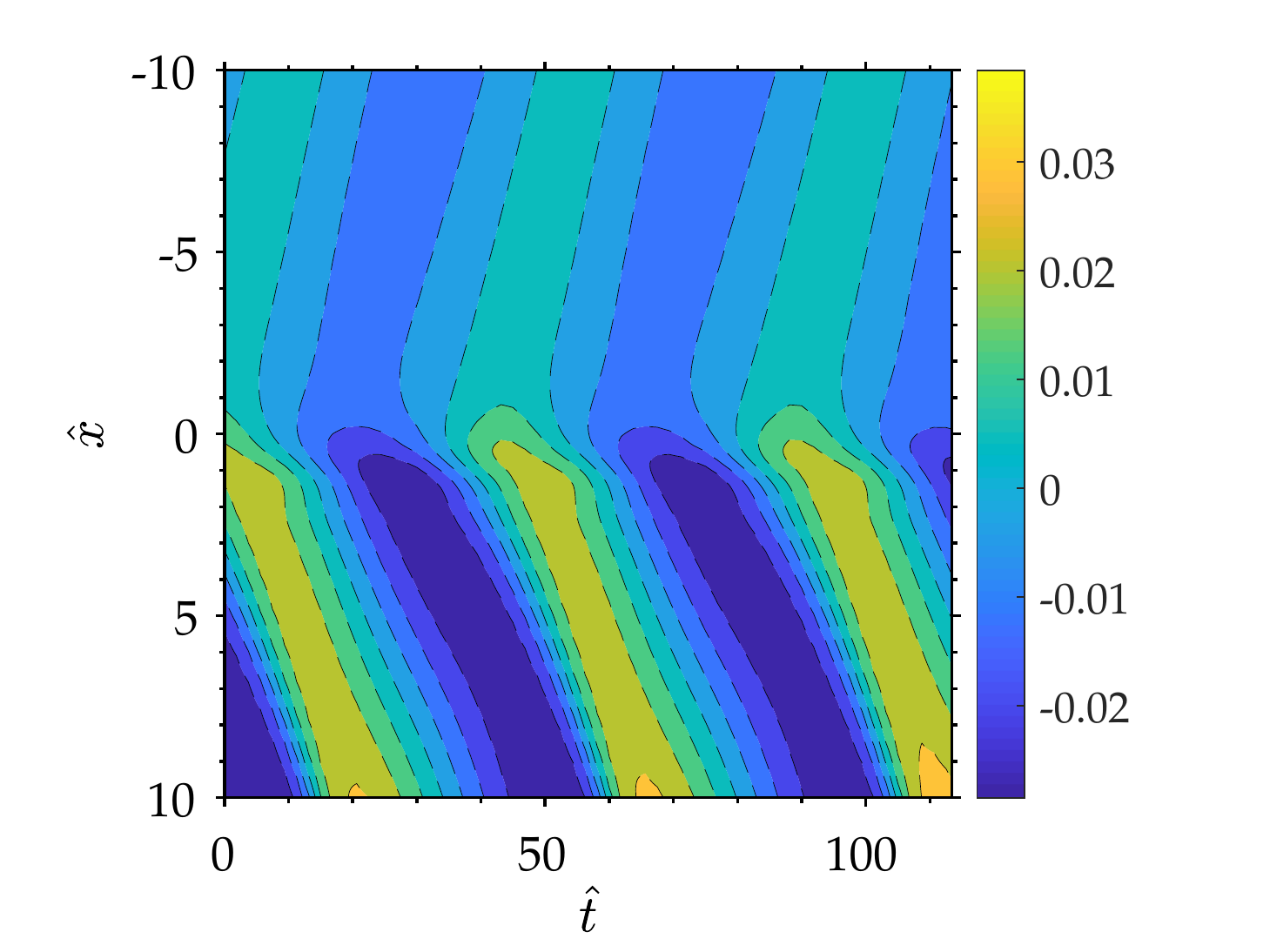} &  
\includegraphics[trim={0.4cm 0 2cm 0},clip,width=4.4cm]{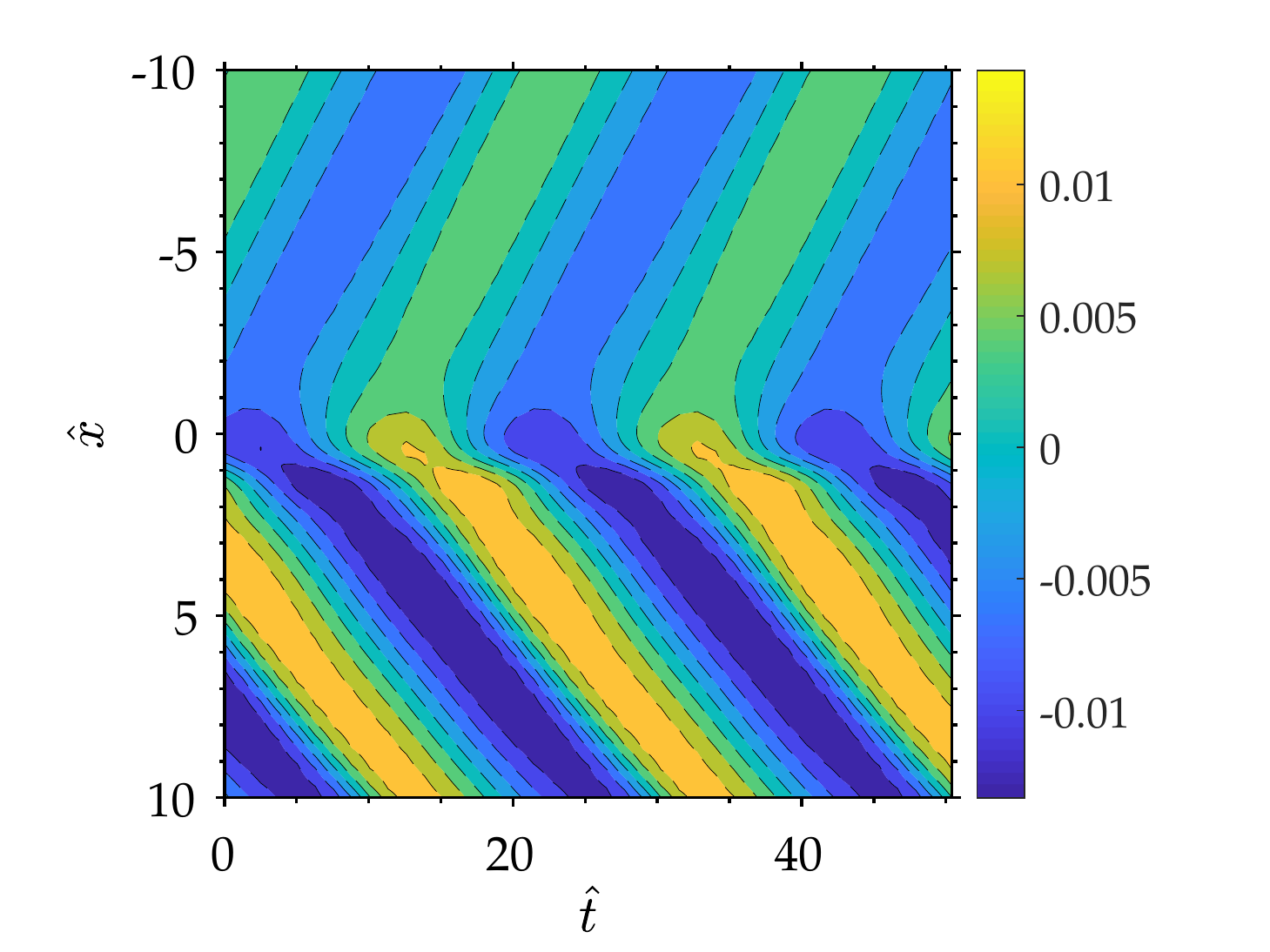} &  
\includegraphics[trim={0.4cm 0 2cm 0},clip,width=4.4cm]{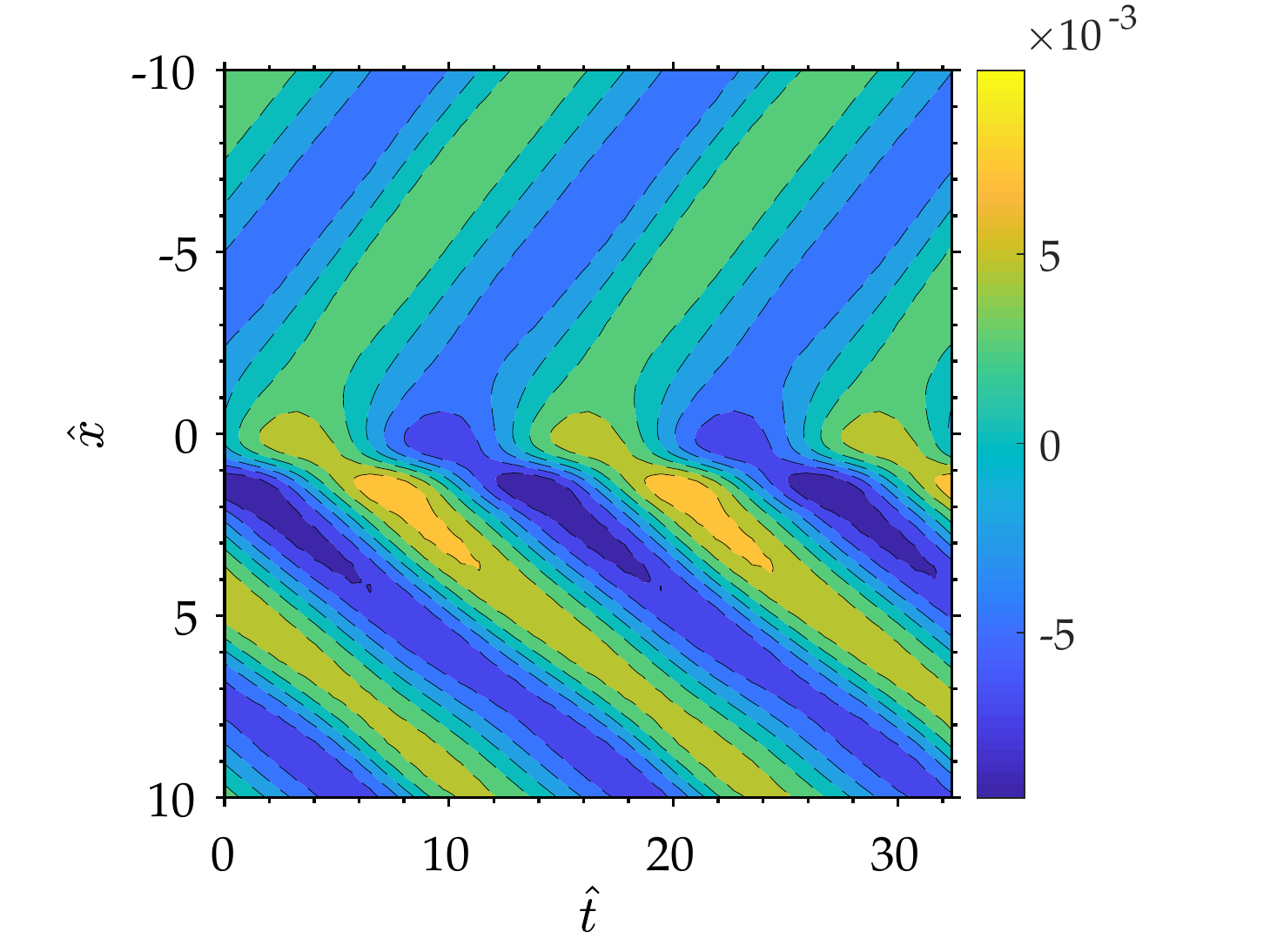} 
\end{tabular}
\vspace{-4mm}
\caption{Contour maps of the coating thickness $\tilde{h}(\hat{x},\hat{t})$, zero-mean shifted in time, as a function of dimensionless space and time. Four kinds of jet perturbation are considered: harmonic oscillations (first row), upward-biased (second row) and downward biased (third row) oscillations and jet pulsations (fourth row).The perturbation frequencies are taken as $\hat{f}=0.02$ (first column), $\hat{f}=0.05$ (second column) and $\hat{f}=0.08$ (third column). All cases refer to galvanizing conditions with $\Pi_g=1.2$ and $\delta=554$.}
\label{OSCI_ZINC}
\end{figure}

\begin{figure*}
\centering
\includegraphics[height=4.3cm]{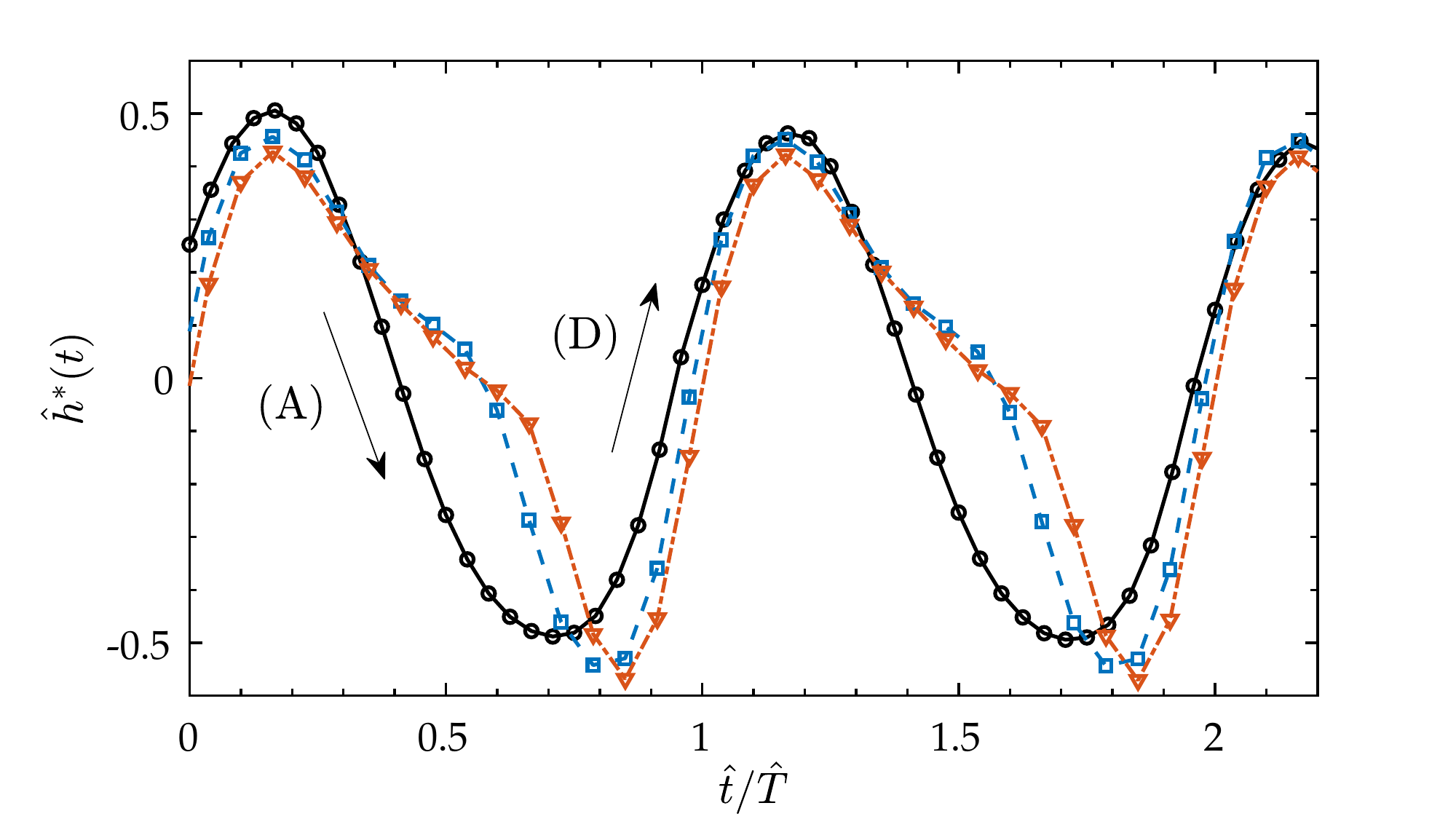}
\includegraphics[height=4.3cm]{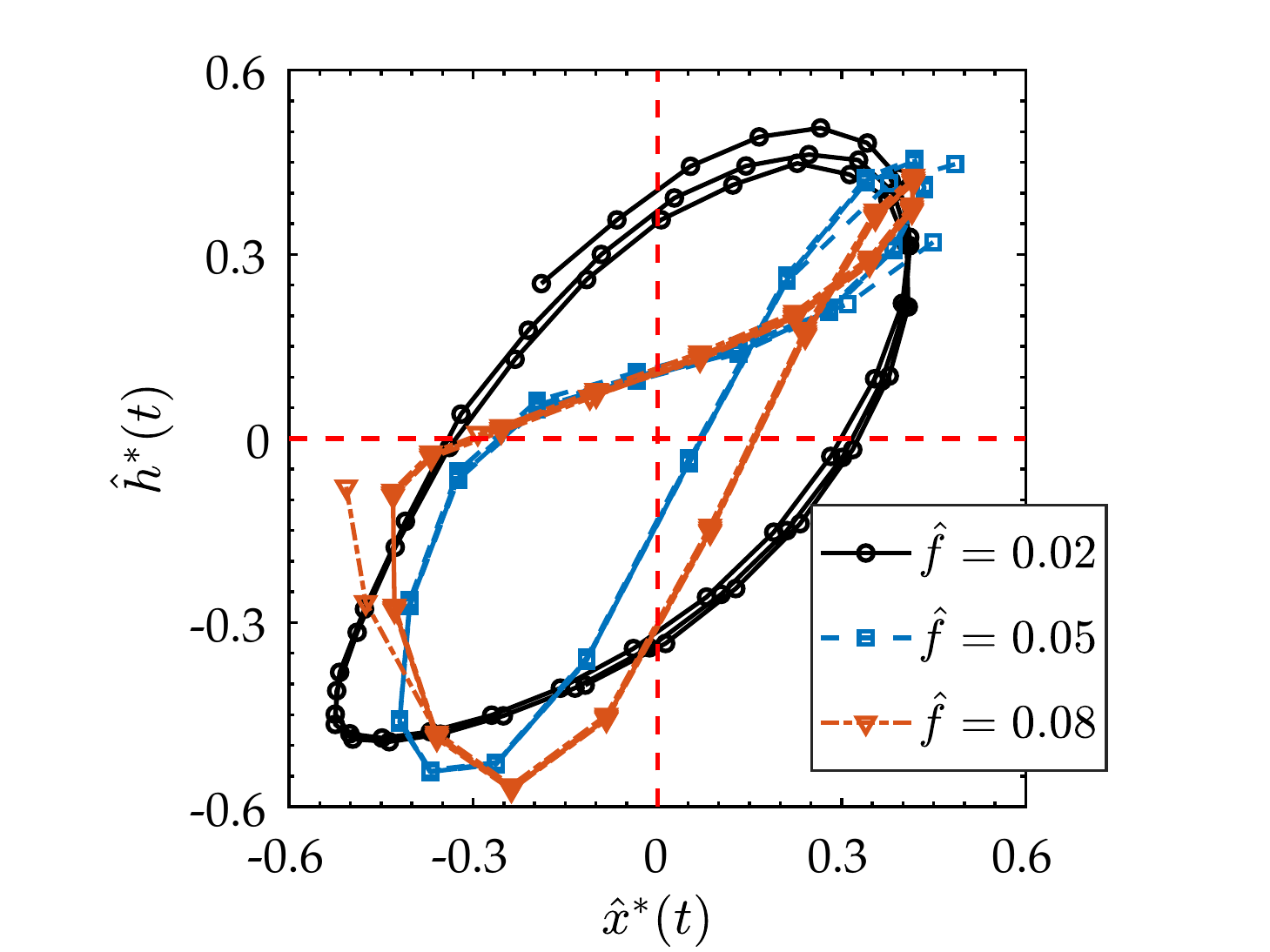}\\
\hspace{14mm}a)\hspace{65mm} b)
\includegraphics[height=4.3cm]{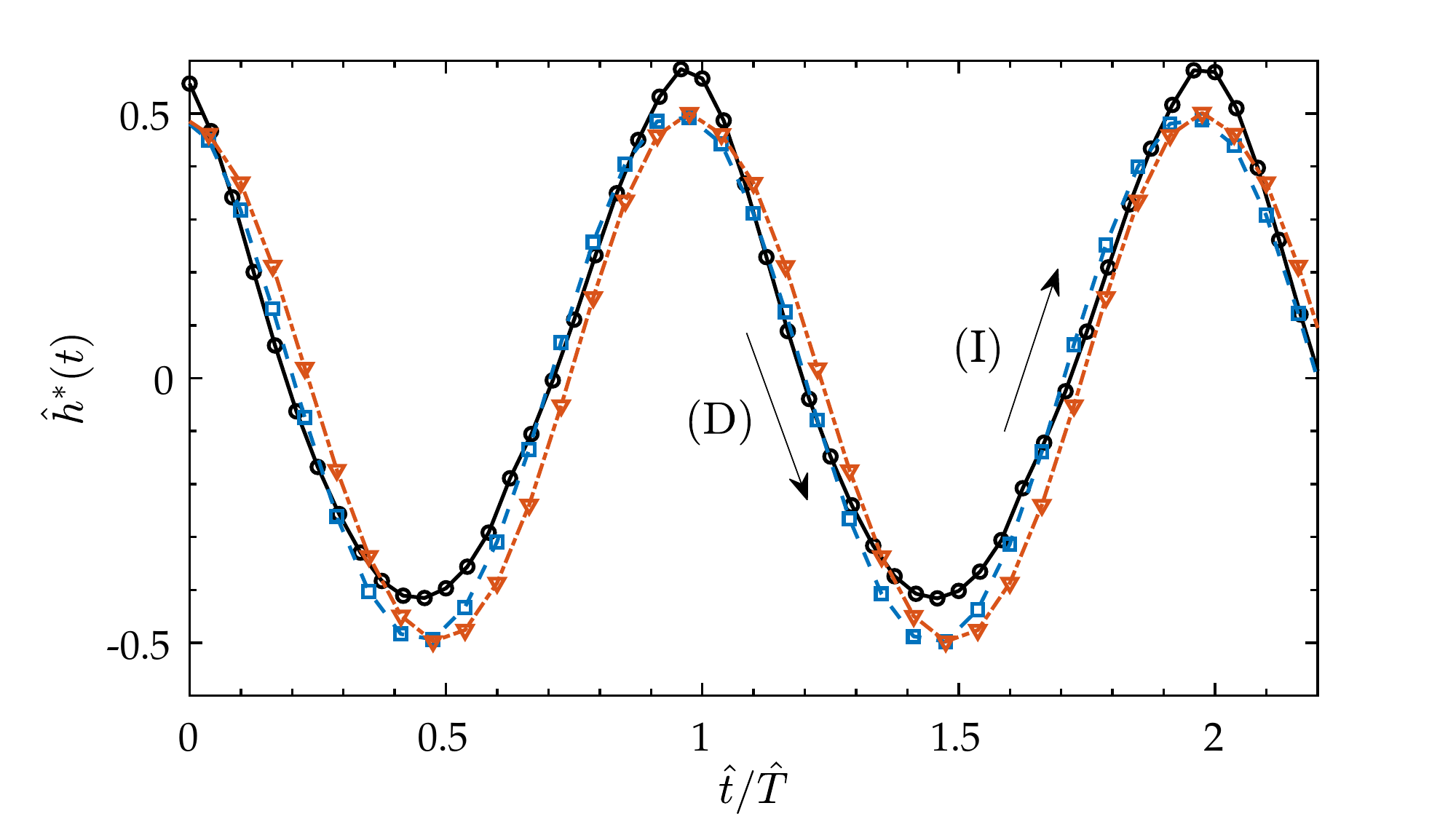}
\includegraphics[height=4.3cm]{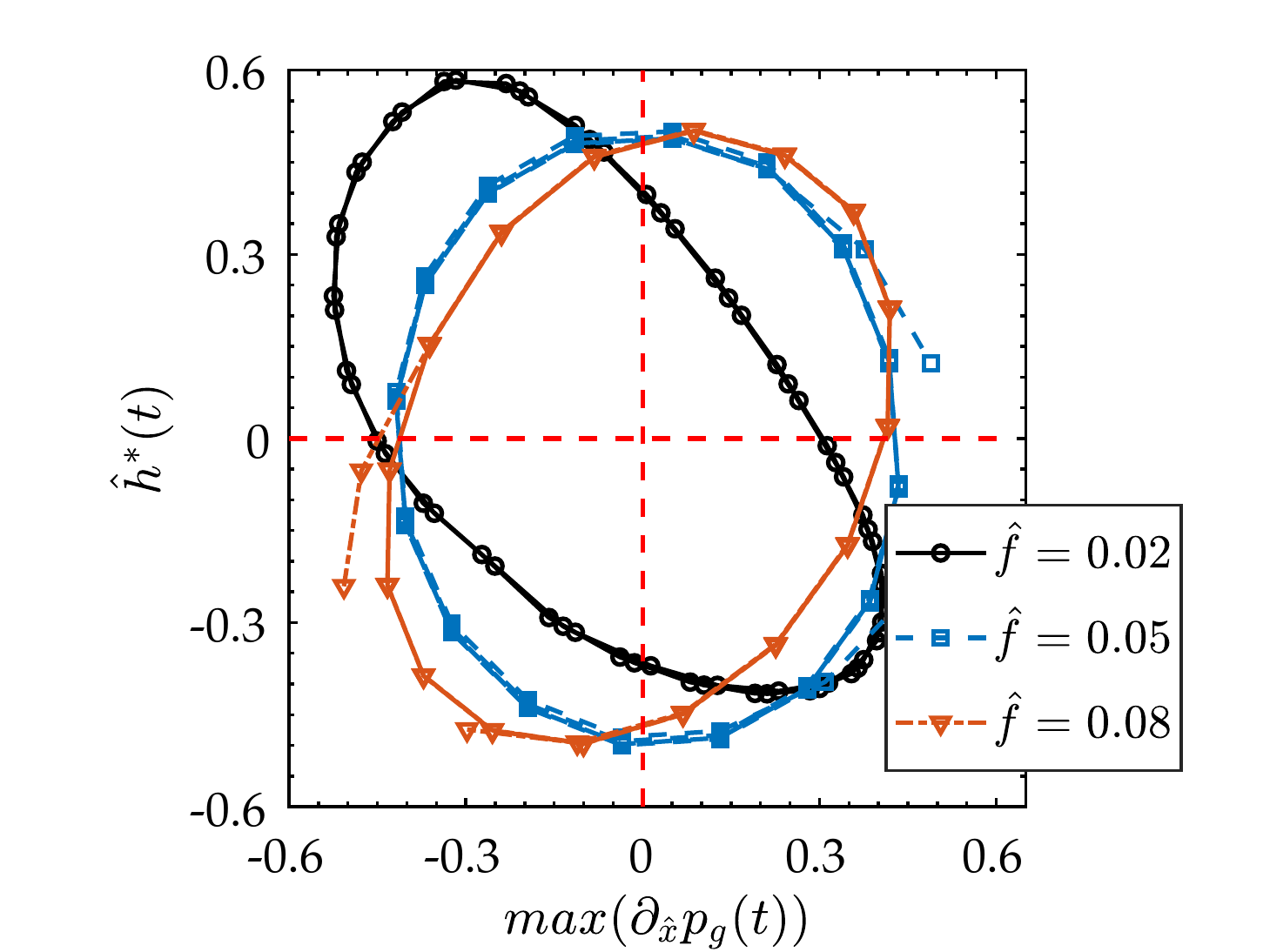}\\
\hspace{14mm}c)\hspace{65mm} d)
\caption{figures (a) and (c) show the film thickness at the impact point $\hat{h}^*(\hat{t})$ as a function of the period normalized time for three dimensionless frequencies, considering jet oscillations (a) and jet pulsations (c). figures (b) and (d) show the phase portrait linking $\hat{h}^*(t)$ to the jet disturbance: this is the time varying location of the impact $x^*(\hat{t})$ in case of an oscillation (in b) and the time varying maxima $\max(\partial_{\hat{x}\hat{p}_g})$ in case of a pulsation (in d). }
\label{ORBITS_H}
\end{figure*}

The liquid film response to harmonic oscillation (first row of figure \ref{OSCI_ZINC}) is discussed first. 
At $\hat{f}=0.02$, the coating thickness is characterized by wave peaks of $\tilde{h}\approx 0.1$. The characteristic lines tracing the propagation of the waves clearly show that these originate below the average impact point, at $\hat{x}\approx -1$, that is in the region normally belonging to the run-back flow. Within the range $\hat{x}\in[-2,2]$, spanned by the jet during the oscillation, regions of liquid film accumulation ($\tilde{h}>0$) and depletion ($\tilde{h}<0$) alternate harmonically as the coating film \emph{follows} the jet oscillation. Within the region intersected by the jet oscillation, the propagation speed of the coating waves is not constant and strongly influenced by the displacement of pressure gradient and shear stress. Outside this region, the wave propagation speed remains constant and approximately equal to the substrate speed for $\hat{x}>2$. {The liquid meniscus follows the displacement of the wiping region and the contour-map of the mean-shifted thickness $\tilde{h}$ is almost symmetric along $\hat{x}=0$}.

\begin{figure*}
\centering
\vspace{1mm}
-----------------Harmonic Oscillation-----------------\\
\vspace{2mm}
\includegraphics[trim={0.4cm 0 1.5cm 0},clip,width=6.7cm]{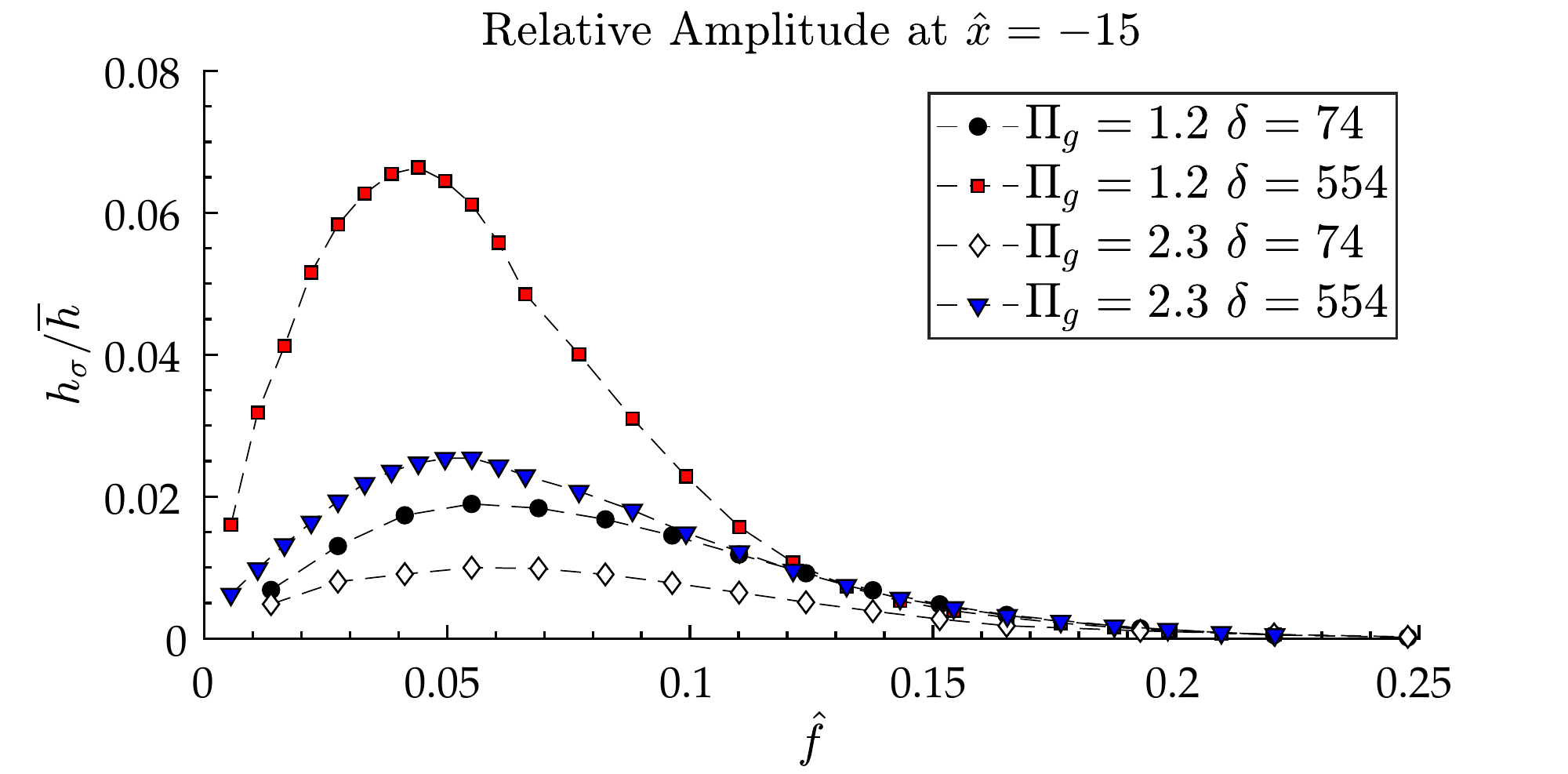}  
\includegraphics[trim={0.4cm 0 1.5cm 0},clip,width=6.7cm]{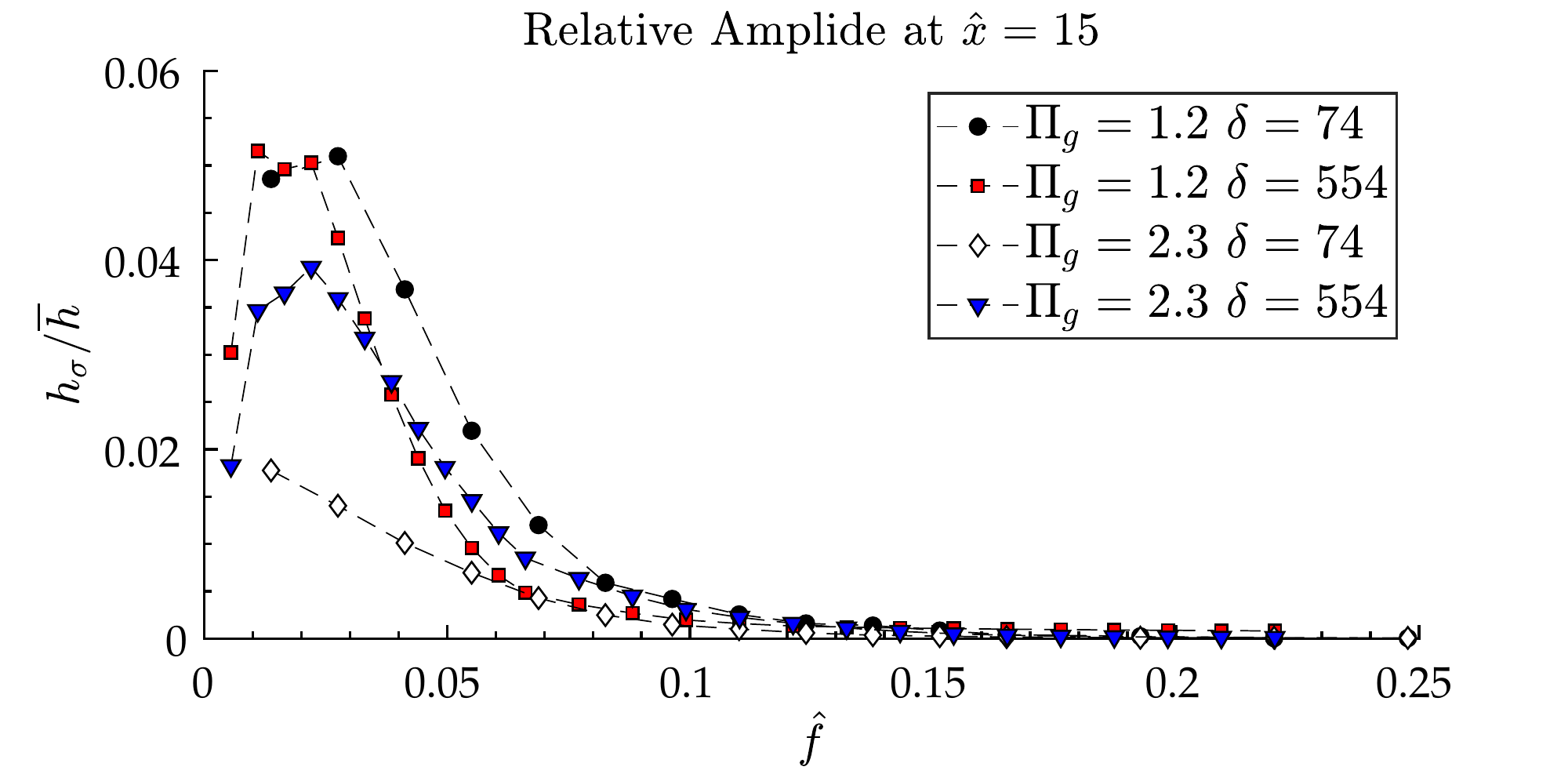} \\
-----------------Upward-Biased Oscillation-----------------\\
\vspace{2mm}
\includegraphics[trim={0.4cm 0 1.5cm 0},clip,width=6.7cm]{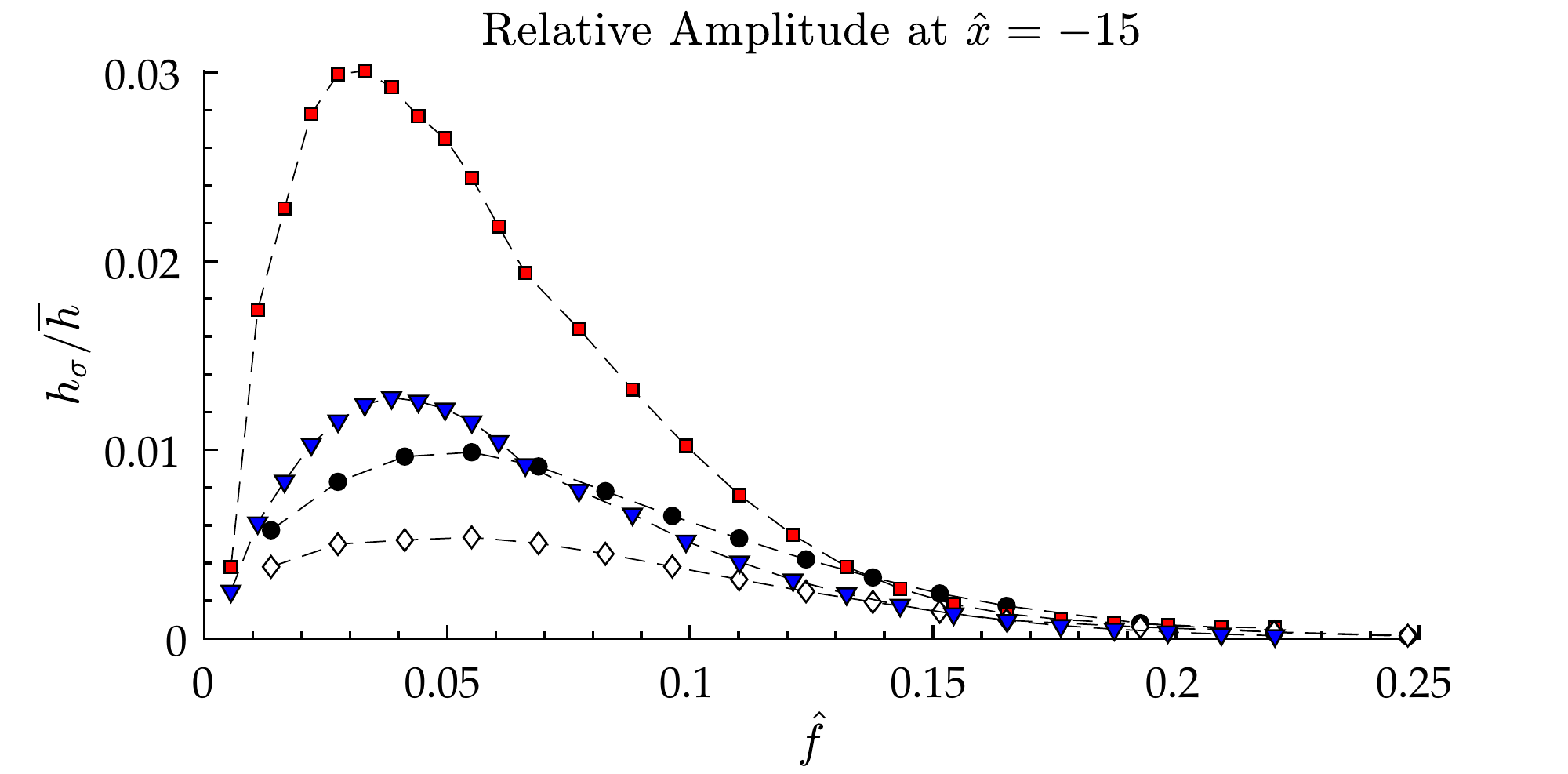} 
\includegraphics[trim={0.4cm 0 1.5cm 0},clip,width=6.7cm]{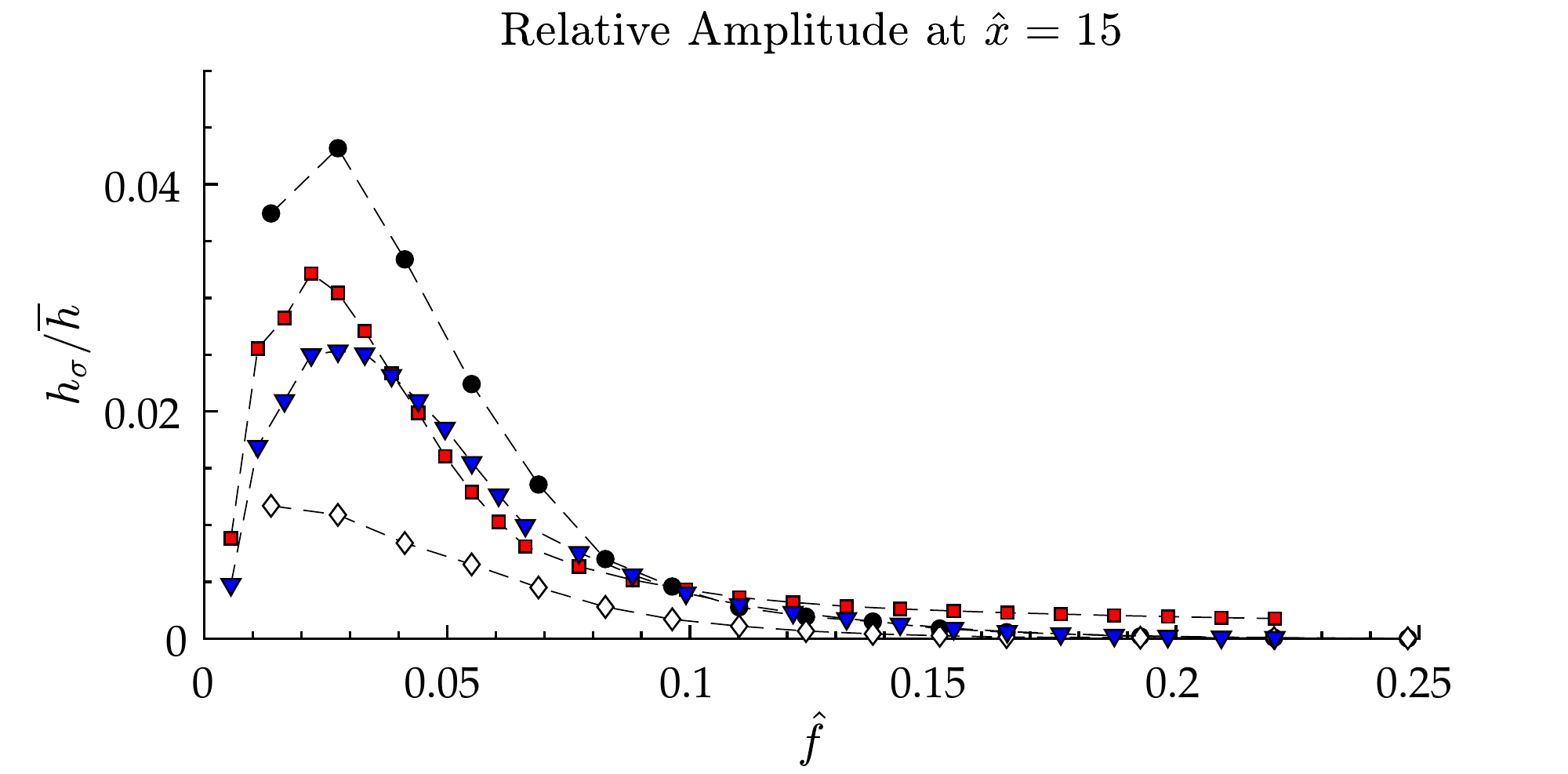} \\
-----------------Downward-Biased Oscillation-----------------\\
\vspace{2mm}
\includegraphics[trim={0.4cm 0 1.5cm 0},clip,width=6.7cm]{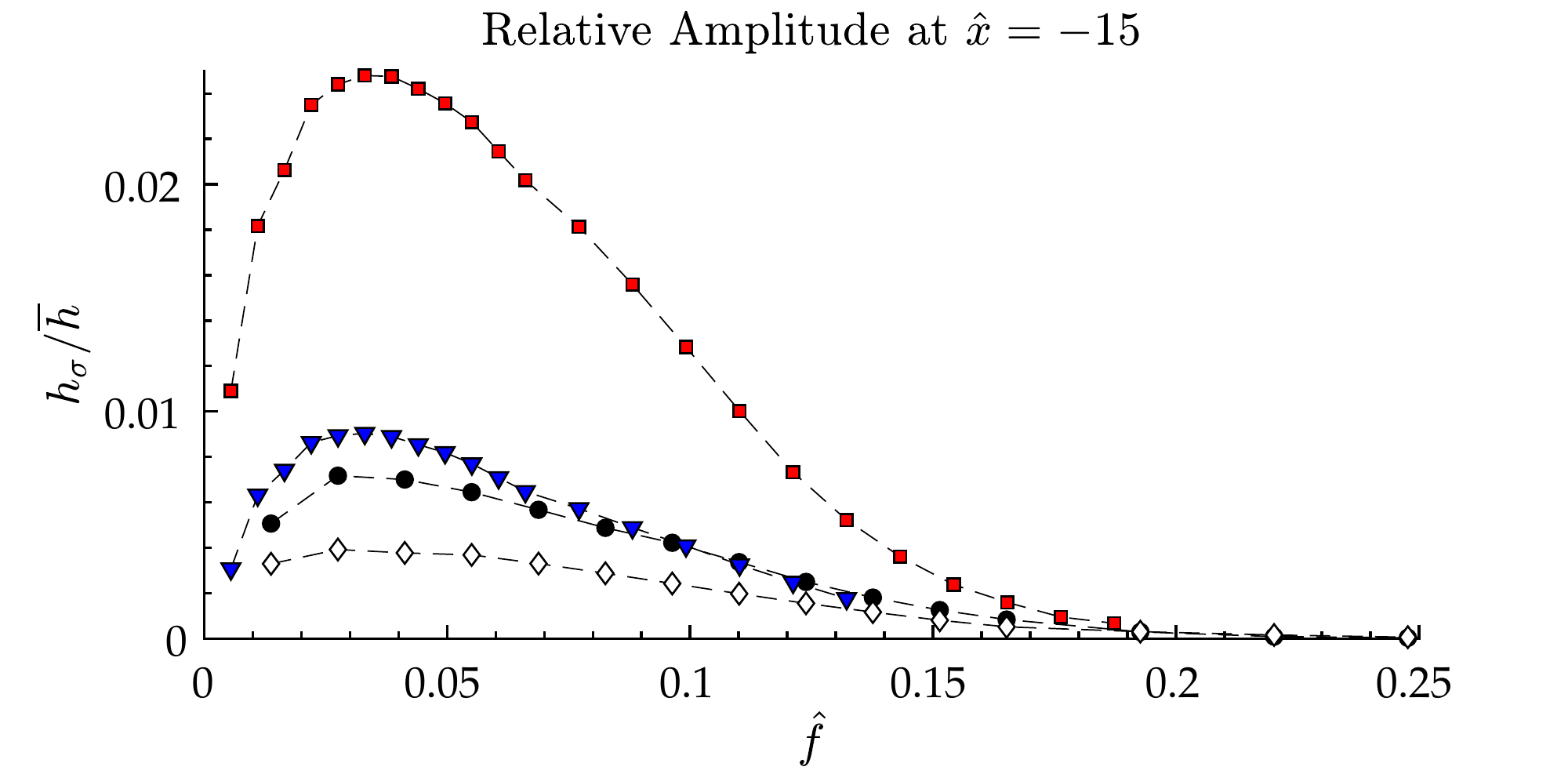}  
\includegraphics[trim={0.4cm 0 1.5cm 0},clip,width=6.7cm]{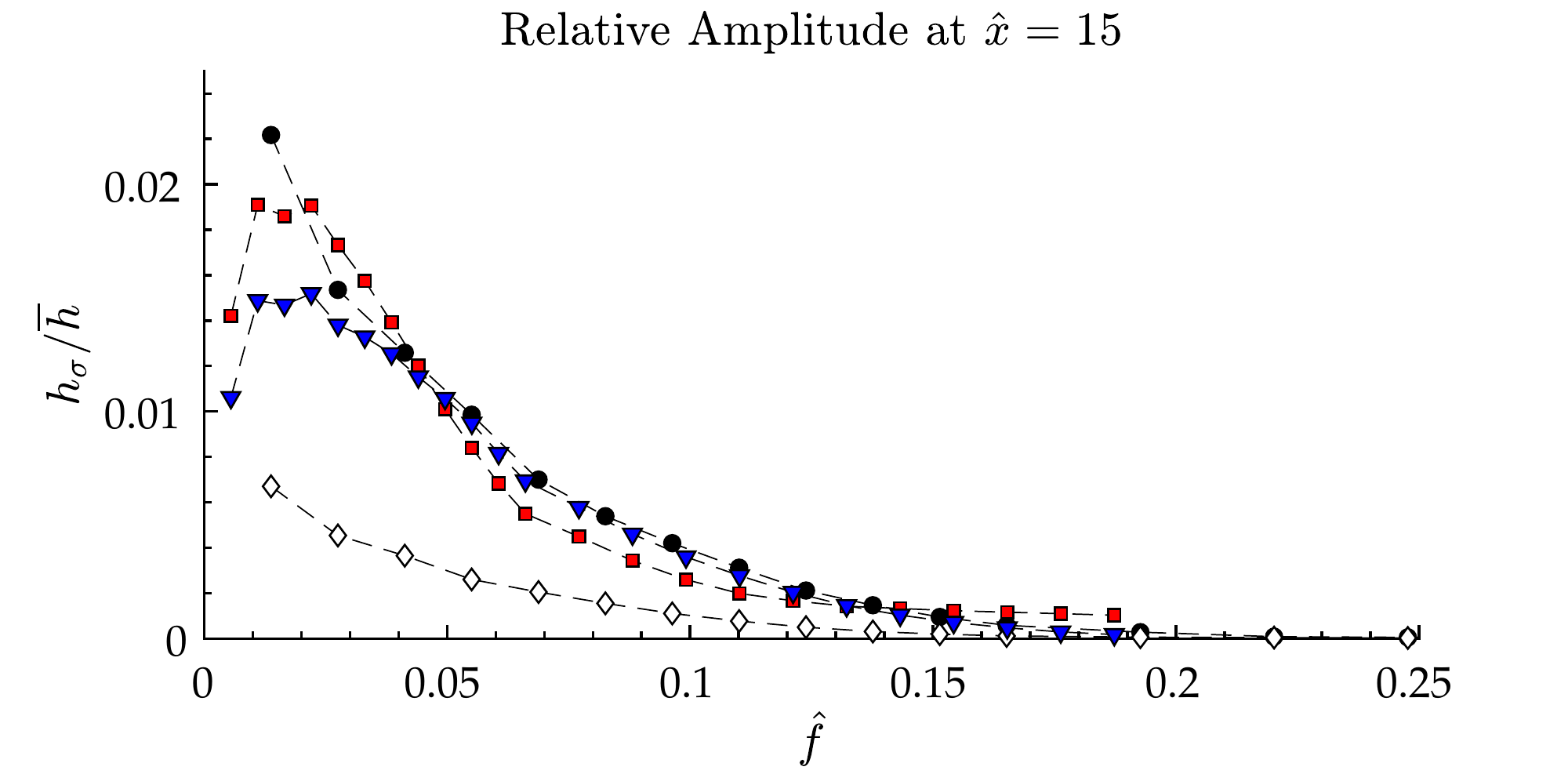}\\
-----------------Jet Pulsation-----------------\\
\vspace{2mm}
\includegraphics[trim={0.4cm 0 1.5cm 0},clip,width=6.7cm]{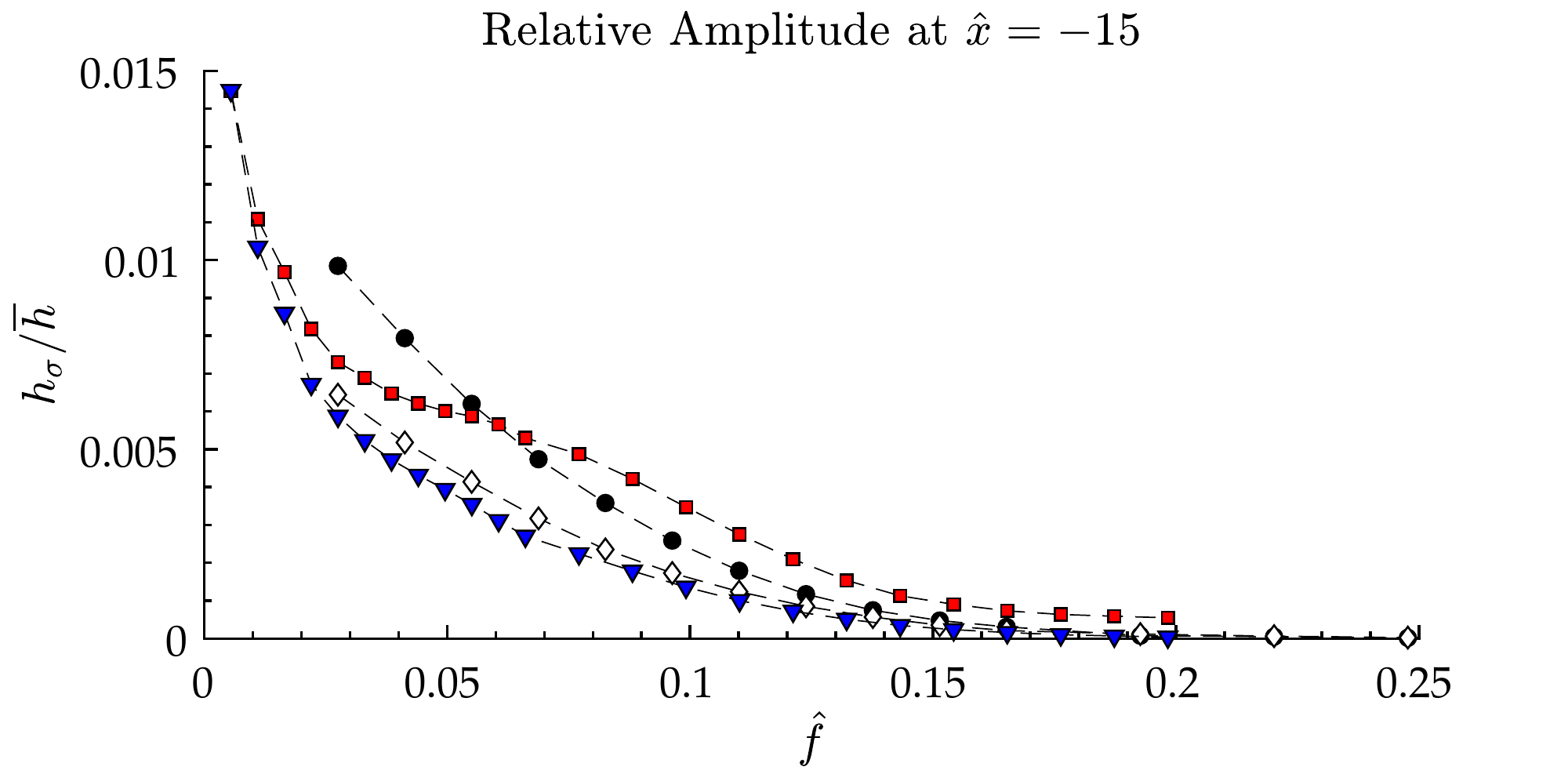} 
\includegraphics[trim={0.4cm 0 1.5cm 0},clip,width=6.7cm]{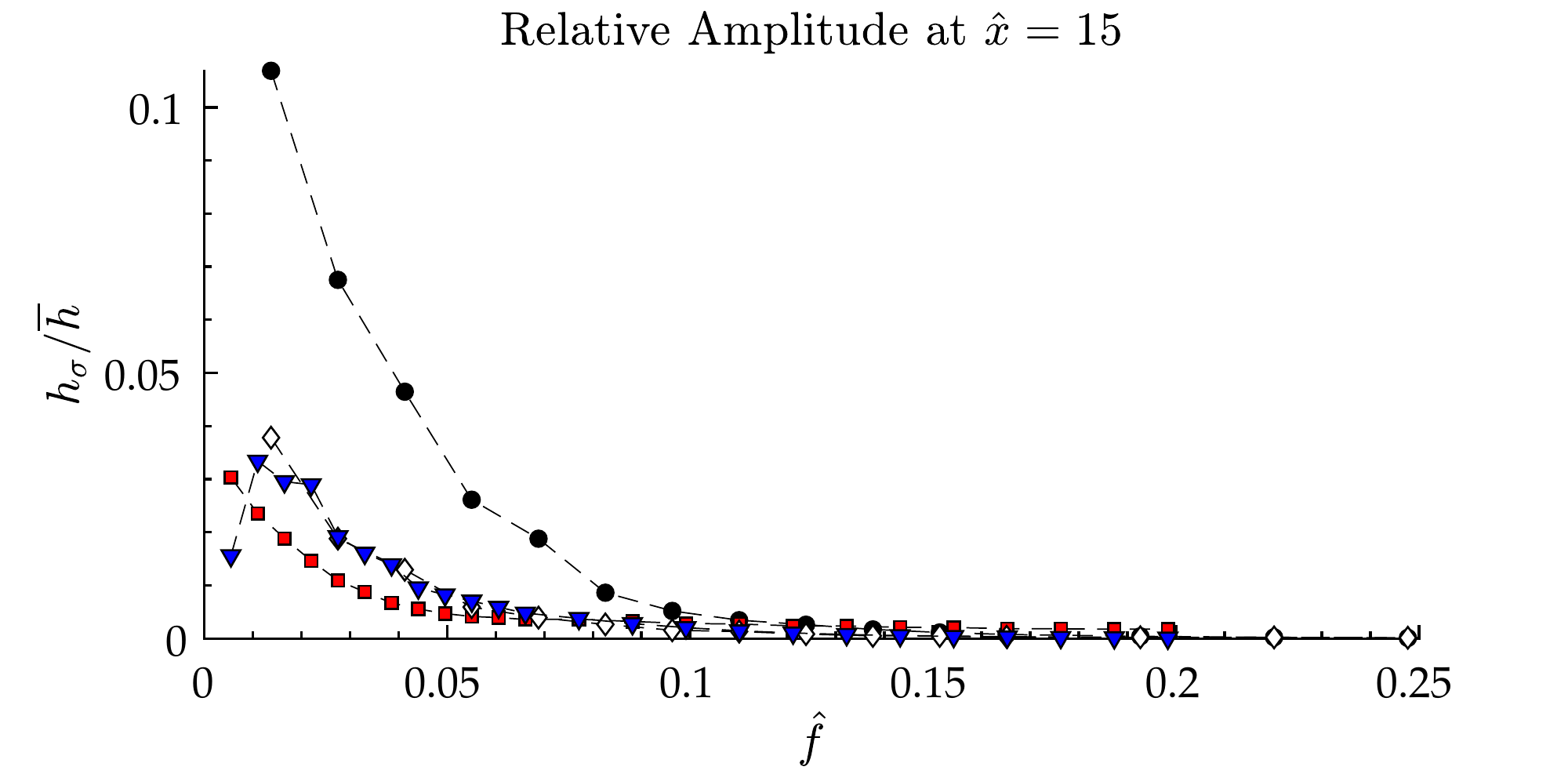}
\caption{Amplitude of the coating waves in terms of $\hat{h}_{\sigma}/\overline{h}$ as a function of the dimensionless perturbation frequency for four wiping conditions, indicated in the legend, and for the four jet perturbations previously considered. The curves on the left column are computed in the final coating region (at $\hat{x}=-15$) while the curves on the right are computed in the run back flow region (at $\hat{x}=15$). All the tests refer to galvanizing conditions.}
\label{TRANSF_COMP}
\end{figure*}

This is not the case at higher frequencies ($\hat{f}=0.05$ and $\hat{f}=0.08$), in which the disturbances in the run-back flow appear comparatively much lower and the waveform of the liquid changes significantly. To better analyze the difference in the wave formation mechanism, 
{we consider the response of the liquid film thickness $\hat{h}^*(t)$ in the impact point $x^*(t)$, defined as the point in which the gas pressure is maximum.}
Figure \ref{ORBITS_H}a and b shows, respectively, the evolution of $\hat{h}^*(t)$ a function of the dimensionless time, scaled by the wave period, and its phase portrait with $x^*(t)$, describing the oscillation of the impact point. For plotting purposes, the curves in figure \ref{ORBITS_H}a are shifted to have matching wave peaks, while the phase portraits are constructed by mean-shifting both signals and normalizing with respect to their peak to peak amplitude.

At the frequency of $\hat{f}=0.02$, the film thickness at the impact point remains almost sinusoidal and has a constant phase delay of approximately $\pi/4$. This phase delay is due to the response time of the liquid film as the jet moves towards the run-back flow, encounters a region of higher thickness, and imposes a thickness reduction. At $\hat{f}=0.05$ and $\hat{f}=0.08$, the response $h^*(t)$ is no longer sinusoidal and the film does not have enough time to allow for the wiping: during the ascending phase of the jet oscillation (denoted as A in figure \ref{ORBITS_H}a), a portion of un-wiped liquid is dragged from the run-back flow region and pushed towards the final coating region, from which it continues its upwards evolution at the speed of the substrate. As a result, the wave peak reached in the case of $\hat{f}=0.05$ is about twenty times higher than in the case of $\hat{f}=0.02$. These results confirm the existence of the mechanisms for the wave formation originally postulated in the previous experimental investigation by the authors \citep{Mendez2019} and which hereinafter referred to as mechanisms A.

The non-harmonic cases in the second and third rows of figure \ref{OSCI_ZINC} show that both the shape and the maximum amplitude of the resulting waves are strongly influenced by the waveform of the jet oscillation. Nevertheless, the key observation from the harmonic case applies: the characteristic lines through which the waves evolve (both towards the runback flow and towards the final coating film) originates \emph{below} the mean wiping point.

It is now instructive to consider the pulsating test cases in the last row of figure \ref{OSCI_ZINC}. In this case, as the impingement point is fixed in time, the waves originate at the wiping line $\hat{x}=0$ and move at a constant speed in both downwards and upwards directions. figure \ref{ORBITS_H}c) and d) show, respectively, the time evolution of the thickness at the impact point and its phase portrait with the maximum pressure gradient during the jet pulsation. The same plotting adjustments of figure \ref{ORBITS_H}a and \ref{ORBITS_H}b in terms of shifting and normalization are adopted. Regardless of the pulsation frequency considered, the thickness at the impinging point remains overall sinusoidal, with the phase delay converging towards $\pi/2$, which is a perfect quadrature. This second mechanism of wave formation is herein referred to as mechanism B.

To conclude this section, figure \ref{TRANSF_COMP} collects the frequency response of the film coating in both the final coating region and the runback flow region for the four perturbations considered and for four combinations of wiping number $\Pi_g$ and rescaled Reynolds number $\delta$. All the test cases refer to galvanizing conditions, and the selected pairs $(\Pi_g$,$\delta)$ are indicated in the legend. For each of the wiping conditions, the plot shows the dependency of the wave amplitude, measured in terms of the standard deviation $h_{\sigma}$ to average $\overline{h}$ ratio, over the dimensionless frequency. For the final coating film, these curves are computed at a location $\hat{x}=-15$ while $\hat{x}=15$ is considered for the runback flow.

As expected from the previous analysis of the contour-maps, the amplitude of the coating waves in the case of an oscillating jet (mechanism A) is significantly larger than in the case of a pulsating jet (mechanism B). Moreover, while the mechanism A shows a region of strong receptivity in the range of dimensionless frequencies $\hat{f}=[0.03-0.08]$, the liquid film behaves as a low pass filter with respect to the mechanism B. Regardless of the mechanism, no coating waves can be expected for perturbations at $\hat{f}>0.2$. 

By comparing the modulation curves for the two regions of the coating flow, the lower portion of the receptivity band, (say $\hat{f}\approx[0.03-0.05]$) appears of significant interest. This range of frequency is also strongly present in the runback flow, while the second portion (say $\hat{f}\approx[0.05-0.08]$) is more attenuated. In the coupling mechanism described in previous studies \citep{Mendez2019,Mendez2017,Pfeiler2017}, the jet oscillation is sustained by the waves in the runback coating flow. These waves appear to be the major cause of the unstable interaction between the two flows.

Concerning the range of possible frequencies, it is interesting to report that $\hat{f}=0.05$ --at which maximum amplitude of the coating waves can be expected-- corresponds to wavelengths of the order of $\lambda=25-35$ \SI{}{mm} depending on the wiping conditions. This range is in agreement with undulation defects observed in several galvanizing lines \citep{Pfeiler2017}. Moreover, the results show that the undulation amplitude produced by an oscillating jet (mechanism A) increases at larger substrate speeds --a fact also in line with industrial observations-- while it remains rather insensitive for a pulsating jet (mechanisms B).

As to the role of the wiping number, its impact on the coating wave amplitude is of more difficult interpretation, especially if one considers that the wiping of liquids with low kinematic viscosity such as liquid zinc or water is also strongly influenced by the shear stress, as shown in figures \ref{RES_1}. While increasing the wiping number results in a thicker runback flow, the main role of the shear stress is that of smoothing the wiping meniscus and thus the transition from the final coating film to the runback flow. The smoother is this meniscus, the smaller is the thickness gradient encountered by the jet during oscillation, and hence the lower the impact of mechanism A.
These results also show that the interaction between the gas and the liquid film is strongly influenced by the waveform of a possible jet oscillation: while the receptivity range of the film remains unaffected, the amplitude of the undulation is significantly larger in a harmonic oscillation than in the non-harmonic ones.

Finally, it is interesting to compare the dimensionless modulation function obtained in the galvanizing conditions with the ones obtained for the wiping of water. Only the response to harmonic oscillation is considered, and shown in figure \ref{RESP_WATER} for four wiping conditions. Despite the largely different dimensionless numbers controlling the process, the maximum undulation amplitude in the final coating film is produced in a similar range of dimensionless frequencies $\hat{f}=[0.03-0.08]$ -- i.e. the band dominated by the mechanism A. Given the largely different properties of the two liquids, these results show that the Skhadov-like scaling well describes the physical phenomena governing the maximum receptivity of the liquid film.

\begin{figure}
\centering
\includegraphics[trim={0.4cm 0 1.5cm 0},clip,width=6.7cm]{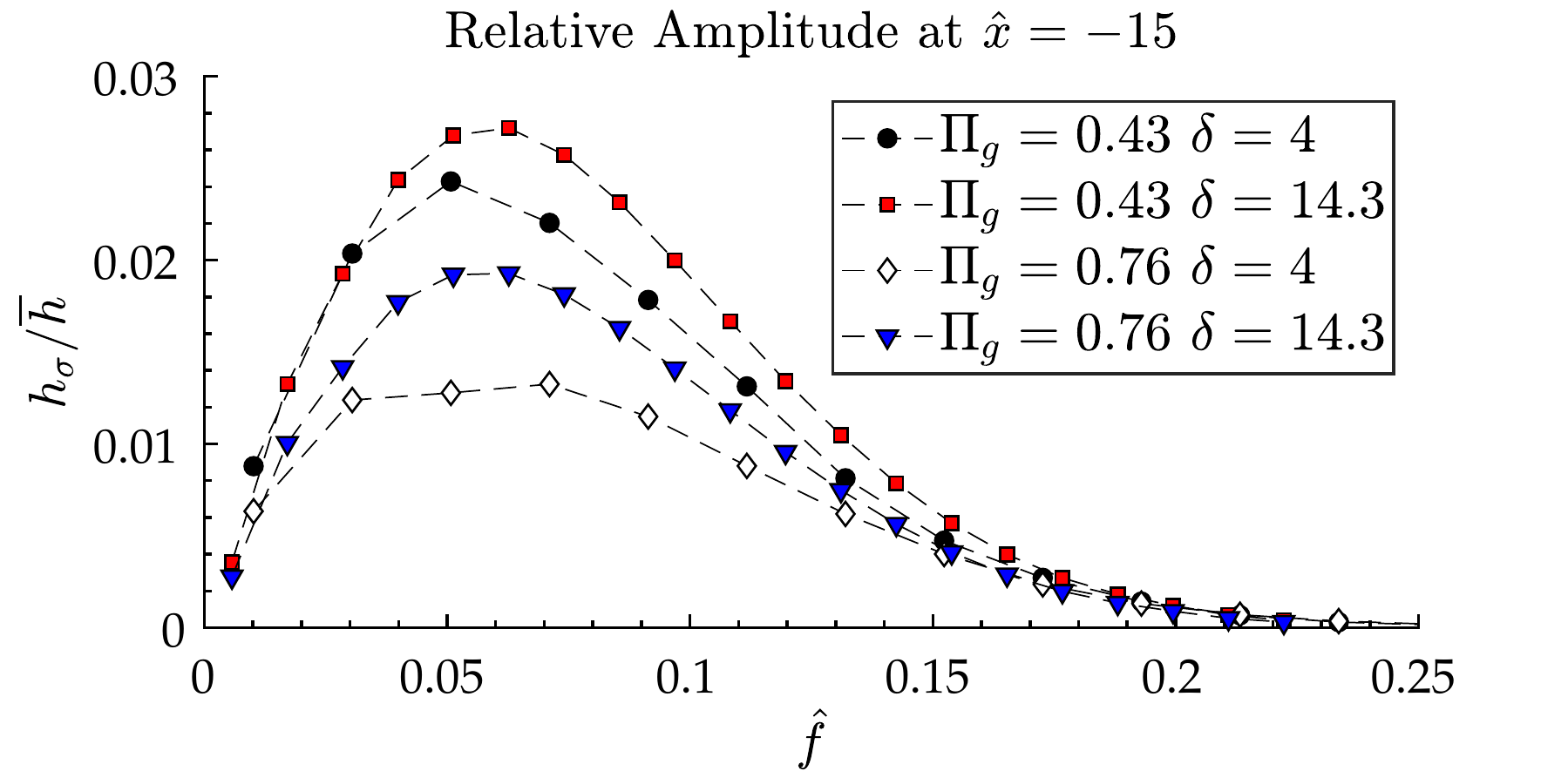} 
\includegraphics[trim={0.4cm 0 1.5cm 0},clip,width=6.7cm]{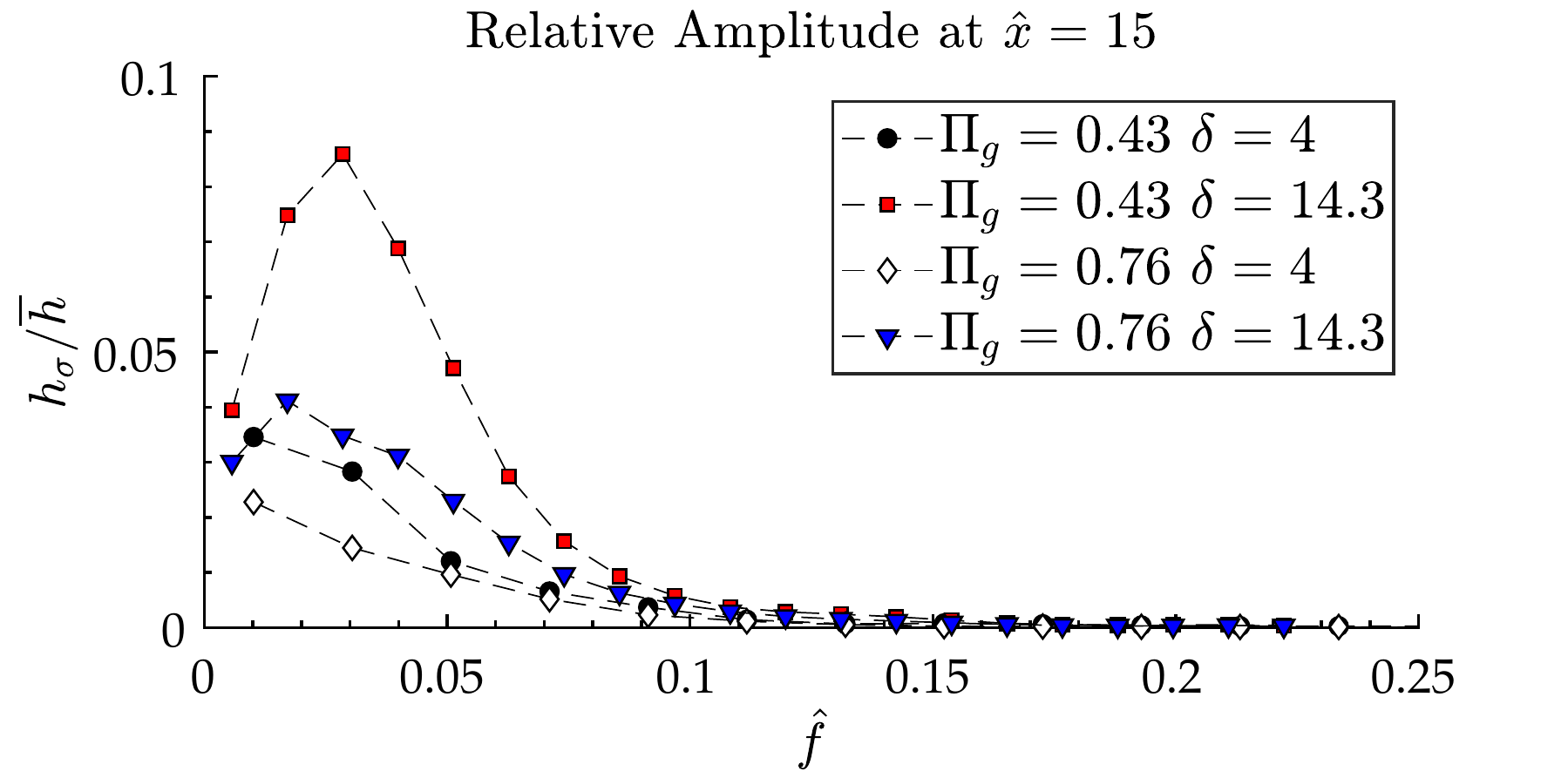} \\
\caption{Same as in figure \ref{TRANSF_COMP} but considering the wiping of water instead of liquid zinc, and harmonic oscillations only.}
\label{RESP_WATER}
\end{figure}

\begin{figure}
\includegraphics[trim={0.4cm 0 1.2cm 0},clip,width=6.8cm]{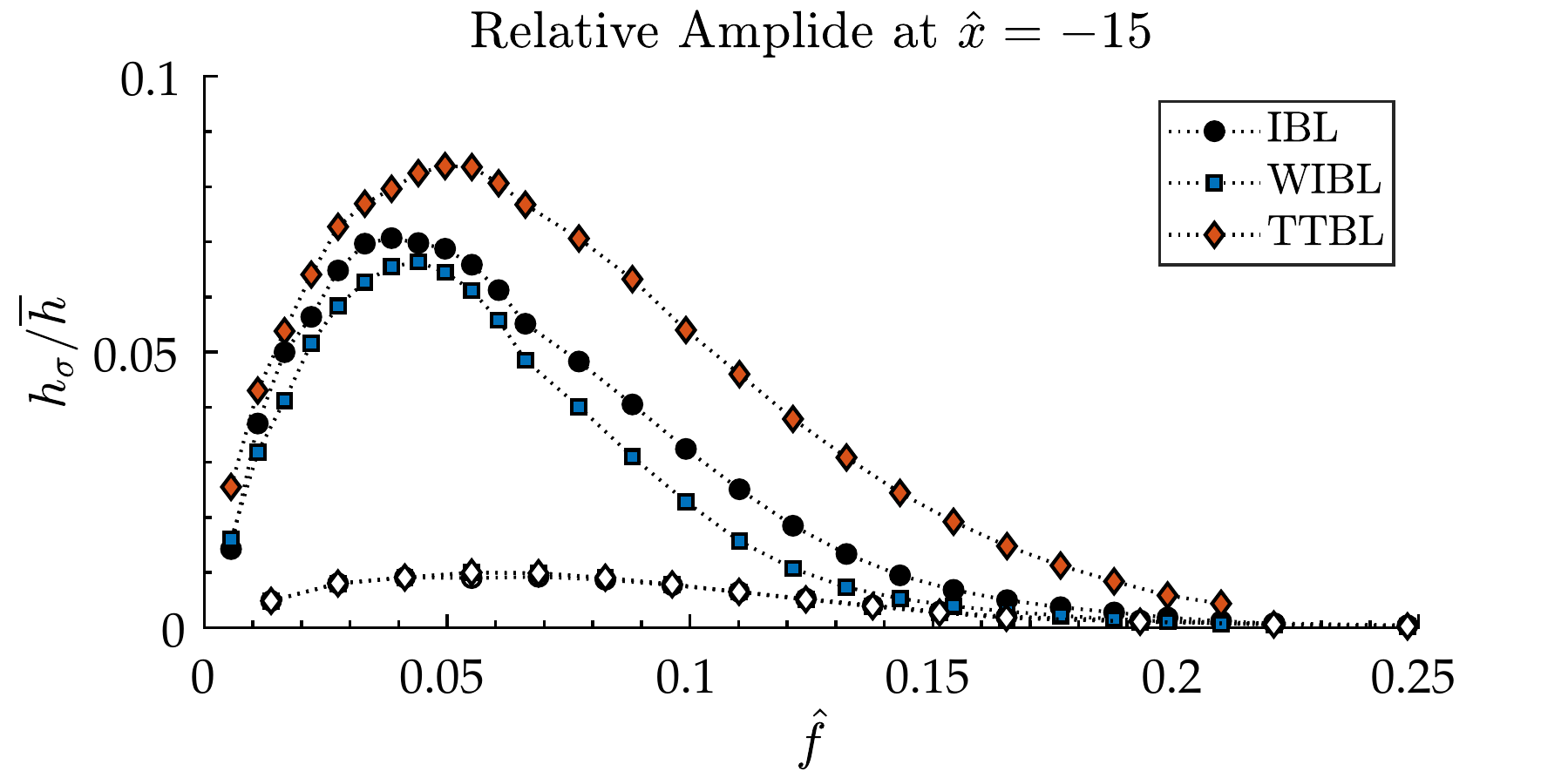}
\includegraphics[trim={0.4cm 0 1.2cm 0},clip,width=6.8cm]{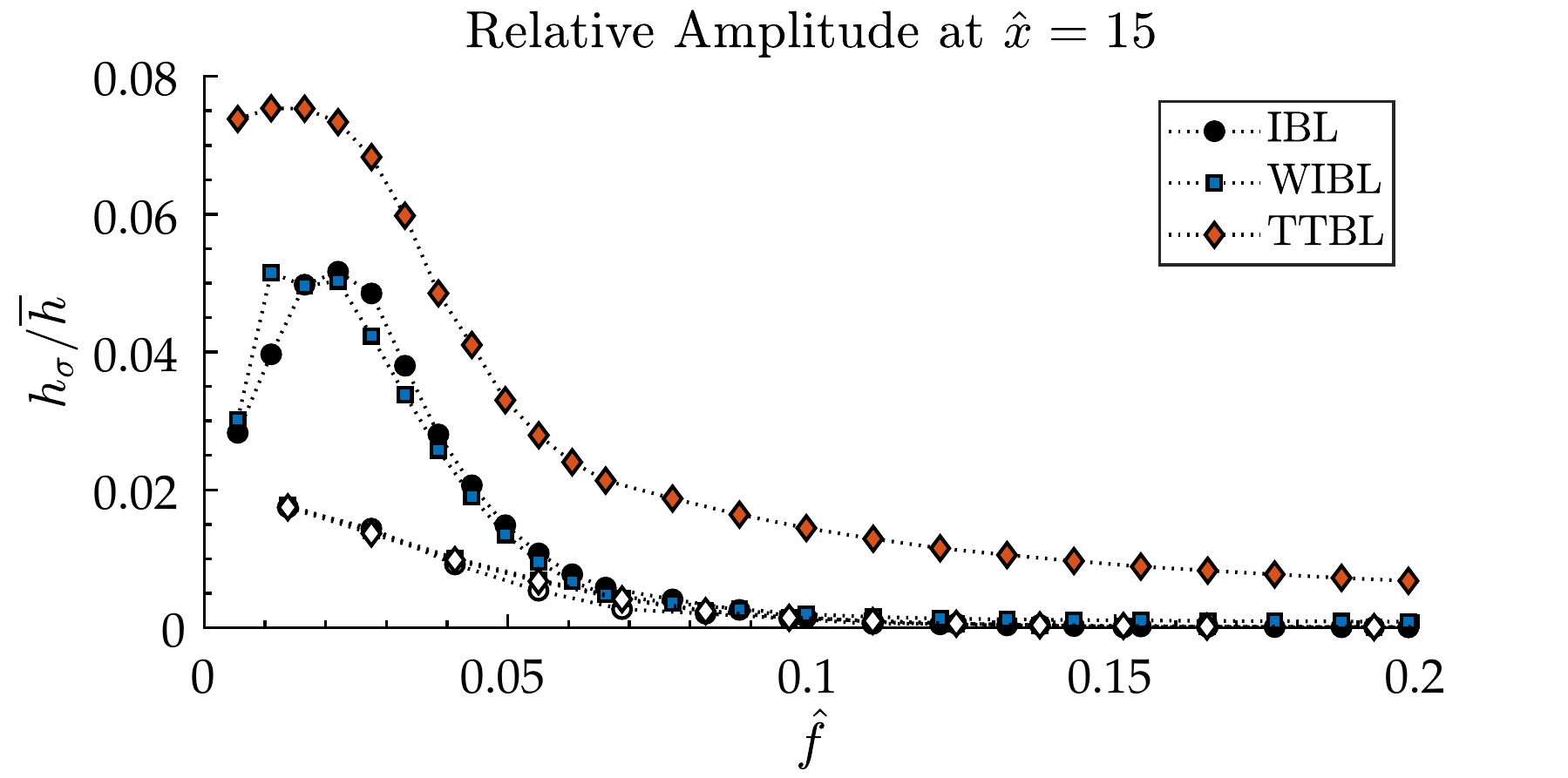}\\
\caption{Same plot as the first row of \ref{TRANSF_COMP}, considering the two extreme cases $\Pi_g=1.2$, $\delta=554$ (largest response, filled markers) and $\Pi_g=2.3$, $\delta=74$ (smallest response, empty markers) comparing the results from the IBL, the WIBL and the TTBL models.}
\label{COMP_NS_SS}
\end{figure}

\subsection{The influence of the modeling strategy}\label{RES_D}

It is finally of interest to analyze the impact of the modeling strategy on the results previously obtained by comparing the IBL, the WIBL and the TTBL models. We here consider the response of the liquid film to harmonic jet oscillations in galvanizing conditions. Figure \ref{COMP_NS_SS} compares the frequency response from the three models on the two extreme cases from the first row of figure \ref{TRANSF_COMP}: these are $\Pi_g=1.2$ and $\delta=554$ (largest response, filled markers) and $\Pi_g=2.3$, $\delta=74$ (smallest response, empty markers). The impact of the model becomes more pronounced as larger waves are considered, with the TTBL predicting significantly larger waves both in the final coating film and in the run back flow. It is nevertheless interesting to observe that the range of maximum receptivity of the liquid film remains unvaried, with the three curves showing the same qualitative behavior.

\begin{figure*}
\centering
\vspace{2mm}
\includegraphics[trim={2.6cm 0 2.6cm 0},clip,width=8.4cm,angle=270,origin=c]{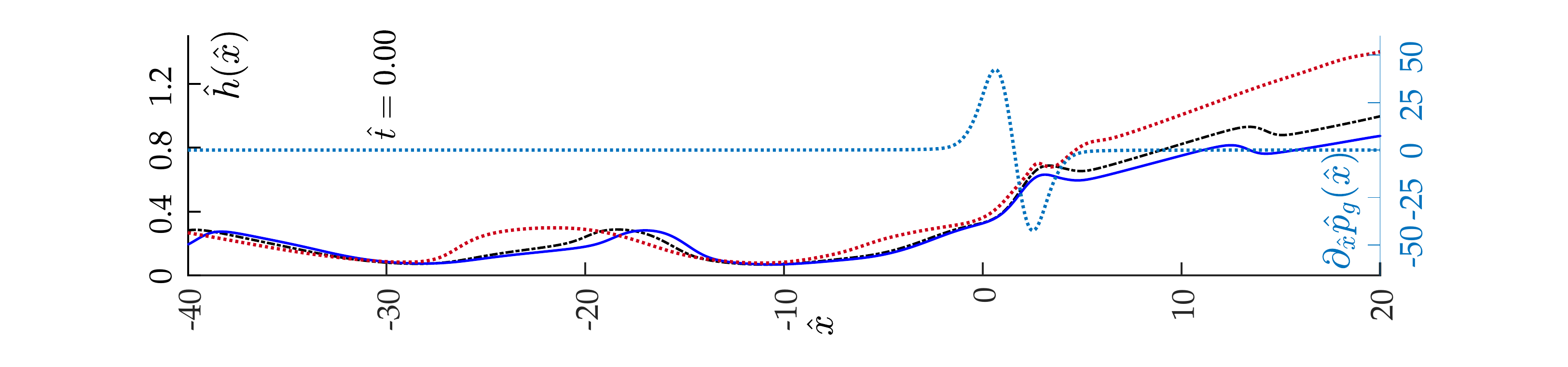}
\includegraphics[trim={2.6cm 0 2.6cm 0},clip,width=8.3cm,angle=270,origin=c]{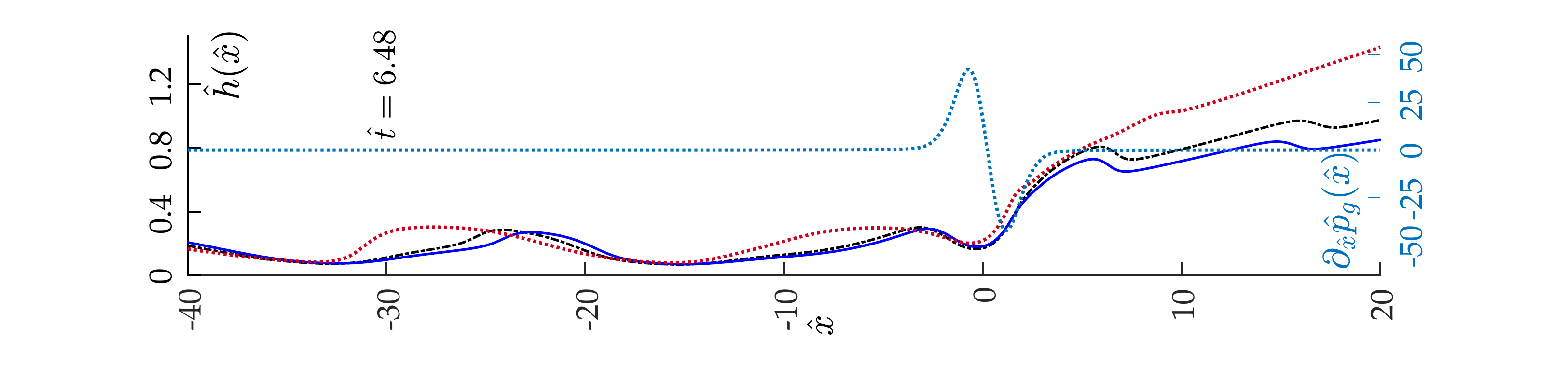}
\includegraphics[trim={2.6cm 0 2.6cm 0},clip,width=8.3cm,angle=270,origin=c]{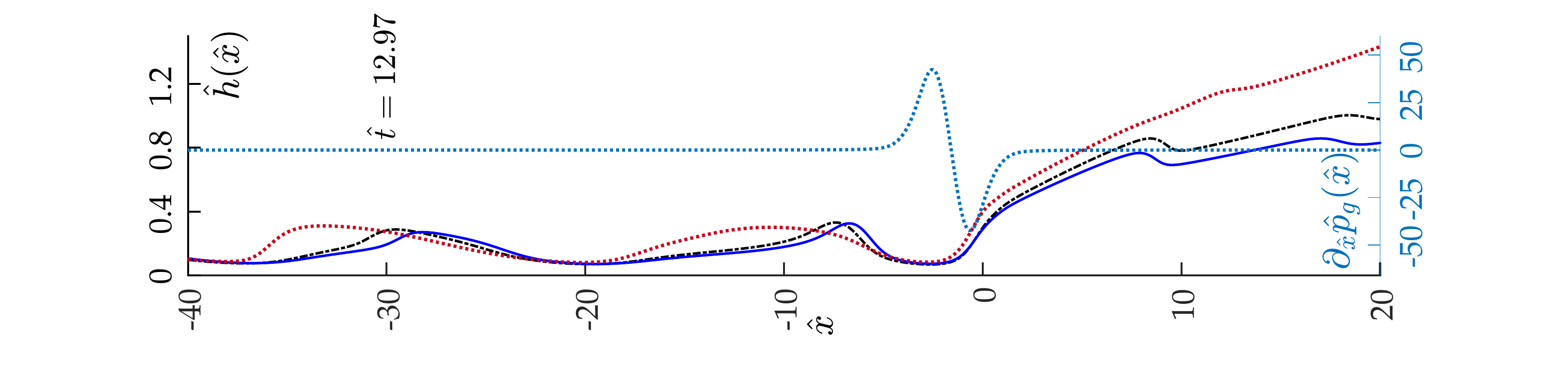}
\includegraphics[trim={2.6cm 0 2.6cm 0},clip,width=8.3cm,angle=270,origin=c]{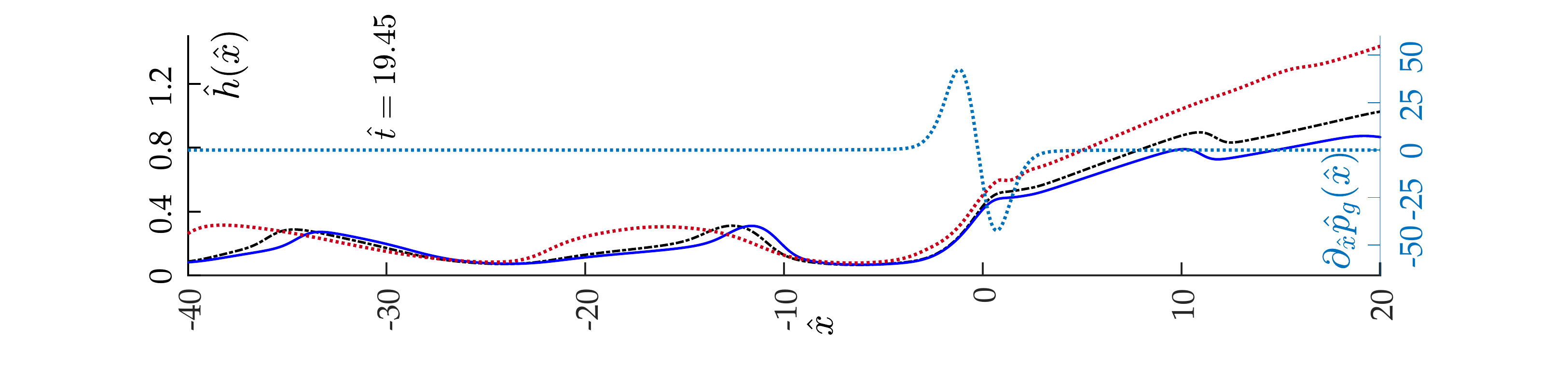}
\includegraphics[trim={2.6cm 0 2.6cm 0},clip,width=8.3cm,angle=270,origin=c]{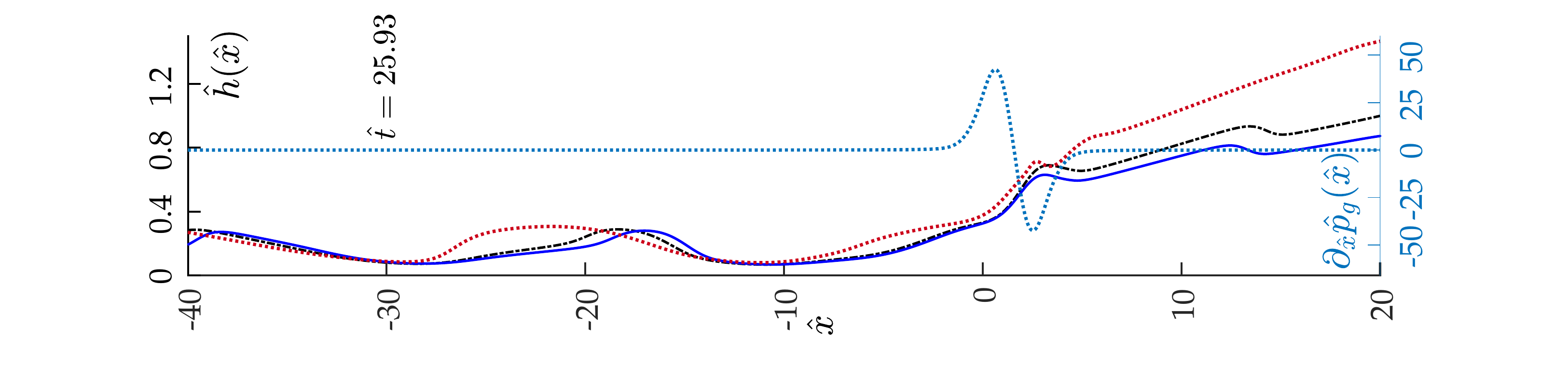}\\
\vspace{1mm}
\includegraphics[width=1.1cm]{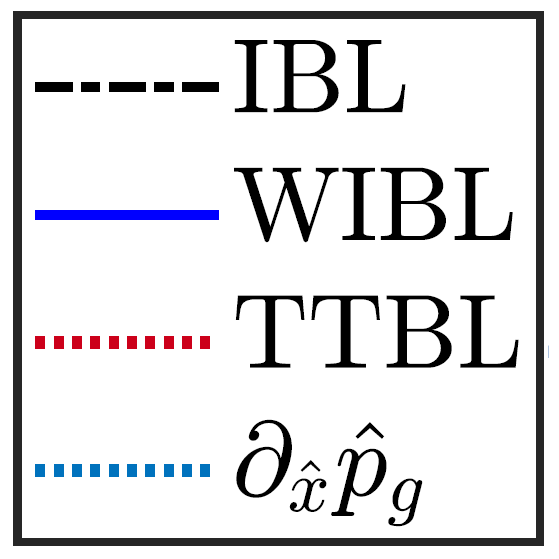}\\
\vspace{2mm}
\includegraphics[trim={2.6cm 0 2.6cm 0},clip,width=8.3cm,angle=270,origin=c]{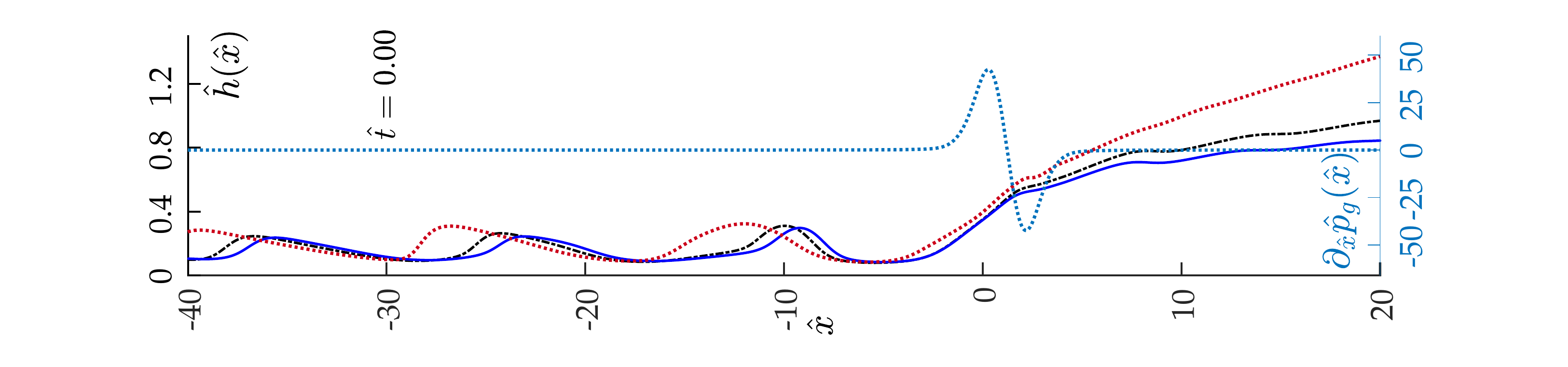}
\includegraphics[trim={2.6cm 0 2.6cm 0},clip,width=8.3cm,angle=270,origin=c]{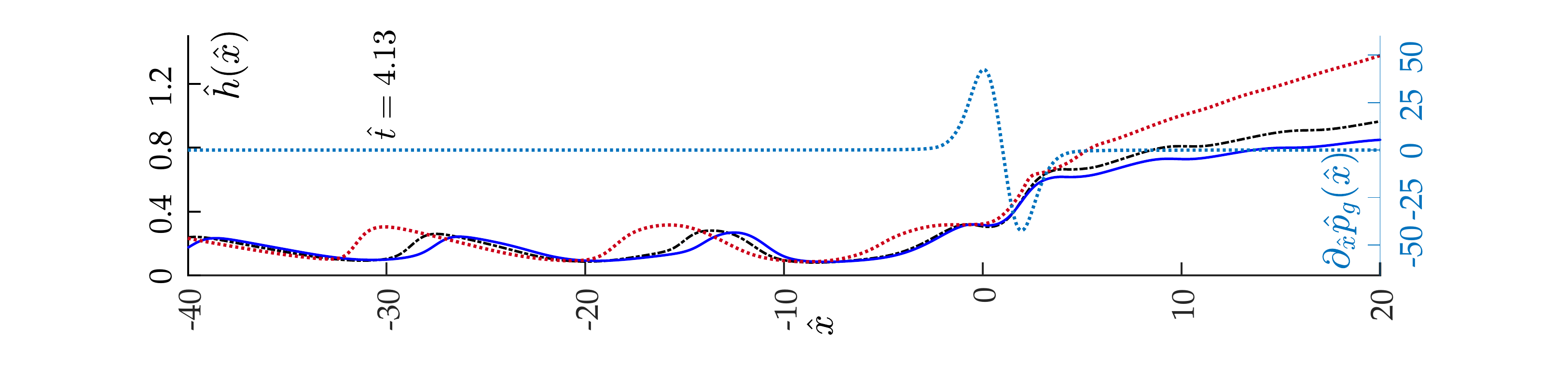}
\includegraphics[trim={2.6cm 0 2.6cm 0},clip,width=8.3cm,angle=270,origin=c]{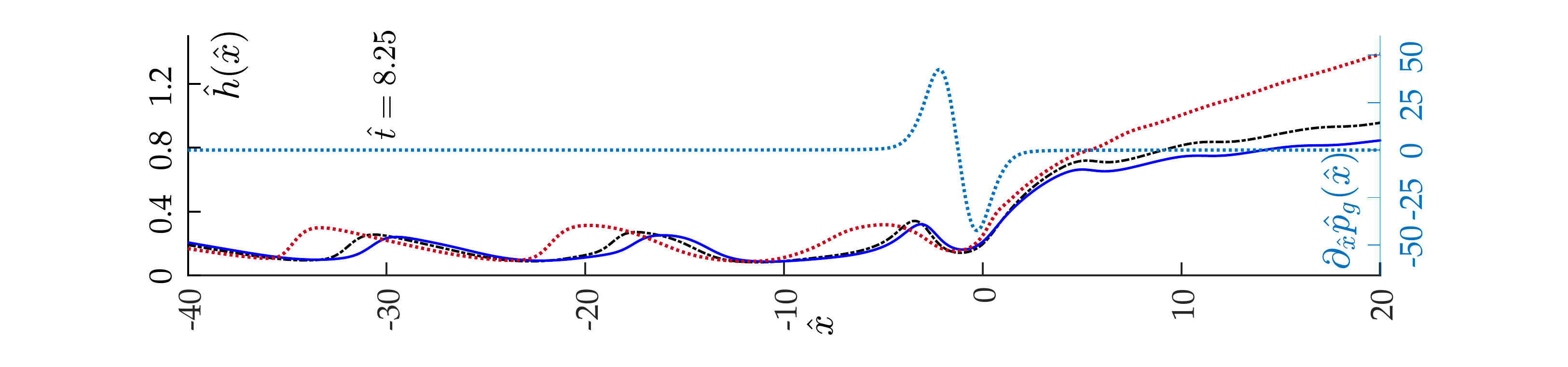}
\includegraphics[trim={2.6cm 0 2.6cm 0},clip,width=8.3cm,angle=270,origin=c]{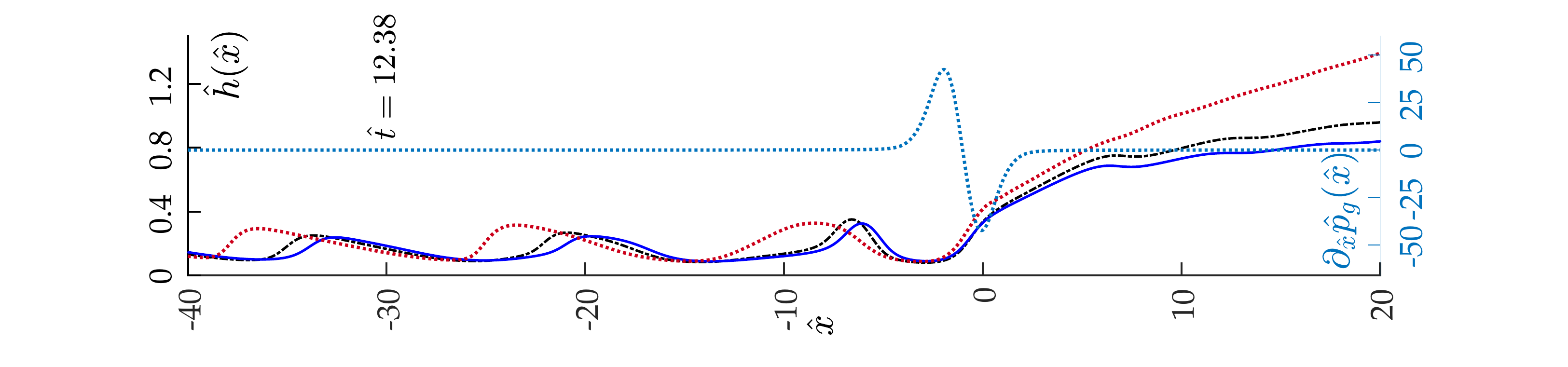}
\includegraphics[trim={2.6cm 0 2.6cm 0},clip,width=8.4cm,angle=270,origin=c]{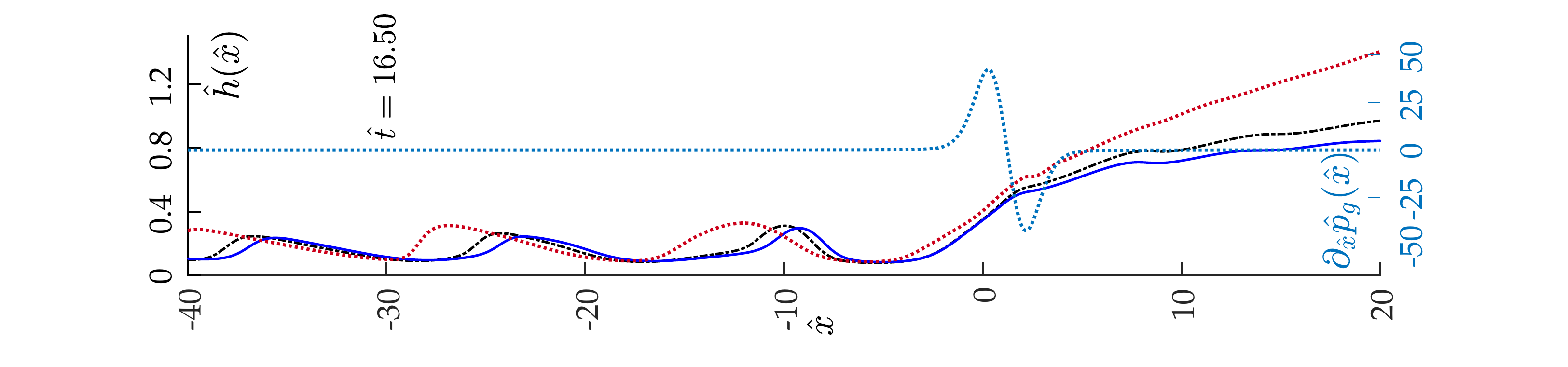}
\caption{Sequence of snapshots for test cases with $\Pi_g=1.2$ and $\delta=554$ and an harmonic oscillation of the impinging jet at $\hat{f}=0.04$ (top) and $\hat{f}=0.06$ (bottom). Each snapshot plots the film thickness for IBL (dashed black line), WIBL (continuous blue line) and TTBL (dotted red line),  together with the pressure gradient imposed by the impinging jet (light blue dotted line). The thickness axis is on the top of each plot; the pressure gradient axis is on the bottom. The time step is indicated in each plot. {Animations of both cases are provided as supplemental Movies 1 and 2.}} 
\label{SNAPS_LF}
\end{figure*}
 
Figure \ref{SNAPS_LF} shows five snapshots of the film thickness computed by the three models, together with the instantaneous pressure gradient profile, for two test cases with oscillation frequency $\hat{f}=0.04$ (on the top) and $\hat{f}=0.06$ (on the bottom) and wiping at $\Pi_g=1.2$ and $\delta=553$. These yields the largest waves in the final coating film mostly originated by mechanism A described in the previous section.

The sequence of five snapshots captures one period of this mechanism, starting from its most downward position (first column). In the second column, the effect of the wiping is visible while the snapshot in the third column captures the formation of the wave in the final coat, as a local minimum of thickness is dragged upward. In the snapshots of the fourth and fifth columns, the jet is in its descending phase, impinging on a much thicker film, and the mechanism A begins its next period.

It is interesting to observe that these waves become strongly asymmetric soon after their formation, with a steeper gradient on their tail. This asymmetry changes as the wave evolves downstream under the action of the shear stress, which imparts higher advection velocity to regions of higher thickness. While the differences between the laminar models are minor, the turbulence model leads to largely different shapes of the waves in the final coating film. Since in this region the TTBL model recovers the IBL, this difference is linked to the discrepancy in the prediction of the thickness of the run-back flow, where the TTBL yields a much thicker film.

\begin{figure*}
\centering
\includegraphics[height=8cm]{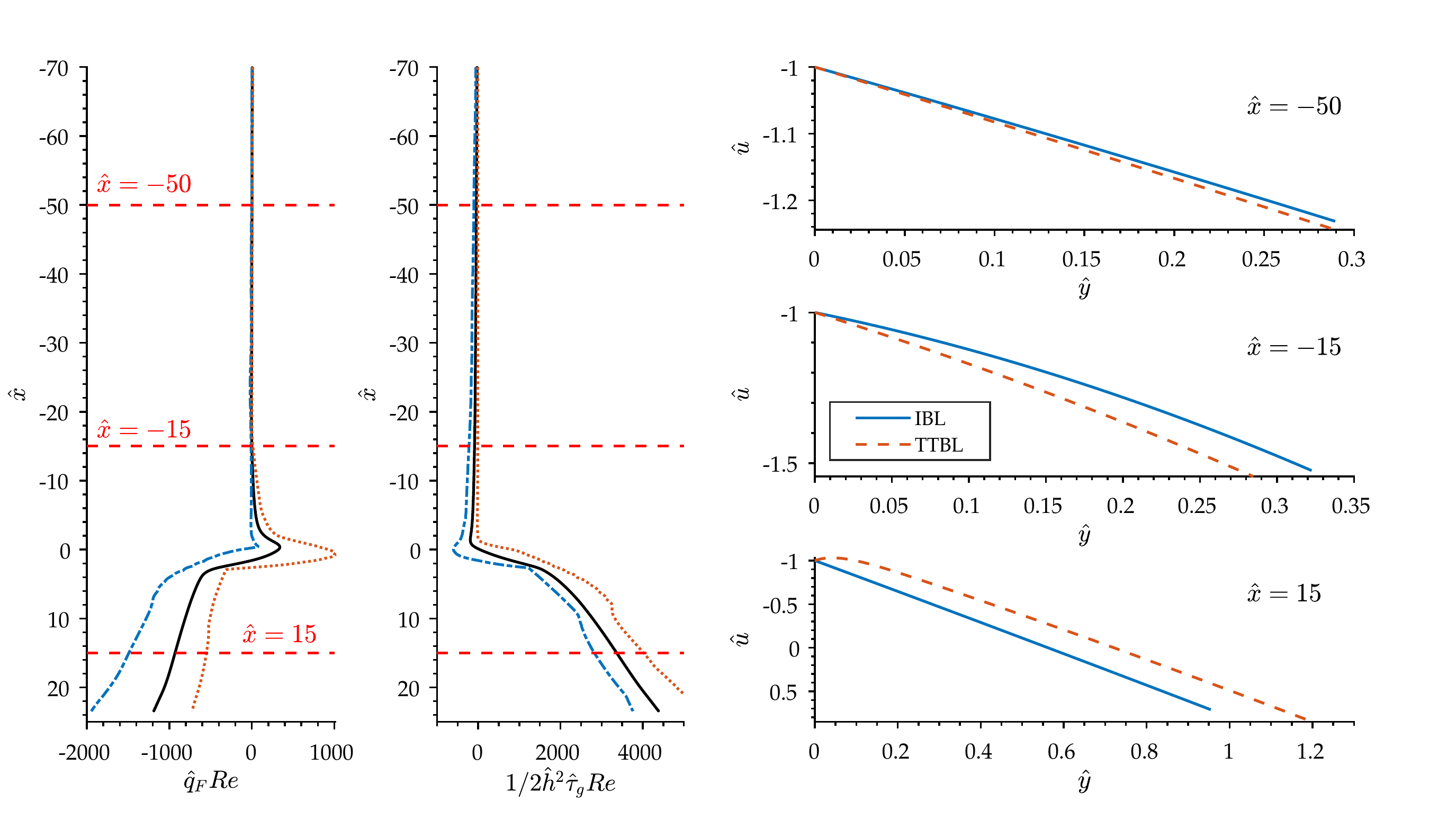}
\caption{The plots on the left shows the minima, mean and maximum distribution for the flow rate contributions $\hat{q}_F$ and $\frac{1}{2}\hat{h}^2\hat{\tau}_g$, multiplied by the global Reynolds $Re$: the absolute value of these quantities leads to the Reynolds numbers in \eqref{Reynoldss}. The plots on the right show the velocity profiles from the IBL and the TTBL under wave peaks in the locations $\hat{x}=-50,-15,15$. }
\label{PROFS_LAST}
\end{figure*}

Focusing on the influence of the turbulent modeling, figure \ref{PROFS_LAST} further analyzes the test case with $\hat{f}=0.04$ in the upper row of figure \ref{SNAPS_LF}. The first plot on the left shows the maxima, minima, and mean distribution of the term $\hat{q}_F Re$,  the absolute value of which controls the transition to turbulence. This term is negligible from $\hat{x}>15$, suggesting that a laminar model is appropriate to simulate the final coating flow. On the other hand, this term is large in the run-back flow and --most importantly-- negative: as described by \eqref{qF}, this is a consequence of the large and positive contribution of the shear stress ($1/2 \hat{h}^2\hat{\tau}_g Re$), depicted in the figure on the right. Mechanism A for the wave formation propagates the discrepancy in the runback flow towards the final coat.

Finally, the plots on the right of figure \ref{PROFS_LAST} show three instantaneous velocity profiles extracted on a wave peak located at $\hat{x}=[-50,-15,15]$ for both the IBL and the TTBL models. Far from the impact point, in the final coating, the velocity profile appears almost linear due to the thin thickness of the layer. Close to the impact point, in the run-back flow region, the velocity profile appears almost linear due to the strong influence of the gas shear stress, although the discrepancy between the two models increases. Finally, it is worth reporting that in this condition, the re-scaled Reynolds number is too high to allow for reconstructing the velocity profile in the WIBL model {as its first order corrections, designed for $\delta \ll 1$, become nonphysical. In particular, at large $\delta$, the coefficients $a_j$ with $j>0$ in table \ref{As} lead to extremely large contributions to the velocity profile. It appears thus surprising that the predicted film response is still closely matching the IBL, as the inconsistency in the large $a_j$'s is somewhat mitigated by the shear stress term in \eqref{LAST_E}}. To conclude, the mechanisms for wave formation revealed in this work, and their range of maximum receptivity are qualitatively independent of the models implemented.

\section{Conclusions}\label{Conclu}

We have presented an extension of classical low-dimensional models for falling liquid films to the jet wiping problem, tailoring accordingly the a Skhadov-like scaling. This process consists of using an impinging gas jet to control the thickness of a liquid film on a moving substrate and is characterized by an unstable interaction between the gas and the liquid film that limits the achievable coating uniformity.

The investigated integral models allow for simulating this complex interaction in industrially relevant conditions with minor computational costs and thus enable insights on the process dynamics that would otherwise not be possible using high fidelity simulations. The proposed models extend the modelling strategies commonly used in falling liquid film to a more complex scenario that includes the motion of the substrate and the presence of imposed pressure gradient and shear stress distribution --in this work simulating an unstable impinging jet. The extended models are the self-similar Integral Boundary Layer (IBL) and the Weighted Integral Boundary Layer (WIBL) were extensively described and framed along with the classic zero-order formulation encountered in the literature of the jet wiping process. Moreover, an extension of the IBL model, referred to as the Transition and Turbulent Boundary Layer (TTBL) model, has been proposed to account for the impact of turbulence in the liquid film response.

The numerical implementation of these models has been successfully validated via DNS simulations using the VOF method in OpenFoam, considering the simplified test case of a pulsating liquid film evolving along with a moving interface. These models were then used to study the response of the liquid film to various kinds of perturbation in the jet flow, including harmonic and non-harmonic oscillations and pulsations. The analysis of the relative influence of all the terms in the equations reveals that the nonlinear advection term dominates over a wide range of wiping conditions and frequency of the perturbation. {Moreover, by analyzing the wiping process in galvanizing conditions (using zinc) and in laboratory conditions (using water), it is shown that the Skhadov-like scaling reveals an interesting similarity of the frequency response of the liquid coat. In the simplest case of a film flowing over an upward-moving surface, the similarity between the two configurations applies to the entire film evolution.}

Two main mechanisms for the formation of waves in the coating film downstream of the wiping region were identified. The first mechanism, referred to as mechanism A, is inherently linked to the presence of a wiping meniscus and to the fact that, during its oscillation, the jet drives upwards liquid from the thicker region below the wiping point. The amplitude and shape of the coating waves produced by this mechanism were shown to be linked to the waveform of the jet oscillation. The second mechanism, referred to as mechanism B, is related to the local variation of the pressure gradient and shear stress, as a result of unsteadiness in the jet flow. The dimensionless transfer function for both mechanisms has been presented, and the region of highest receptivity of the film have been identified. In particular, while the liquid film behaves as a low pass filter against the mechanisms B, a region of strong receptivity to the mechanism A is found both upstream and downstream the wiping point in the range of dimensionless frequencies $\hat{f}=0.03-0.05$. In galvanizing conditions, these correspond to wavelength in the range $\lambda=25-35$ \SI{}{mm}, in good agreement with industrial observations.

Finally, the comparison between the IBL and WIBL methods shows good qualitative agreement at the highest wiping numbers $\Pi_g$, highest rescaled Reynolds number $\delta$ and highest perturbation frequency $\hat{f}$. On the other hand, the TTBL predicts a much larger thickness in the run-back flow. As mechanism A is triggered, this results in different wave shapes in the final coating. Nevertheless, these discrepancies do not alter the main results concerning the mechanisms on the undulation formation nor their range of large receptivity of the liquid film over the investigated wiping conditions. Although the interaction between liquid film and the impinging jet flow is characterized by coupling phenomena that do not fit in the simplified one-way coupling framework, the mechanisms revealed by this study certainly play an essential role in the stability of the jet wiping process.

\section*{Acknowledgement}

The authors gratefully acknowledge the financial support from ArcelorMittal and the essential contributions of Martin Buszyk and Davide Ninni. They took part in the development of the first IBL solver, implemented in the VKI software package BLEW (Boundary LayEr Wiping), during their short training program at the von Karman Institute. B. Scheid thanks the F.R.S.-FNRS for the financial support. Anne Gosset would like to thank the CESGA (Centro de Supercomputación de Galicia) for the computational resources.

\appendix

\section{From NS to  equation (1)}\label{A1}

The long-wave formulation is based on the assumption that the streamwise reference scale $[x]$ is $[x]\gg[h]$ and thus a film parameter $\varepsilon=[h]/[x]\ll1$ allows ordering the importance of all the terms.
The continuity equation, under the assumption of incompressibility, can be scaled as:

\begin{equation}
\Biggr\{ \frac{U_p}{[x]}\Bigg\} \partial_{\hat x} \hat{u}+\Biggr\{ \frac{\varepsilon U_p }{[h]}\Biggr\}\partial_{\hat y} \hat{v}=0\,, \end{equation} and thus yields \eqref{C} provided that $[v]=\varepsilon [u]=\varepsilon U_p$, having taken $[u]=U_p$. Taking $[t]=[x]/U_p$, the momentum equation in the streamwise coordinate can be scaled as:

\begin{equation}
\Biggr\{\frac{U_p^2}{[x]}\Biggr\} \Bigl(\partial_{\hat t}\hat{u}+ \hat {u}\, \partial_{\hat x} \hat{u} +  \hat {v}\partial_{\hat y} \hat{u}\Bigr)=-  \Biggr\{ \frac{[p]}{\rho_l [x]}\Biggr\} \partial_{\hat x}\, \hat{p}_l+\Biggr\{\nu_l \frac{U_p}{[x]^2} \Biggr\} \partial_{\hat x \hat x} \hat{u}+\Biggr\{\nu_l \frac{U_p}{[h]^2} \Biggr\} \partial_{\hat y\hat y} \hat{u} +g\,.
\end{equation}

Multiplying both sides by $[h]^2/\nu\,U_p$, taking the reference pressures $[p]$, thickness $[h]$ from table \ref{Scaling_Table} and neglecting terms of $\mathcal{O}(\varepsilon^2)$ yield \eqref{Mx}. The same procedure on the momentum equation along the cross-stream direction $y$ leads to \eqref{My}. The kinematic boundary conditions \eqref{BCKin} can be obtained scaling by $\varepsilon U_p$, while the full force balance at the interface scales as

\begin{equation}
\label{NN}
\bigl(\hat p_l -\hat p_g\bigr) \Bigl\{\rho_l g [x]\Bigr\}- 2\mu_l\,\mathbf{n}\cdot \mathbf{E}_l\cdot \mathbf{n}-2\,\frac {[h] \sigma_l}{[x]^2} \,\hat{\kappa}=0\,\quad \rm{in} \quad y=h\,,
\end{equation} where the second term accounts for the viscous term in the normal direction and involves the normal unit vector $\mathbf{n}=(-\partial_{ x} {h}\,,1)^T/\sqrt{1+(\partial_{x} h)^2}$ and the symmetric part of the rate of deformation tensor $\mathbf{E}_l=1/2\,\bigl(\nabla \mathbf{v}+\nabla \mathbf{v}^T\bigr)$, with $\nabla \mathbf{v}=(u,v)$ the velocity field; the third term accounts for the surface tension, involving the mean curvature $\hat \kappa=-1/2\,\nabla\cdot \hat{\mathbf{n}}$. Expanding the viscous term in \eqref{NN} yields:

\begin{equation}
\mathbf{n}\cdot\mathbf{E}_l\cdot \mathbf{n}=\frac{( \partial_y u + \partial_x v) \partial_x h- \partial_y v- \partial_x u (\partial_x h)^2 }{1+(\partial_x h)^2} \,\quad \mbox{in} \quad \hat{y}=\hat{h}\,.
\end{equation}

Scaling this expression and neglecting terms in $\mathcal{O}(\varepsilon^2)$, the contribution of normal viscous stresses reads

\begin{equation}
2\mu_l\,\frac{U_p}{[h]}\,{\hat {\mathbf{n}}\cdot\hat{\mathbf{E}}_l\cdot\hat{\mathbf{n}}}=2\mu_l\,\frac{U_p}{[h]}\,\varepsilon\,\Bigl ( \partial_{\hat y} \hat v-\partial_{\hat{x}}\hat{h} \, \partial_{\hat y}\hat u  \Bigr) \,\,\quad \mbox{in} \quad \hat{y}=\hat{h}\,.
\end{equation}

Introducing this result in \eqref{NN}, observing that at $\mathcal{O}(\varepsilon)$ the dimensionless curvature becomes $\hat{\kappa}=1/2\,\partial_{\hat x\hat x}\,\hat h$ and dividing by the reference pressure $[p]=\rho\,g\,[x]$ give

\begin{equation}
\label{N2}
\bigl(\hat p_l -\hat p_g\bigr)- \varepsilon^2\,{\hat{\mathbf{n}}\cdot\hat{\mathbf{E}}_l\cdot\hat{\mathbf{n}}}+\varepsilon^3\,\frac{l_c^2}{[h]^2} \,\partial_{\hat x \hat x}\hat{h}=0\,,
\end{equation} where $l_c=\sqrt{\sigma/\rho\,g}$ is the capillary length. It is from this equation that the choice of the streamwise length scale $[x]$ --hence film parameter $\varepsilon$-- is taken. Following the scaling approach proposed by \citet{Shkadov1971}, this choice is made such that the surface tension contribution remains of leading order. Therefore:

\begin{equation}
\varepsilon=\biggl(\frac{[h^2]}{l_c^2}\biggr)^{1/3}=\biggl(\frac{\mu\,U_p}{\sigma}\biggr)^{1/3}=Ca^{1/3}
\end{equation}

By construction, then, the contribution of elongational viscosity is neglected and \eqref{NN} reduces to \eqref{P_BC}. This approximation is valid for $Ca^{1/3}\ll1$ and the weight of the viscous term becomes $Ca^{2/3}$. Finally, concerning the force balance in the tangential direction, the full equation reads:

\begin{equation}
\hat{\mathbf{n}} \cdot\bigl ( 2\mu_l\mathbf{E}_l\bigr) \cdot\hat{\mathbf{t}} =\tau_g\Bigl |_h\,,\,\quad \rm{in} \quad \hat{y}=\hat{h}\,,
\end{equation} where $\hat{\mathbf{t}}=(1,\partial_x h)^T/\sqrt{1+(\partial_x h)^2}$ is the tangential unit vector. Expanding the matrix multiplication and dividing the result by $U_p/[h]$ the scaled form of this equation becomes:

\begin{equation}
\label{TT}
\bigl (1-\varepsilon^2(\partial_{\hat{x}} \hat h)^2\bigr )(\varepsilon^2 \partial_{\hat{x}} \hat v  +\partial_{\hat y} \hat u)+2 \varepsilon^2 \partial_{\hat x }\hat h \Bigl (\partial_{\hat y}\hat {v}-\partial_{\hat x} \hat u \Bigr )=\hat{\tau}_g\,\,\quad \textit{in} \quad \hat{y}=\hat{h}\,,
\end{equation} having introduced the reference shear stress $[\tau]=\mu\,[u]/[h]$.
To the leading order $\mathcal{O}(\varepsilon)$, this equation simplifies to \eqref{Tau_BC}.

\section{Details of the WIBL Model Derivation}\label{TABS}

After introducing the velocity profile \eqref{Vel_PROF}, the fourth-order polynomial in $\overline{y}=y/h$ in \eqref{GRAND_POLY} is obtained. Setting all the coefficients of the polynomial to zero the linear system in \eqref{System} is derived. The vector components on the RHS of \eqref{System}, denoted as $\mathcal{G}=\{G_0,\cdots,G_4\}$, is shown in table \ref{Gs}. Observe that all the variables in this and in the following tables are dimensionless and the $\hat{}$ are dropped to ease the notation. Moreover, the derivation is carried out with an arbitrary scaling of the strip velocity, such that $u(y=0)=-\gamma$.  This allows to retreive the model for the jet wiping if $\gamma=1$ and the classical models for falling liquid films if $\gamma=0$. 

Setting all the coefficients $G_i$ equal to zero, the solution of the system in \eqref{System} yields the coefficients of the expansion $\mathcal{A}=\{a_0,\dots a_4\}$ in table \ref{As}. These have a functional dependency on $a_0$ for all the terms weighted by the re-scaled Reynolds number $\delta=\varepsilon Re$.  Observe that in the falling film case, with $\gamma=0$ and $\partial_{\hat{x}}\hat{p}_g=\hat{\tau}_g=0$, the equations in tables \ref{Gs} and \ref{As} recovers equations (6.43) and (6.44) in \citealt{Kalliadasis2012}.

The coefficients in table \eqref{As} are introduced in the contribution $\hat{q}_F$ of the flow rate \eqref{Q}, obtaining \eqref{Q_0} from table \ref{Q_A}. From this, the first coefficient $a_0$ is isolated in \eqref{A_0NS}. Introducing this expression in the shear stress term \eqref{D_Tau} gives the results in \eqref{LAST_E}, having introduced $a_0= {3q}/{h}-{3}/{2}h\tau_g+3\gamma+\mathcal{O}(\varepsilon)$ inside the parentheses and truncating all the terms in $\mathcal{O}(\varepsilon^2)$. Equation \eqref{LAST_E} (which for $\gamma=1$ is \eqref{LAST_E2}) can be finally introduced in \eqref{I2} to close the WIBL model.

\begin{table*}
\begin{minipage}{1\textwidth}
\begin{align}
 G_0&=\partial_{ x  x  x}\,h-\partial_{ x}p_g+1 \\
  G_1&=\varepsilon Re\,\biggl[  \frac{ a_0 \,  \partial_{ t} h  }{h } -  \partial_{ t} a_0 - h \,  \partial_{ t} {{\tau}_g}+\gamma \biggl( h \,  \partial_{ x} {{\tau}_g}- \frac{ a_0 \,  \partial_{ x} h  }{h }+  \partial_{ x} a_0\biggr)\biggr]     \\
   G_2&=\varepsilon Re\,\biggl[ -\frac{ {{h }^{2}}\, {{\tau}_g}   \partial_{ x} {{\tau}_g}  }{2}-\frac{ a_0 \, h \,  \partial_{ x} {{\tau}_g}  }{2}+\frac{ a_0 \, {{\tau}_g}   \partial_{ x} h  }{2}+\frac{ {a^{2}_0}\,  \partial_{ x} h  }{2 h }-\frac{ a_0 \,  \partial_{ t} h  }{h }+\frac{  \partial_{ t} a_0  }{2}\\&\nonumber\frac{ h \, {{\tau}_g}\partial_{ x} a_0  }{2}-\frac{ a_0 \,  \partial_{ x} a_0  }{2}+\gamma\biggl(\frac{ a_0 \,  \partial_{ x} h  }{h } - \frac{  \partial_{ x} a_0  }{2}\biggr)\biggr]  \\
     G_3&=\varepsilon Re\,\biggl( -\frac{2  a_0 \, {{\tau}_g}  \partial_{ x} h  }{3}-\frac{2  {a^{2}_0}\, \partial_{ x} h  }{3 h }+\frac{ h \, {{\tau}_g}  \partial_{ x} a_0  }{3}+\frac{ a_0 \, \partial_{ x} a_0  }{3}\biggr)\\
     G_4&=\varepsilon Re\,\biggl(\frac{ {a^{2}_0}\,  \partial_{ x} h  }{6 h }-\frac{ a_0 \,  \partial_{ x} a_0  }{12}\biggr)
\end{align}
\hrule
\end{minipage}
\caption{$\mathcal{G}=\{G_0,\cdots,G_4\}$ vector components for the system in \eqref{System}.}
\label{Gs}
\end{table*}

\begin{table*}
\begin{minipage}{1\textwidth}
\begin{align}
    a_0&={h }^{2}\left ( 1+ \partial_{x x x} h  - \partial_{x} {p_g} \right ) +\varepsilon Re\Bigl[-\frac{ {{h }^{4}} {{\tau}_g}  \partial_{ x} {{\tau}_g}  }{6}-\frac{ a_0  {{h }^{3}}  \partial_{ x} {{\tau}_g}  }{6}-\frac{ {{h }^{3}}  \partial_{ t} {{\tau}_g}  }{2} -\frac{ {{h }^{2}}  \partial_{ t} a_0  }{3} +\\&\nonumber\frac{ {{a}_0^{2}} h   \partial_{ x} h  }{30}+\frac{ a_0  h   \partial_{ t} h  }{6}- \frac{ {{h }^{3}} {{\tau}_g}   \partial_{ x} a_0  }{12}-\frac{ a_0  {{h }^{2}}  \partial_{ x} a_0  }{10}+\gamma\biggl(\frac{ {{h }^{2}}  \partial_{ x} a_0  }{3}+\frac{ {{h }^{3}}  \partial_{ x} {{\tau}_g}  }{2}-\frac{a_0  h   \partial_{ x} h }{6}\biggr)\biggr]\\
  a_1&=\varepsilon Re \biggl[-\frac{ {{h }^{4}} {{\tau}_g} \partial_{ x} {{\tau}_g}  }{12}-\frac{ a_0  {{h }^{3}}  \partial_{ x} {{\tau}_g}  }{12}-\frac{ {{h }^{3}}  \partial_{ t} {{\tau}_g}  }{4}+\frac{ {a^{2}_0} h   \partial_{ x} h  }{60}-\frac{ {{h }^{2}}  \partial_{ t} a_0  }{6}-\frac{ {{h }^{3}} {{\tau}_g}   \partial_{ x} a_0  }{24}+\\&\nonumber\frac{ a_0  h   \partial_{ t} h  }{12}-\frac{ a_0  {{h }^{2}}  \partial_{ x} a_0  }{20}+\gamma\biggl(\frac{ {{h }^{2}}  \partial_{ x} a_0  }{6}+\frac{ {{h }^{3}}  \partial_{ x} {{\tau}_g}  }{4}-\frac{ a_0  h   \partial_{ x} h  }{12}\biggr)\biggr]   \\
   a_2&=\varepsilon Re\biggl[-\frac{ {{h }^{4}} {{\tau}_g}   \partial_{ x} {{\tau}_g}  }{18}-\frac{ a_0  {{h }^{3}}  \partial_{ x} {{\tau}_g}  }{18}+\frac{ {a^{2}_0} h   \partial_{ x} h  }{90}+\frac{ {{h }^{2}}  \partial_{ t} a_0  }{18}-\frac{ a_0  h   \partial_{ t} h  }{9}-\frac{ {{h }^{3}} {{\tau}_g} \partial_{ x} a_0  }{36}\\&\nonumber-\frac{ a_0  {{h }^{2}}  \partial_{ x} a_0  }{30}+\gamma\biggr(\frac{ a_0  h   \partial_{ x} h  }{9}-\frac{ {{h }^{2}}  \partial_{ x} a_0  }{18}\biggr)\biggr]\\
     a_3&=\varepsilon Re \left(-\frac{ {{\tau}_g} a_0  {{h }^{2}}  \partial_{x} h  }{24}-\frac{ {a^{2}_0} h   \partial_{x} h  }{30}+\frac{ {{\tau}_g} {{h }^{3}}  \partial_{x} a_0  }{48}+\frac{ a_0  {{h }^{2}}  \partial_{x} a_0  }{60}\right )\\
     a_4&=\varepsilon Re \left(\frac{ {a^{2}_0} h   \partial_{x} h  }{150}-\frac{ a_0  {{h }^{2}}  \partial_{x} a_0 }{300}\right )
\end{align}
\hrule
\end{minipage}
\caption{ Full solution for the coefficients $\mathcal{A}=\{a_0,\dots a_4\}$ in \eqref{Vel_PROF}, obtained as vector $\mathcal{A}=\Gamma^{-1}\,\mathcal{G}$ in \eqref{System}.}
\label{As}
\end{table*}

\begin{table*}
\begin{minipage}{0.99\textwidth}
\begin{align}
   \label{Q_0}
   q&=\frac{{{h }^{2}} {{\tau}_g} }{2}+\frac{a_0 h }{3}-\gamma h+\varepsilon Re \biggl [-\frac{7  {{h }^{5}} {{\tau}_g}   \partial_{ x} {{\tau}_g}  }{360}-\frac{7  a_0  {{h }^{4}}  \partial_{ x} {{\tau}_g}  }{360}-\frac{ {{h }^{4}}  \partial_{t} {{\tau}_g}  }{24}-\frac{ {{h }^{3}}  \partial_{t} a_0  }{45} \\&\nonumber+\frac{ {{a_0}^{2}} {{h}^{2}}  \partial_{ x} h }{504}+ \frac{ a_0  {{h }^{2}}  \partial_{t} h  }{360}-\frac{ {{h }^{4}} {{\tau}_g}   \partial_{ x} a_0  }{120}-\frac{3  a_0  {{h }^{3}}  \partial_{ x} a_0  }{280}- \frac{ a_0  {{h }^{3}} {{\tau}_g}   \partial_{ x} h  }{360}\\&\nonumber+\gamma\biggl( \frac{ {{h }^{3}}  \partial_{ x} a_0}{45}- \frac{ a_0  {{h }^{2}}  \partial_{ x} h  }{360}+\frac{ {{h }^{4}}  \partial_{ x} {{\tau}_g}  }{24}\biggr)\biggr]\\
   \label{A_0NS}
      a_0&=\frac{3q}{h}-\frac{3}{2}h\tau_g+3\gamma+\varepsilon Re \biggl [\frac{7  {{h }^{4}} {{\tau}_g}   \partial_{ x} {{\tau}_g}  }{120}+\frac{7  a_0  {{h }^{3}}  \partial_{ x} {{\tau}_g}  }{120}+\frac{ {{h }^{3}}  \partial_{ t} {{\tau}_g}  }{8}+\frac{ {{h }^{2}}  \partial_{ t} a_0  }{15}\\&\nonumber+\frac{ a_0  {{h }^{2}} {{\tau}_g}   \partial_{ x} h  }{120}-\frac{ {{a^{2}_0}} h   \partial_{ x} h  }{168}-\frac{ a_0  h   \partial_{ t} h  }{120}+\frac{ {{h }^{3}} {{\tau}_g}   \partial_{ x} a_0  }{40}+\frac{9  a_0  {{h}^{2}}  \partial_{ x} a_0  }{280}\\&\nonumber+\gamma\biggl(\frac{ a_0  h   \partial_{ x} h  }{120}-\frac{ {{h }^{2}}  \partial_{ x} a_0  }{15}-\frac{ {{h }^{3}}  \partial_{ x} {{\tau}_g}  }{8}\biggr)\biggr]\\
      \label{LAST_E}
      \Delta \tau&=\frac{3}{2}\tau_g-\frac{3\,q}{h^2}-\frac{3\gamma}{h}+\varepsilon Re \biggl [-\frac{19  {{h }^{3}}\, {{\tau}_g}   \partial_{x} {{\tau}_g}  }{3360}-\frac{17  h \, q \,  \partial_{x} {{\tau}_g}  }{560}-\frac{ {{h }^{2}}\,  \partial_{t} {{\tau}_g}  }{40}+\frac{12  {{q }^{2}}\,  \partial_{x} h  }{35 {{h }^{2}}}\\&\nonumber-\frac{ h \, {{\tau}_g}   \partial_{x} q  }{56}-\frac{18  q \,  \partial_{x} q  }{35 h }-\frac{  \partial_{t} q  }{5}-\frac{ {{h }^{2}}\, {{{{\tau}_g}}^{2}}  \partial_{x} h}{112}-\frac{ q \, {{\tau}_g}   \partial_{x} h  }{280}\\&\nonumber+\gamma\biggl(\frac{6  q \,  \partial_{x} h  }{35 h }-\frac{4   \partial_{x} q  }{35}-\frac{3  h \, {{\tau}_g} \partial_{x} h  }{140}- \frac{3  {{h }^{2}}\,  \partial_{x} {{\tau}_g}  }{560}+\gamma\frac{  \partial_{x} h  }{35}\biggr) \biggr]
\end{align}
\hrule
\end{minipage}
\caption{ Full expression for $q$ using the coefficients $\mathcal{A}=\{a_0,\dots a_4\}$ in table \ref{As} in \eqref{Q}, together with the resulting expression of $a_0$ and the final result on the shear stress term $\Delta \tau$ for the WIBL model of the jet wiping process.}
\label{Q_A}
\end{table*}

To retrieve the WIBL model for a falling liquid film (see \citealt{Kalliadasis2012}, eq. 6.51), it suffices to introduce $\gamma=0$ and $\partial_{\hat{x}}\hat{p}_g=\hat{\tau}_g=0$ in \eqref{LAST_E} and observe that in this case the advection term in \eqref{ADV_eps} simplifies to $\mathcal{F}=6 q^2/(5h)$. Then, \eqref{I2} becomes  (6.51) in \cite{Kalliadasis2012}. 

\section{Implemented Numerical Schemes}\label{A_N}

The high order fluxes $\mathbf{F}^{H}$ are derived from the two step Lax-Wendroff scheme (LxW) while the low order fluxes $\mathbf{F}^{L}$ are taken from the two-step Lax Friedrich scheme (LxF). The flux terms in the high order scheme are

\begin{equation}
\mathbf{F}^{H} = 
\begin{cases} \mathbf{F}^{+} &=\mathbf{F}(\mathbf V_{i}^{k}\,,\mathbf V_{i+1}^{k})= \mathbf{F}(\mathbf V_{i+{1}/{2}}^{k+{1}/{2}})
  \\[5pt]
\mathbf{F}^{-}&=\mathbf{F}(\mathbf V_{i}^{k}\,,\mathbf V_{i-1}^{k})= \mathbf{F}(\mathbf V_{i-{1}/{2}}^{k+{1}/{2}})
\end{cases}
\end{equation}

The flux terms in the lower order scheme adds diffusive terms and reads

\begin{equation}
\mathbf{F}^{L} = 
\begin{cases} \mathbf{F}^{+} &=\mathbf{F}(\mathbf V_{i}^{k}\,,\mathbf V_{i+1}^{k})=\frac 1 2 \mathbf{F}(\mathbf V_{i+{1}/{2}}^{k+{1}/{2}})+\frac{\Delta x}{2 \Delta t} (\mathbf V_{i+1/2}^{k+1/2}-\mathbf V_{i}^{k})
 \\[5pt]
\mathbf{F}^{-}&=\mathbf{F}(\mathbf V_{i}^{k}\,,\mathbf V_{i-1}^{k})=\frac 1 2 \mathbf{F}(\mathbf V_{i-{1}/{2}}^{k+{1}/{2}})+\frac{\Delta x}{2 \Delta t} (\mathbf V_{i}^{k}-\mathbf V_{i-1/2}^{k+1/2})\\
\end{cases}
\end{equation}

Both schemes use the midpoint solutions in time:

\begin{subequations}
 \label{Midpoints}
\begin{equation}
 \mathbf V_{i-\frac{1}{2}}^{k+\frac{1}{2}} = \frac{1}{2}( \mathbf V_{i}^{k}+  \mathbf V_{i-1}^{k})+\frac{\Delta t}{2\Delta x}\Biggl [
\mathbf{F}( \mathbf V_{i}^{k})-\mathbf{F}( \mathbf V_{i-1}^{k})\Biggr]
 + \frac{1}{2}\Delta  t \mathbf{S}_{i+\frac{1}{2}}^{k}
\end{equation}
\begin{equation}
 \mathbf V_{i-\frac{1}{2}}^{k+\frac{1}{2}} = \frac{1}{2}( \mathbf V_{i}^{k}-  \mathbf V_{i-1}^{k})+\frac{\Delta t}{2\Delta x}\Biggl [
\mathbb{F}( \mathbf V_{i}^{k})-\mathbf{F}( \mathbf V_{i-1}^{k})\Biggr]
 + \frac{1}{2}\Delta  t \mathbf{S}_{i-\frac{1}{2}}^{k}
\end{equation}
\end{subequations}

\bibliographystyle{jfm}
\bibliography{Mendez_et_al_Jet_wiping_2020}

\end{document}